\pdfoutput=1
\documentclass[aps,prd,twocolumn,letterpaper,superscriptaddress,floatfix,showpacs]{revtex4}

\usepackage{epsfig}
\usepackage{rotating}
\usepackage{float}
\usepackage{floatflt}
\usepackage{multirow}
\usepackage{bigstrut}
\usepackage{amssymb,amsmath}
\usepackage{slashbox}
\usepackage{graphicx}
\usepackage{xspace}
\usepackage{subfigure}

\usepackage[displaymath]{lineno}

\usepackage[overload]{textcase}
\usepackage{textcomp}

\usepackage{pstricks}

\usepackage{verbatim}
\usepackage{colortab}

\usepackage{url}

\usepackage{fncylab}

\usepackage[sort&compress]{natbib}

\usepackage{supertabular}

\begin{document}
\graphicspath{{./Pdf/}}



\newcommand\munit{\ensuremath{\mathrm{GeV}/c^2}\xspace}
\newcommand\punit{\ensuremath{\mathrm{GeV}/c}\xspace}
\newcommand\mgunit{\ensuremath{\mathrm{eV}/c^2}\xspace}
\newcommand\mkunit{\ensuremath{\mathrm{keV}/c^2}\xspace}
\newcommand\lumunit{\ensuremath{{\rm cm^{-2}\ s^{-1}}}\xspace}
\newcommand\dedxunit{\ensuremath{\frac{\rm keV}{\rm cm}}\xspace}

\newcommand{\et}{\ensuremath{E\!_T}\xspace}
\newcommand{\met}{\ensuremath{E\!\!\!\!/_T}\xspace}
\newcommand{\mett}{\mbox{${E\!\!\!\!/_T}$}}
\newcommand{\metx}{\mbox{${E\!\!\!\!/_{T}~\!\!\!\!^{x}}$}}  
\newcommand{\mety}{\mbox{${E\!\!\!\!/_{T}~\!\!\!\!^{y}}$}}  
\newcommand{\mmetx}{\mbox{Mean $(E\!\!\!\!/_{T}~\!\!\!\!^{x})$}}
\newcommand{\mmety}{\mbox{Mean $(E\!\!\!\!/_{T}~\!\!\!\!^{y})$}}
\newcommand{\mpt}{\ensuremath{p\!\!\!\!/_T}\xspace}
\newcommand{\summet}{\mbox{$\sigma(E\!\!\!\!/_T)$}}
\newcommand{\summetx}{\mbox{$\sigma(E\!\!\!\!/_{T}~\!\!\!\!^{x})$}}
\newcommand{\summety}{\mbox{$\sigma(E\!\!\!\!/_{T}~\!\!\!\!^{y})$}}
\newcommand{\sumet}{\ensuremath{\Sigma {\rm E_{T}}}\xspace}
\newcommand{\sumpt}{\ensuremath{\Sigma {p_{T}}}\xspace}
\newcommand{\sumpttrk}{\ensuremath{\Sigma {p_{\rm T}^{\mathrm{trk}}}}\xspace}
\newcommand{\dedx}{\ensuremath{\frac{\mathrm{d}E}{\mathrm{d}x}}\xspace}

\newcommand{\chisq}{\ensuremath{\chi^{2}}\xspace}
\newcommand{\mN}{\ensuremath{m_{
{\none}}}\xspace}
\newcommand{\tauN}{\ensuremath{\tau_{
{\none}}}\xspace}

\newcommand{\tcorr}{\ensuremath{t_{\mathrm{corr}}}\xspace}
\newcommand{\tarr}{\ensuremath{t^{\gamma}_{\mathrm{corr}}}\xspace}

\newcommand{\etc}{\mbox{$\Sigma {\rm E_{T}^{Corrected}}$}}
\newcommand{\setc}{\mbox{$\sqrt{\Sigma {\rm E_{T}^{Correc
ted}}}$}}
\newcommand{\Zee}{\mbox{$Z \rightarrow ee$}\xspace}
\newcommand{\Wjets}{\mbox{$W+\mathrm{jets}$}\xspace}
\newcommand{\Zmumu}{\mbox{$Z \rightarrow \mu\mu$}\xspace}
\newcommand{\Wenu}{\ensuremath{W \rightarrow e\nu}\xspace}
\newcommand{\Wenuj}{\ensuremath{W \rightarrow e\nu+\mathrm{jets}}\xspace}
\newcommand{\Wnotrack}{{\tt W\_NOTRACK}\xspace}
\newcommand{\Metpem}{MET\_PEM\xspace}
\newcommand{\Zeg}{Z \rightarrow e\gamma}

\newcommand{\gmetjets}{\ensuremath{\gamma+\mathrm{jet}+\met}\xspace}
\newcommand{\gmetjet}{\ensuremath{\gamma+\mathrm{jet}+\met}\xspace}
\newcommand{\ggmet}{\ensuremath{\gamma\gamma+\met}\xspace}
\newcommand{\gmet}{\ensuremath{\gamma+\met}\xspace}
\newcommand{\eeggmet}{\ensuremath{ee\gamma\gamma\met}\xspace}

\newcommand{\gt}{\mbox{$>$}}
\newcommand{\lt}{\mbox{$<$}}

\newcommand{\gev} {\rm \,GeV}
\newcommand{\gevt} {\rm \,GeV^2}
\newcommand{\ipb}{\mbox{${\rm pb}^{-1}$}}
\newcommand{\dphi}{\ensuremath{\Delta\phi(\met,\rm{jet})}\xspace}
\newcommand{\dphicosm}{\ensuremath{\Delta\phi(\mu\mathrm{-stub},\gamma)}\xspace}

\newcommand{\pt}{\ensuremath{p_{_T}}\xspace}
\newcommand{\tzero}{\ensuremath{t_{0}}\xspace}
\newcommand{\zzero}{\ensuremath{z_{0}}\xspace}
\newcommand{\etg}{\ensuremath{E_{T}^{\gamma}}}
\newcommand{\etjet}{\ensuremath{E_{T}^{\mathrm{jet}}}}
\newcommand{\etajet}{\ensuremath{\eta^{\mathrm{jet}}}}

\newcommand{\chargino}{\mbox{$\widetilde{\chi}_{1}^{\pm}$}}
\newcommand{\ncl}{\mbox{$\rm N_{95\%}$}}
\newcommand{\sigexp}{\ensuremath{\sigma^{\mathrm{exp}}_{95}}\xspace}
\newcommand{\sigobs}{\ensuremath{\sigma^{\mathrm{obs}}_{95}}\xspace}
\newcommand{\sigmc}{\ensuremath{\sigma_{\mathrm{Signal\ MC}}}\xspace}
\newcommand{\sigprod}{\ensuremath{\sigma_{\mathrm{prod}}}\xspace}

\newcommand{\grav}{\ensuremath{\widetilde{G}}\xspace}
\newcommand{\none}{\ensuremath{\widetilde{\chi}_1^0}\xspace}
\newcommand{\ntwo}{\ensuremath{\widetilde{\chi}_2^0}\xspace}
\newcommand{\conep}{\ensuremath{\widetilde{\chi}_1^{+}}\xspace}
\newcommand{\conem}{\ensuremath{\widetilde{\chi}_1^{-}}\xspace}
\newcommand{\conepm}{\ensuremath{\widetilde{\chi}_1^{\pm}}\xspace}
\newcommand{\conemp}{\ensuremath{\widetilde{\chi}_1^{\mp}}\xspace}
\newcommand{\stau}{\ensuremath{\widetilde{\tau}}\xspace}
\newcommand{\decay}{\mbox{$p\bar{p}\rightarrow X\rightarrow \none\none$}\xspace}
\newcommand{\nonetogG}{\ensuremath{\none\rightarrow\gamma\grav}\xspace}
\newcommand{\Mm}{M_{m}}
\newcommand{\etal}{{\em et al.}\xspace}

\newcommand{\ag}{\ensuremath{\alpha}\xspace}
\newcommand{\bg}{\ensuremath{\beta}\xspace}
\newcommand{\dg}{\ensuremath{^{\circ}}\xspace}
\newcommand{\rphi}{$(r,\phi)$\xspace}
\newcommand{\rz}{$r,z$}

\newcommand{\mum}{\ensuremath{\mu{\rm m}}\xspace}
\newcommand{\mus}{\ensuremath{\mu{\rm s}}\xspace}

\def\bi{\begin{itemize}}
\def\ei{\end{itemize}}
\def\bc{\begin{center}}
\def\ec{\end{center}}
\def\and{\/\mbox{and}}
\def\egg{e\gamma\gamma}
\def\egga{e\gamma_{13}\gamma_{13}}
\def\eggb{e\gamma_{25}\gamma_{25}}
\def\eg{e\gamma}
\def\g{\gamma}
\def\mgg{\mu\gamma\gamma}
\def\mgga{\mu\gamma_{13}\gamma_{13}}
\def\mggb{\mu\gamma_{25}\gamma_{25}}
\def\mga{\mu\gamma_{13}}
\def\mgb{\mu\gamma_{25}}
\def\mg{\mu\gamma}
\def\m{\mu}
\def\etog{e\rightarrow \gamma}
\def\eetoeg{ee\rightarrow e\gamma}
\def\etoj{e\rightarrow jet}

\def\bvec#1{\vec{\bf #1}}
\newcommand{\rd}{{\rm d}}
\def\invpb{\mbox{pb$^{-1}$}\xspace}
\def\invfb{\mbox{fb$^{-1}$}\xspace}
\def\invnb{\mbox{nb$^{-1}$}\xspace}

\def\pythia{{\sc pythia}\xspace}
\def\prospino{{\sc prospino2}\xspace}

\def\Journal#1#2#3#4{{#1}\textbf{~#2}, #4 (#3)}
\def\PrePrint#1{\mbox{hep-ph/#1}}
\def\PRL{\rm Phys. Rev. Lett.}
\def\PRD{{\rm Phys. Rev.} D}

\newcommand{\pbar}{\ensuremath{\overline{p}}}
\newcommand{\ppbar}{\ensuremath{p\overline{p}}\xspace}
\newcommand{\cdf}{CDF~II}
\newcommand{\vev}{\text{\it vev}}
\newcommand{\sign}{\ensuremath{\mathrm{sign}}}

\newcommand{\tevE}{\ensuremath{\sqrt{s} = 1.96~\TeV}}
\newcommand{\SORaby}{\ensuremath{\mathrm{MSO}_{10}\mathrm{SM}}}

\affiliation{Institute of Physics, Academia Sinica, Taipei, Taiwan 11529, Republic of China} 
\affiliation{Argonne National Laboratory, Argonne, Illinois 60439} 
\affiliation{University of Athens, 157 71 Athens, Greece} 
\affiliation{Institut de Fisica d'Altes Energies, Universitat Autonoma de Barcelona, E-08193, Bellaterra (Barcelona), Spain} 
\affiliation{Baylor University, Waco, Texas  76798} 
\affiliation{Istituto Nazionale di Fisica Nucleare Bologna, $^t$University of Bologna, I-40127 Bologna, Italy} 
\affiliation{Brandeis University, Waltham, Massachusetts 02254} 
\affiliation{University of California, Davis, Davis, California  95616} 
\affiliation{University of California, Los Angeles, Los Angeles, California  90024} 
\affiliation{University of California, San Diego, La Jolla, California  92093} 
\affiliation{University of California, Santa Barbara, Santa Barbara, California 93106} 
\affiliation{Instituto de Fisica de Cantabria, CSIC-University of Cantabria, 39005 Santander, Spain} 
\affiliation{Carnegie Mellon University, Pittsburgh, PA  15213} 
\affiliation{Enrico Fermi Institute, University of Chicago, Chicago, Illinois 60637} 
\affiliation{Comenius University, 842 48 Bratislava, Slovakia; Institute of Experimental Physics, 040 01 Kosice, Slovakia} 
\affiliation{Joint Institute for Nuclear Research, RU-141980 Dubna, Russia} 
\affiliation{Duke University, Durham, North Carolina  27708} 
\affiliation{Fermi National Accelerator Laboratory, Batavia, Illinois 60510} 
\affiliation{University of Florida, Gainesville, Florida  32611} 
\affiliation{Laboratori Nazionali di Frascati, Istituto Nazionale di Fisica Nucleare, I-00044 Frascati, Italy} 
\affiliation{University of Geneva, CH-1211 Geneva 4, Switzerland} 
\affiliation{Glasgow University, Glasgow G12 8QQ, United Kingdom} 
\affiliation{Harvard University, Cambridge, Massachusetts 02138} 
\affiliation{Division of High Energy Physics, Department of Physics, University of Helsinki and Helsinki Institute of Physics, FIN-00014, Helsinki, Finland} 
\affiliation{University of Illinois, Urbana, Illinois 61801} 
\affiliation{The Johns Hopkins University, Baltimore, Maryland 21218} 
\affiliation{Institut f\"{u}r Experimentelle Kernphysik, Universit\"{a}t Karlsruhe, 76128 Karlsruhe, Germany} 
\affiliation{Center for High Energy Physics: Kyungpook National University, Daegu 702-701, Korea; Seoul National University, Seoul 151-742, Korea; Sungkyunkwan University, Suwon 440-746, Korea; Korea Institute of Science and Technology Information, Daejeon, 305-806, Korea; Chonnam National University, Gwangju, 500-757, Korea} 
\affiliation{Ernest Orlando Lawrence Berkeley National Laboratory, Berkeley, California 94720} 
\affiliation{University of Liverpool, Liverpool L69 7ZE, United Kingdom} 
\affiliation{University College London, London WC1E 6BT, United Kingdom} 
\affiliation{Centro de Investigaciones Energeticas Medioambientales y Tecnologicas, E-28040 Madrid, Spain} 
\affiliation{Massachusetts Institute of Technology, Cambridge, Massachusetts  02139} 
\affiliation{Institute of Particle Physics: McGill University, Montr\'{e}al, Canada H3A~2T8; and University of Toronto, Toronto, Canada M5S~1A7} 
\affiliation{University of Michigan, Ann Arbor, Michigan 48109} 
\affiliation{Michigan State University, East Lansing, Michigan  48824}
\affiliation{Institution for Theoretical and Experimental Physics, ITEP, Moscow 117259, Russia} 
\affiliation{University of New Mexico, Albuquerque, New Mexico 87131} 
\affiliation{Northwestern University, Evanston, Illinois  60208} 
\affiliation{The Ohio State University, Columbus, Ohio  43210} 
\affiliation{Okayama University, Okayama 700-8530, Japan} 
\affiliation{Osaka City University, Osaka 588, Japan} 
\affiliation{University of Oxford, Oxford OX1 3RH, United Kingdom} 
\affiliation{Istituto Nazionale di Fisica Nucleare, Sezione di Padova-Trento, $^u$University of Padova, I-35131 Padova, Italy} 
\affiliation{LPNHE, Universite Pierre et Marie Curie/IN2P3-CNRS, UMR7585, Paris, F-75252 France} 
\affiliation{University of Pennsylvania, Philadelphia, Pennsylvania 19104} 
\affiliation{Istituto Nazionale di Fisica Nucleare Pisa, $^q$University of Pisa, $^r$University of Siena and $^s$Scuola Normale Superiore, I-56127 Pisa, Italy} 
\affiliation{University of Pittsburgh, Pittsburgh, Pennsylvania 15260} 
\affiliation{Purdue University, West Lafayette, Indiana 47907} 
\affiliation{University of Rochester, Rochester, New York 14627} 
\affiliation{The Rockefeller University, New York, New York 10021} 

\affiliation{Istituto Nazionale di Fisica Nucleare, Sezione di Roma 1, $^v$Sapienza Universit\`{a} di Roma, I-00185 Roma, Italy} 

\affiliation{Rutgers University, Piscataway, New Jersey 08855} 
\affiliation{Texas A\&M University, College Station, Texas 77843} 
\affiliation{Istituto Nazionale di Fisica Nucleare Trieste/\ Udine, $^w$University of Trieste/\ Udine, Italy} 
\affiliation{University of Tsukuba, Tsukuba, Ibaraki 305, Japan} 
\affiliation{Tufts University, Medford, Massachusetts 02155} 
\affiliation{Waseda University, Tokyo 169, Japan} 
\affiliation{Wayne State University, Detroit, Michigan  48201} 
\affiliation{University of Wisconsin, Madison, Wisconsin 53706} 
\affiliation{Yale University, New Haven, Connecticut 06520} 
\author{T.~Aaltonen}
\affiliation{Division of High Energy Physics, Department of Physics, University of Helsinki and Helsinki Institute of Physics, FIN-00014, Helsinki, Finland}
\author{J.~Adelman}
\affiliation{Enrico Fermi Institute, University of Chicago, Chicago, Illinois 60637}
\author{T.~Akimoto}
\affiliation{University of Tsukuba, Tsukuba, Ibaraki 305, Japan}
\author{M.G.~Albrow}
\affiliation{Fermi National Accelerator Laboratory, Batavia, Illinois 60510}
\author{B.~\'{A}lvarez~Gonz\'{a}lez}
\affiliation{Instituto de Fisica de Cantabria, CSIC-University of Cantabria, 39005 Santander, Spain}
\author{S.~Amerio$^u$}
\affiliation{Istituto Nazionale di Fisica Nucleare, Sezione di Padova-Trento, $^u$University of Padova, I-35131 Padova, Italy} 

\author{D.~Amidei}
\affiliation{University of Michigan, Ann Arbor, Michigan 48109}
\author{A.~Anastassov}
\affiliation{Northwestern University, Evanston, Illinois  60208}
\author{A.~Annovi}
\affiliation{Laboratori Nazionali di Frascati, Istituto Nazionale di Fisica Nucleare, I-00044 Frascati, Italy}
\author{J.~Antos}
\affiliation{Comenius University, 842 48 Bratislava, Slovakia; Institute of Experimental Physics, 040 01 Kosice, Slovakia}
% ...

\collaboration{CDF Collaboration\footnote{With visitors from $^a$Universiteit Antwerpen, B-2610 Antwerp, Belgium, 
$^b$Chinese Academy of Sciences, Beijing 100864, China, 
$^c$University of Bristol, Bristol BS8 1TL, United Kingdom, 
$^d$University of California Irvine, Irvine, CA  92697, 
$^e$University of California Santa Cruz, Santa Cruz, CA  95064, 
$^f$Cornell University, Ithaca, NY  14853, 
$^g$University of Cyprus, Nicosia CY-1678, Cyprus, 
$^h$University College Dublin, Dublin 4, Ireland, 
$^i$University of Edinburgh, Edinburgh EH9 3JZ, United Kingdom, 
$^j$Universidad Iberoamericana, Mexico D.F., Mexico, 
$^k$University of Manchester, Manchester M13 9PL, England, 
$^l$Nagasaki Institute of Applied Science, Nagasaki, Japan, 
$^m$University de Oviedo, E-33007 Oviedo, Spain, 
$^n$Queen Mary, University of London, London, E1 4NS, England, 
$^o$Texas Tech University, Lubbock, TX  79409, 
$^p$IFIC(CSIC-Universitat de Valencia), 46071 Valencia, Spain,
$^x$Royal Society of Edinburgh/Scottish Executive Support Research Fellow, 
}}
\noaffiliation







\pagenumbering{roman}




\title {Search for Heavy, Long-Lived Neutralinos that Decay to Photons at CDF II Using Photon Timing}


\begin{abstract}

We present the results of the first hadron collider search for heavy, 
 long-lived neutralinos that decay via \nonetogG in gauge-mediated supersymmetry
 breaking models. 
Using an integrated luminosity of $570\pm34$~\invpb of $p\bar{p}$ collisions at $\sqrt{s}=1.96$~TeV, we select  
$\gamma$+jet+missing transverse energy candidate events based on
the arrival time of a high-energy photon at the electromagnetic
calorimeter as measured with a timing system that was recently
installed on the CDF~II detector. We find 2~events, 
consistent with the background estimate of 1.3$\pm$0.7 events. 
While our search
  strategy does not rely on model-specific dynamics, we set cross
  section limits 
 and place the
  world-best 95\% C.L. lower limit on the \none\ mass of 101~\munit at
  \tauN~=~5~ns.
\end {abstract}

\maketitle

\pagestyle{headings}
\setlength{\headheight}{36pt}


\pagenumbering{arabic}
\setlength{\headheight}{12pt}
\pagestyle{myheadings}



\long\def\symbolfootnote[#1]#2{\begingroup%
\def\thefootnote{\fnsymbol{footnote}}\footnote[#1]{#2}\endgroup} 

\section{Introduction}
\label{sec:intro}
Models of gauge-mediated supersymmetry (SUSY) breaking (GMSB)~\cite{gmsb2} are 
attractive for several reasons.  Theoretically they  solve the ``naturalness 
problem''~\cite{martin} and provide a low mass (warm) dark matter candidate~\cite{ref2}. 
 From an experimental standpoint they provide a natural 
explanation for the observation of an \eeggmet~\cite{eeggmet,rapdef} candidate event by the CDF 
experiment during Run~I at the Fermilab
Tevatron.  In particular, the photon ($\gamma$)
and missing transverse energy (\met) can be produced by the decay of
the lightest neutralino (\none) into a photon and a weakly 
interacting, stable gravitino (\grav).  While much 
attention has been given to prompt \none$\rightarrow\gamma$\grav decays, 
versions of the model that take into account cosmological constraints favor a 
\grav  with \mkunit mass and a \none with a lifetime that is on the order of 
nanoseconds or more~\cite{feng2}. 

Here we describe 
in detail~\cite{pwt} the first search for heavy,
long-lived neutralinos using photon timing at a hadron collider in 
 the \gmetjet final state where we require at least one jet and at least one photon. 
 The data comprise an integrated luminosity of 570$\pm$34~\invpb\ of $p{\bar p}$ collisions at
$\sqrt{s}=1.96$~TeV from the Tevatron collected with the CDF~II
detector~\cite{CDFII}.  Previous searches for sub-nanosecond ~\cite{ggmet,lep} and  nanosecond-lifetime~\cite{lep} 
 \none$\rightarrow\gamma$\grav decays using non-timing techniques have 
 yielded null results.  The present results extend the sensitivity
 to larger \none lifetimes and masses. 

The structure of this paper is as follows:~the remainder of this section provides a more detailed motivation for the search 
and describes 
the CDF detector, in particular the recently installed timing
system on the electromagnetic calorimeters (the ``EMTiming'' system) that is used
to measure the time of arrival of photons.
Section~\ref{sec:photonid} describes how photons
from heavy, long-lived particles would interact with the detector and how the standard  
identification criteria for prompt photons are modified 
to keep the identification efficiency high for delayed photons.  
The section further describes the 
photon timing measurement.  We describe the data sample in 
Section~\ref{sec:selection} and discuss the event  pre-selection criteria.
 Section~\ref{sec:bkgs} describes the various background sources as well
as the methods of estimating the rate at which they populate the signal region.  After a description and estimation of the acceptance for GMSB events in Section~\ref{sec:acc_eff},  we continue in Section~\ref{sec:optimization}  with a description of the optimization procedure and the expected sensitivity.  
 The data are studied in Section~\ref{sec:results} 
 and  limits are set on  
 GMSB with a model-independent discussion
of the sensitivity.  Section~\ref{sec:conclusion} concludes with the
final results and a discussion of the future prospects for a similar
analysis with more data.

\subsection{Theory and Phenomenology}

Many minimal GMSB models are well specified with a small number of free
parameters. The electroweak symmetry breaking mechanism originates in a 
``hidden sector'' (not further specified in the model) and is mediated to the visible 
scalars and fermions by  messenger fields; for more details
see~\cite{gmsb2} and references therein.
 The free parameters of the minimal GMSB model are as follows:  the messenger mass scale, 
$M_{\rm m}$; the number of messenger fields, $N_{\rm m}$; a parameter 
$\Lambda$ that determines the gaugino and scalar masses; the ratio of the 
neutral Higgs vacuum expectation values, $\tan(\beta)$; and the sign of the 
higgsino mass parameter, $\sign(\mu)$. 
 For models with $N_{\rm m}=1$ and low 
$\tan(\beta)\lesssim 30$ the weakly interacting \grav is the lightest 
supersymmetric particle (LSP) and the next-to-lightest supersymmetric particle 
(NLSP) is the lightest neutralino \none. 
 For models with $N_{\rm m}>1$ or $\tan(\beta)\gtrsim 30$, the NLSP is a slepton (mostly $\tilde{\tau}_1$)~\cite{nlsp}. 
 As there are many GMSB parameter combinations that match this 
phenomenology, representative ``model lines'' have been identified
that allow a good specification of the model
 with only one free parameter that sets 
the particle masses.  This analysis follows line 8 of the Snowmass 
Points and Slopes (SPS 8) proposal~\cite{snowmass} and assumes  
$M_{\mathrm{m}}=2 \Lambda,$  $\tan(\beta)=15 \nonumber,$ ${\rm sgn}(\mu)=1 
\nonumber,$ $N_{\mathrm{m}}=1 \nonumber,$ and R-parity conservation.
In this model the \none
decays via $\none\to\gamma\grav$ with a branching ratio of $\sim$100\%
but leaves the \none mass and lifetime as free parameters.

Non-minimal GMSB models with a non-zero \none lifetime and a
$\sim$1-1.5~\mkunit mass \grav are favored as they
are consistent with current astronomical observations and models of
the early universe that take inflation into account~\cite{baltz}.
 If the 
\grav's are too light ($\lesssim$1~\mkunit),  
they can destroy the nuclei produced during  big bang
nucleosynthesis, leading to a cosmic microwave background that is
different from observations~\cite{feng2}. If they are too heavy
($\gtrsim$1~\mkunit),  
while they are a warm dark matter
candidate~\cite{ref2} and consistent with models of galaxy structure
formation, their density can cause the universe to
overclose. To include the proper GMSB messenger particle decays and lifetimes, 
an additional SUSY breaking scale is included and provides an additional parameter in the model that relates the \none
lifetime  with the \grav and the \none masses. 
 In this formulation~\cite{gmsb2} our parameter choices, SPS 8,  
  favor a lifetime of several nanoseconds for the 100~\munit \none mass range, just above current exclusions~\cite{ggmet,lep}.  

\begin{figure}[tbp]
  \centering
  \subfigure[]{
    \includegraphics[scale=.3]{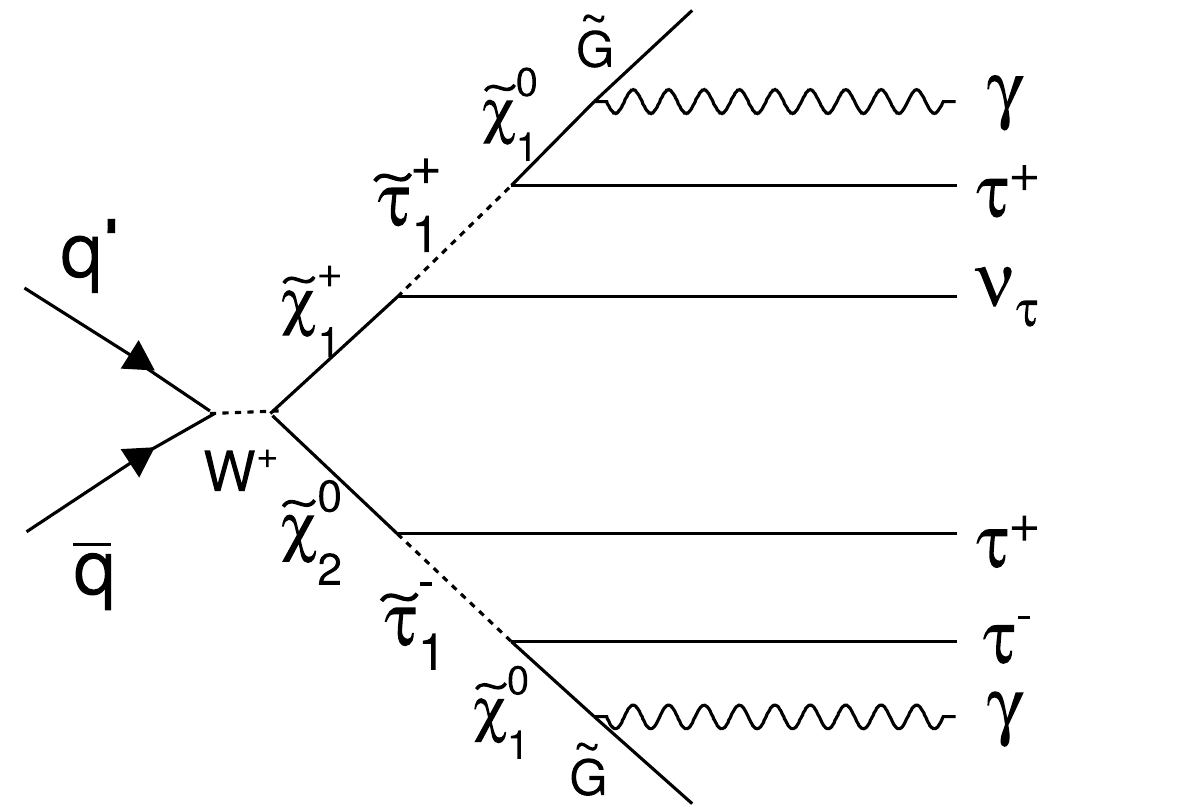}} \\
    \vspace{-20pt}
  \subfigure[]{
    \includegraphics[scale=.3]{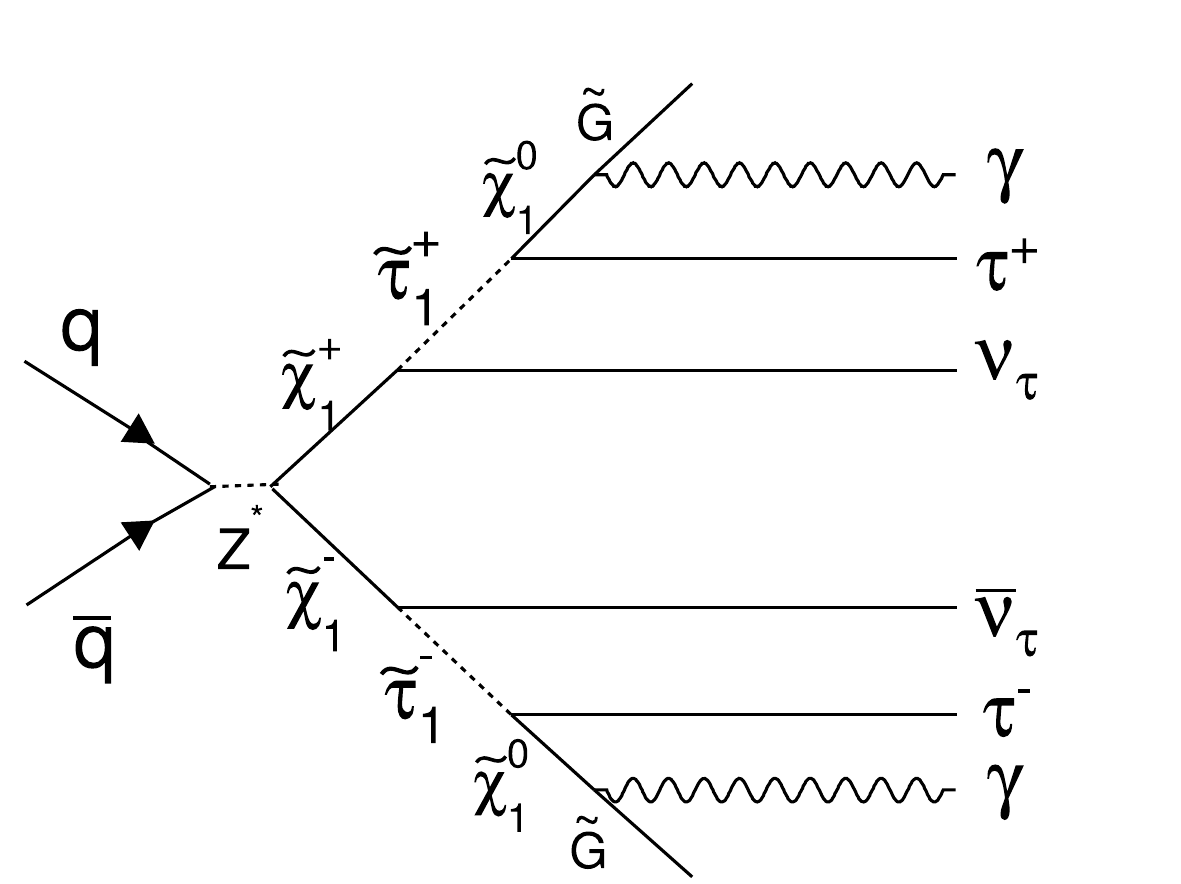}}
  \caption[Feynman diagrams of the dominant tree-production processes
  at the Fermilab Tevatron for the SPS 8 GMSB model line from
  SPS~8.]{\label{fig:process}\small Feynman diagrams of the dominant
    tree-production processes at the Fermilab Tevatron for the SPS~8 GMSB
    model line.  
    The taus and second
    photons, if available, can be identified 
as jets in the detector. Note that only one choice for the charge
    is shown. 
    }
\end{figure}

In $p\bar{p}$ collisions supersymmetric particles are pair-produced due to  
 $R$-parity conservation. We probe a range of $\Lambda$
 not already excluded at 95\% confidence level (C.L.) in previous collider
experiments~\cite{ggmet,lep} where the squarks and gluinos have masses of
$\sim$600-800~\munit and the sleptons and gauginos have masses of
$\sim$100-300~\munit.  At the Tevatron, with $\sqrt{s}=1.96$~TeV,
squarks and gluinos are too heavy to have significant production cross
 sections, hence    
gaugino pair-production dominates~\cite{gmsb2}.
Individually, $\ntwo\conepm$ and $\conep\conem$ production, as shown
in Fig.~\ref{fig:process}, contribute 45\% and 25\%, respectively, of
the total GMSB production cross section (\sigprod).  The rest of the
production is mostly slepton pairs. 
 We note that \sigprod is independent of the \none lifetime. 

This analysis focuses on the
\gmet final state which is expected to be more sensitive to the favored
 nanosecond lifetime scenario~\cite{prospects}.
 To identify GMSB events, we use the CDF~II detector. As shown in Fig.~\ref{fig:process}, each gaugino 
decays (promptly) to a \none in association with taus whose
decays can be identified as jets~\cite{jets}.
 Whether the \none$\rightarrow\gamma$\grav decay occurs
either inside or outside the detector volume depends on the \none
decay length (and the detector size). The \none's and/or the \grav's  leaving
the detector give rise to \met since they
are weakly interacting particles (the neutrinos in the event also
 affect the \met).
Depending on whether one or 
two \none's decay inside the detector,
 the event has the signature of high energy \ggmet or \gmet, often with one or more
additional particles from the heavier sparticle decays. These are identifiable as an additional jet(s) in the detector. We do not require 
the explicit identification of a tau. This has the advantage of reducing the model dependence of our results, making them  applicable to other possible gaugino decay models. 
 A study to see if there is additional sensitivity from adding $\tau$
 identification to the analysis is in progress.  

\begin{figure}[tb]
  \centering  
  \vspace{-3.05in}
  \subfigure[]{
    \includegraphics[scale=0.45]{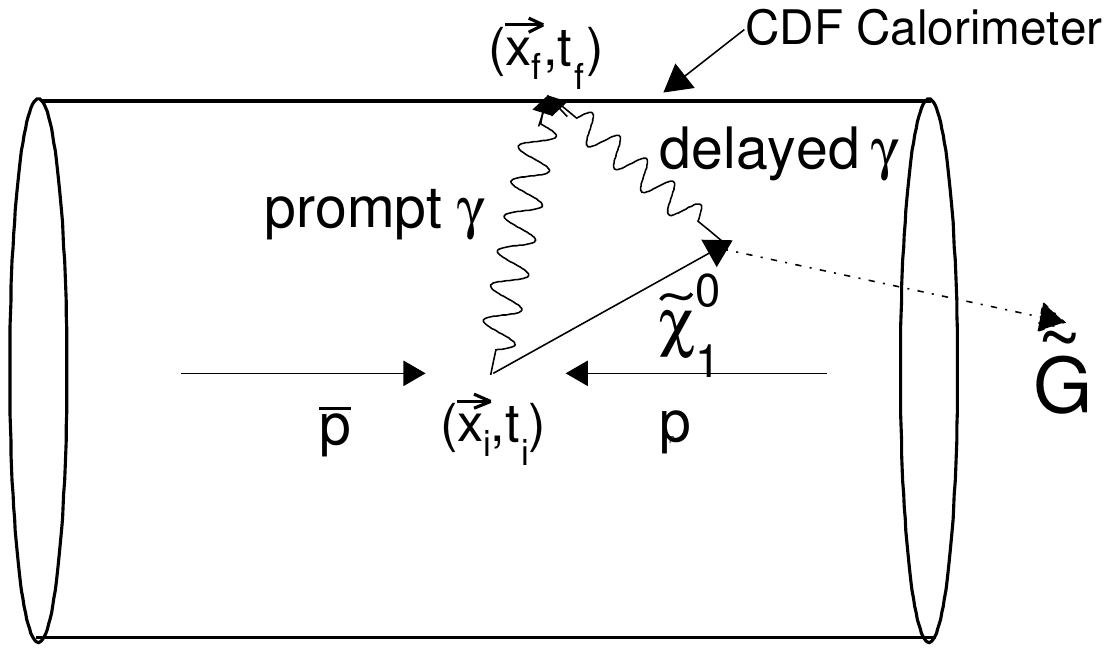}}
 \subfigure[]{
 \includegraphics[scale=.45]{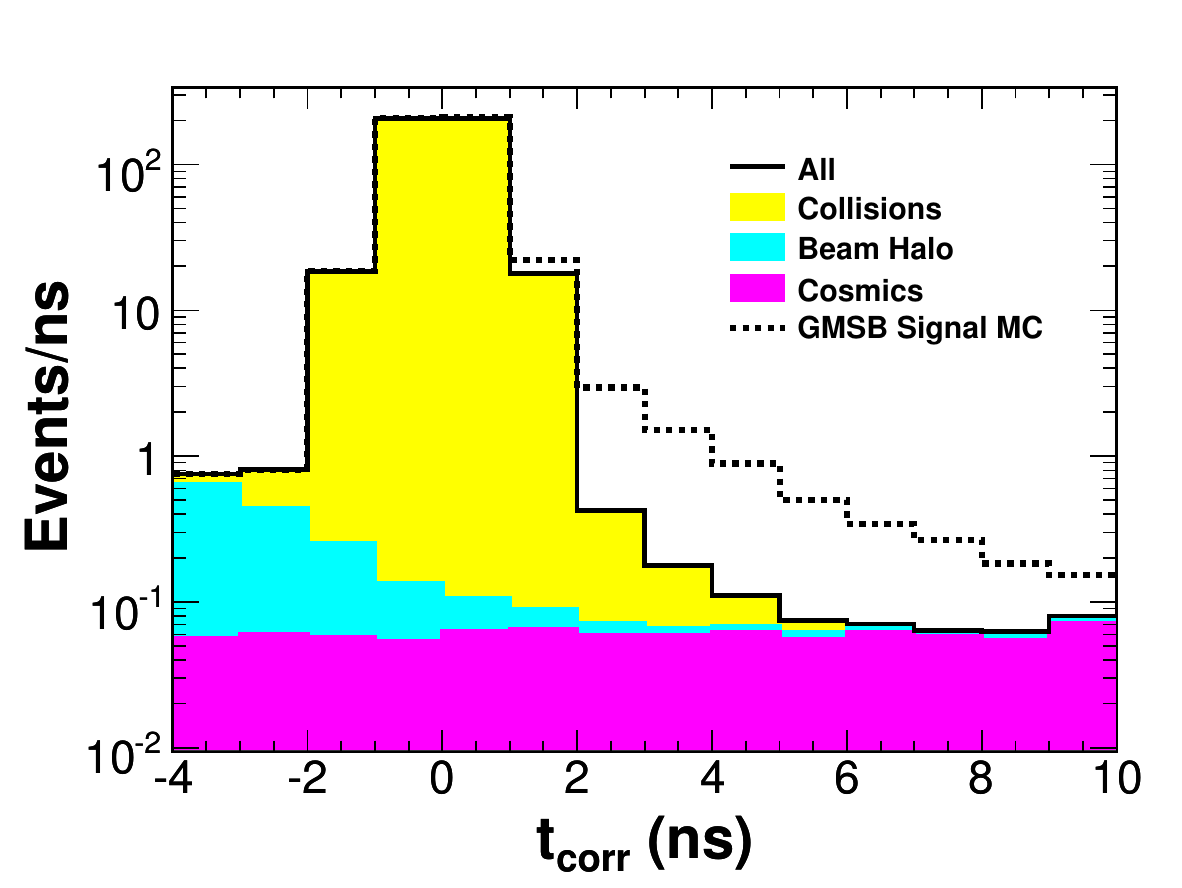}}
\caption [(a)The schematics of a long-lived \none decaying into a \grav
  and a photon inside the CDF detector.]{\label{fig:delta_s}\small (a) The
    schematic of a long-lived \none decaying into a \grav and a
    photon inside the detector. While the \grav leaves undetected
    the photon travels to the detector wall and deposits energy in the
    detector. A prompt photon would travel directly from the collision
    point to the detector walls. Relative to the expected arrival
    time, the photon from the \none would appear ``delayed.''
    (b) The \tcorr distribution for a simulated  
     GMSB signal at an example point of \mN~=~100~\munit and
    \tauN~=~5~ns as well as for standard model and non-collision backgrounds. 
 }
\end{figure}

The arrival time of photons at the
detector allows for a good separation between nanosecond-lifetime
\none's and promptly produced standard model 
 (SM) photons as well as non-collision backgrounds.
Figure~\ref{fig:delta_s}(a) illustrates a \none$\rightarrow\gamma$\grav decay 
 in the CDF detector after a macroscopic decay length. 
A suitable timing separation variable is 
\begin{equation}
  t_{\rm corr}\equiv(t_{f}-t_{i})-\frac{|\vec{x}_{f}-\vec{x}_{i}|}{c},
  \label{eq:delta_s}
\end{equation}
where $t_{f}-t_{i}$ is the time between the collision $t_{i}$ and the
arrival time $t_{f}$ of the photon at the calorimeter, and $|\vec{x}_{f}-\vec{x}_{i}|$ is
the distance between the position where the photon hits the detector and
the collision point. Here, $t_{\rm corr}$ is the photon arrival
time corrected for the collision time and the time-of-flight. 
 Prompt photons will produce $t_{\rm corr}\equiv 0$
while photons from long-lived particles will appear ``delayed''
($t_{\rm corr}>0$), ignoring resolution effects. 
Figure~\ref{fig:delta_s}(b) shows the simulated distribution of $t_{\rm
corr}$ for a  GMSB signal, prompt photons, and non-collision
backgrounds in the detector.  

\subsection{Overview of the Search}
\label{sec:overview}

This search selects photons with a delayed arrival time from a sample
of events with a  high transverse energy (\et) isolated photon, 
large \met, and a high-\et jet 
to identify 
 gaugino cascade decays. 
The background to this search can be separated into two
types of sources: collision and non-collision backgrounds. Collision 
 backgrounds come from SM production, such as strong interaction (QCD) and 
 electroweak processes.  Non-collision backgrounds come
 from photon candidates that are either emitted by cosmic ray
muons as they traverse the detector or are from  
beam related backgrounds that produce an energy deposit in the calorimeter  
that is reconstructed as a photon. 

 The search was performed as a blind analysis, picking the final
 selection criteria  based on the signal and background expectations alone. 
The background rates in the signal region are estimated using  
\tcorr control regions from the same \gmetjet data sample  
 and comparing to the distribution shapes of the various backgrounds. 
A Monte Carlo (MC) simulation is used to model the GMSB event dynamics
and timing in the detector and to estimate the signal expectations. 
Combining these backgrounds and signal event estimates permits a calculation of the most
sensitive combination of event requirements. 
 We note that the jet requirement helps make this search 
 sensitive to any model that produces a
large mass particle decaying to a similar final state.

\subsection{
The CDF II Detector and the EMTiming System}
\label{sec:tools}

The CDF~II detector is a 
general-purpose magnetic spectrometer, whose detailed description
can be found in~\cite{CDFII} and references therein.  The salient 
 components are summarized here.
The magnetic spectrometer consists of tracking devices
inside a \mbox{3-m} diameter, 5-m long superconducting solenoid magnet
that operates at 1.4~T. A set of silicon microstrip detectors 
(SVX) and a 3.1-m long drift chamber (COT) with 96 layers of sense wires
measure the position ($\vec{x}_i$) and time ($t_i$) of the $\ppbar$
interaction and the momenta of charged particles. Muons
from the collision or cosmic rays are identified by a system of drift
chambers situated outside the calorimeters in the region with
pseudorapidity $|\eta|<1.1$. The calorimeter consists of
projective towers ($\delta\phi=15^\circ$ and $\delta\eta\approx0.1$) with electromagnetic and hadronic compartments and 
is divided into a central barrel that surrounds the solenoid coil
(\mbox{$|\eta|<1.1$}) and a pair of end-plugs that cover the region
$1.1<|\eta|<3.6$.  Both calorimeters are used to identify and measure
the energy and position of photons, electrons, jets, and  \mett.  

The electromagnetic calorimeters were recently instrumented with a new
system, the EMTiming system (completed in Fall 2004), which is
described in detail in~\cite{nim} and references therein. The
following features are of particular relevance for the present
analysis. The system measures the arrival time of electrons and photons in
 each tower with  $|\eta|<2.1$ using the
electronic signal from the EM shower in the calorimeter. 
 In the region $|\eta|<1.1$, used in this analysis, 
  photomultiplier tubes (PMTs) on opposite azimuthal sides of the
  calorimeter tower  
convert the scintillation light generated by the shower into an 
analog electric signal.  The energy measurement integrates the charge over a 132~ns timing window around the collision time from $\sim$20~ns before the collision until $\sim$110~ns afterwards.  New electronics inductively branches off
$\sim$15\% of the energy 
 of the anode signal  and sends it to a discriminator. 
 If the signal for a tower is above 2~mV ($\sim$3-4~GeV energy deposit),
 a digital pulse is sent to a  time-to-digital converter (TDC) that
 records  the photon  arrival time and is read out for
each event by the data-acquisition system. The resolution of the time of arrival measurement is $0.50\pm0.01$~ns for the photon energies used in this analysis.

\section{Photon Identification and Timing}
\label{sec:photonid}

The CDF detector has been used for the identification (ID) of
high-energy photons for 
 many years, and a standardized set of ID criteria (cuts) for the region
$|\eta|<1.0$ is now well established.  Each cut 
is designed to separate real, promptly produced 
photons from photons from $\pi^0\to\gamma\gamma$ decays, hadronic jets, electrons, and other backgrounds,  
see~\cite{pwt,ggmet,photoncuts} 
 for more details and the Appendix for a description of the ID variables. 

\begin{figure}[tpb]
\centering 
\vspace{-12pt}
\subfigure[]{
    \includegraphics [scale=0.42]
{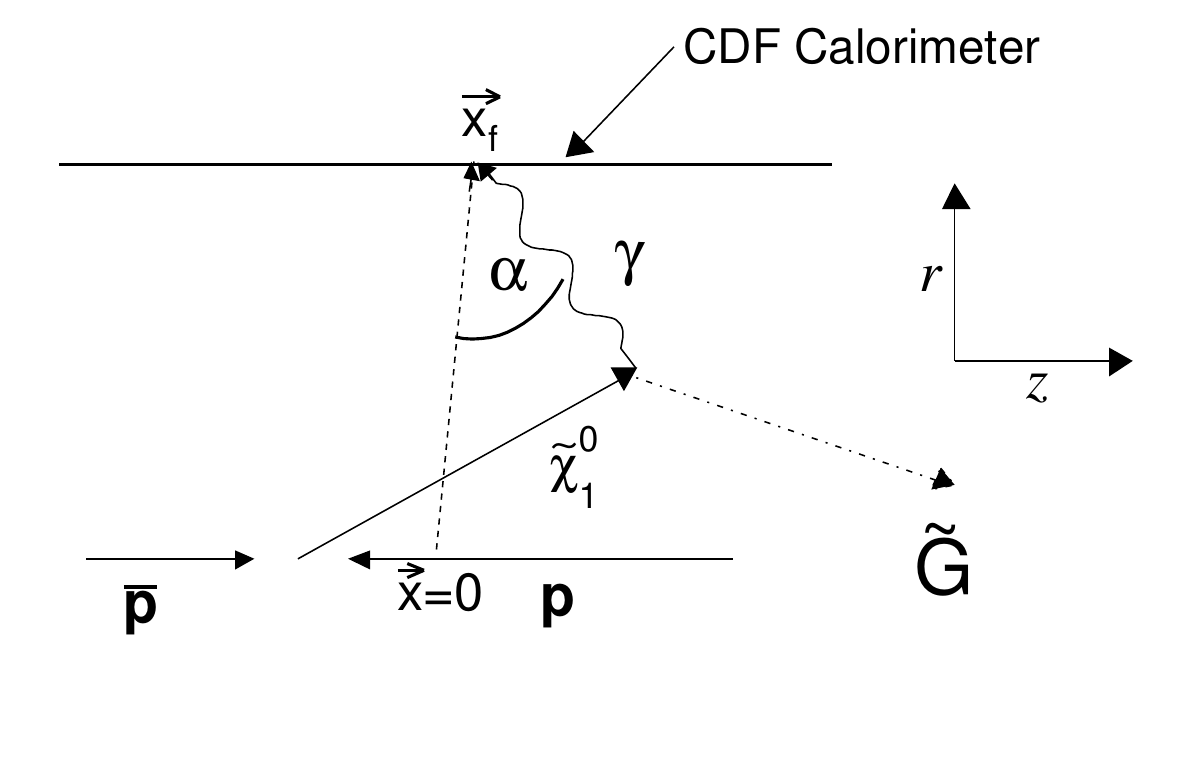}} \\
\vspace{-12pt}
  \subfigure[]{ 
   \includegraphics[scale=0.35]{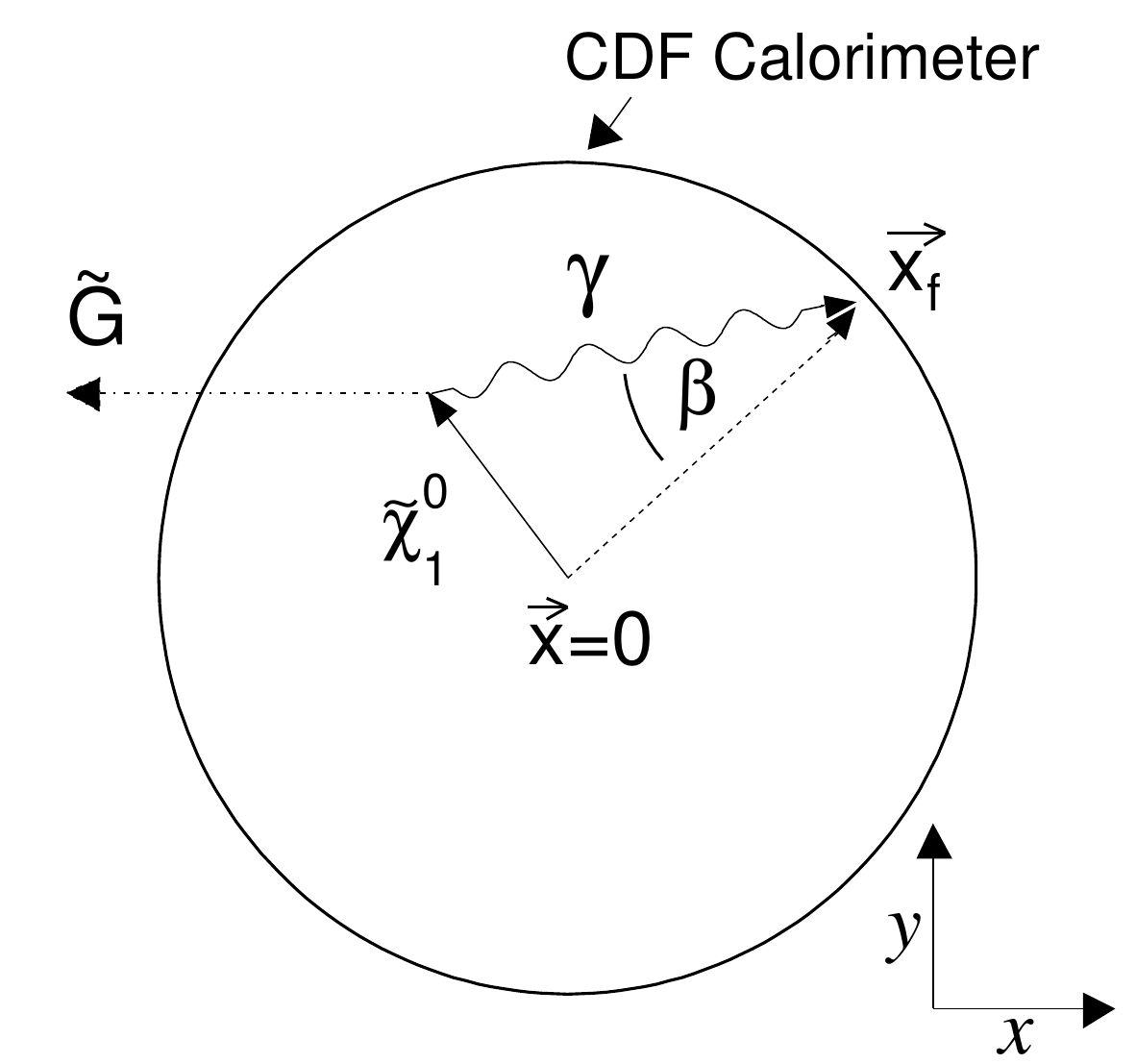}}
  \vspace{-12pt}
\caption[The definitions of the \ag~and \bg~incident angles using
  schematic diagrams of a long-lived \none decaying to a photon and a
  \grav in the CDF detector in the ($r$,$z$)- and the
  ($r$,$\phi$)-planes.] {\small The definitions of the \ag~and
    \bg~incident angles using schematic diagrams of a long-lived \none
    decaying to a photon and a \grav in the CDF detector. 
    The angles $\alpha$ and $\beta$ are the projections of the
    incident angle $\psi$ at the
    front face of the calorimeter in the ($r$,$z$)- and the
    ($r$,$\phi$)-plane, respectively.
\label {fig:alpha}
}

\end{figure}

Unlike photons from SM processes, ``delayed'' photons from long-lived
\none's are not expected to hit the calorimeter coming directly from
the collision point~\cite{prospects}. 
 As shown in Fig.~\ref{fig:delta_s}(a), \none's with a long lifetime and small boost can produce a photon from \nonetogG with a large 
path length from the collision position to the calorimeter (large \tcorr). 
We define the photon incident angle at the face of the EM calorimeter,
$\psi$, as the angle between the momentum
vector of the photon from the \none and the 
vector to the center of the detector.
For convenience we consider the $\psi$ projection onto 
the ($r,z$)-plane and label it $\alpha$, and the $\psi$ projection onto the
($r,\phi$)-plane and label it $\beta$; see Fig.~\ref{fig:alpha}. 
 This distinction is made as the photon ID variable efficiencies vary
 differently between $\alpha$ and $\beta$.    

\begin{figure}[tbp]
 \centering
  \vspace{-7pt}
  \hspace{40pt}
 {\includegraphics [scale=0.44]{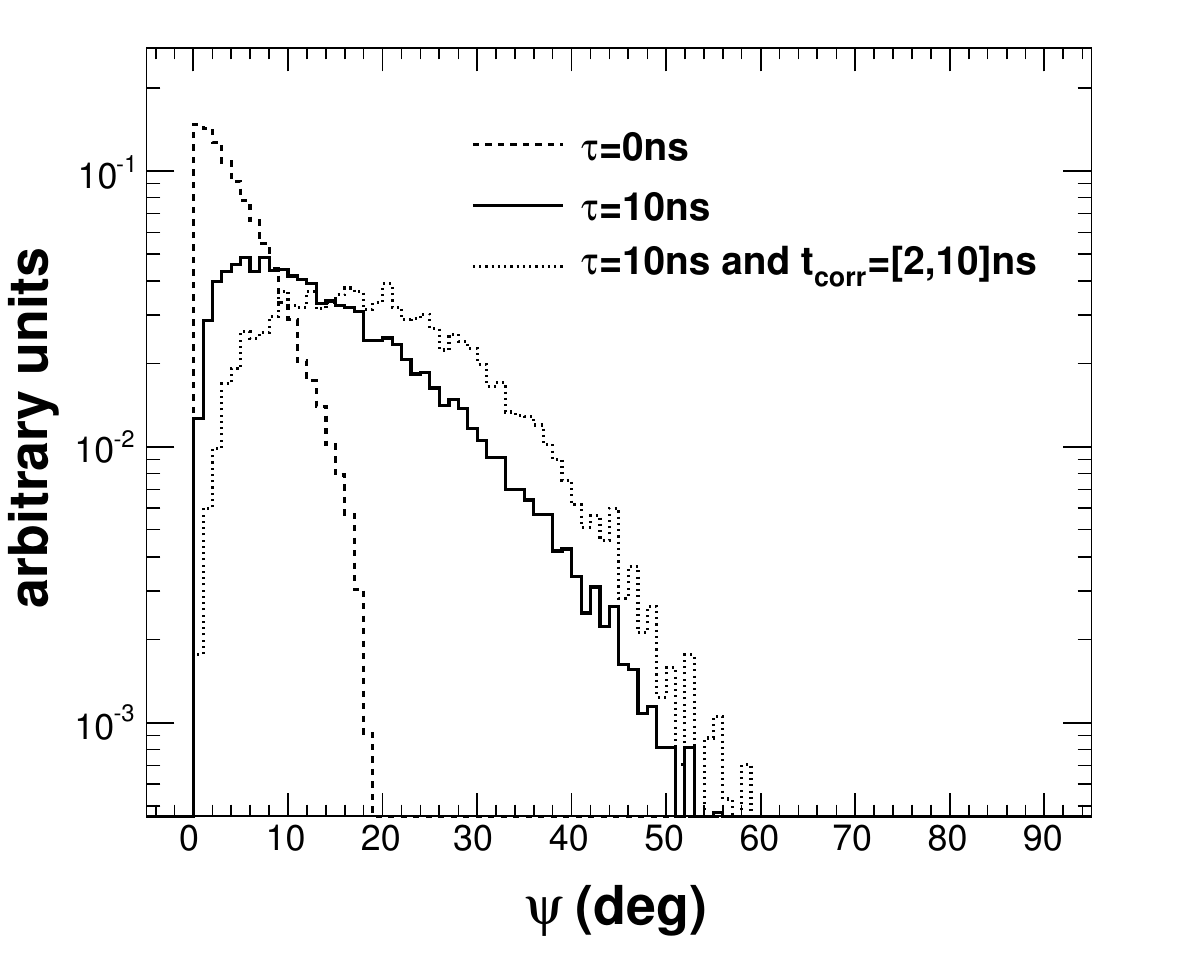}}
\caption[
The distribution of 
  $\beta$ and the total incident angle $\alpha\oplus\beta$ 
for simulated photons from
  \none's in a GMSB model \mN~=~110~\munit).]{\small The
    distribution of the 
    total incident angle $\psi$ at the front face of the
    calorimeter for simulated photons from \none's 
    with \mN~=~110~\munit. ``Prompt'' photons from \none's with a
    lifetime of 0~ns (solid) are compared to photons from \none's with
    a lifetime 10~ns (dashed). The dotted histogram shows the
    distribution for a lifetime of 10~ns for photons with  
    $2\leq\tcorr\leq10$~ns and shows that, as expected, delayed photons can have a significant incident angle.
   }
  \label{fig:delta_exp}
\end {figure}

Figure~\ref{fig:delta_exp} compares the $\psi$ distribution
for prompt, SM-like photons and photons from long-lived \none's.
 Each are simulated as the decay product of a \none 
 with \mN~=~110~\munit  using the {\sc pythia}
 MC generator~\cite{pythia}. 
The  distributions of promptly produced photons~\cite{alpha_explain} 
have a maximum at $\psi$=0\dg and extend to $\sim$18\dg~in \ag while \bg is always 
$\ll1^\circ$ 
as the beam has negligible extent in the $x$-$y$ plane. 
 The most common angle $\psi$  
for a simulated neutralino sample with \tauN~=~10~ns is $\sim$10\dg and extends out
to maximum angles
of $\sim$60\dg~and $\sim$40\dg~in \ag and \bg respectively. For this sample, 
 the majority of photons arrive at angles between 0 and 40\dg~total
incident angle.  The mean of the distribution rises as a function of
\tauN but becomes largely independent of \mN and
\tauN in the range $10<\tauN<35$~ns.
Also shown is the distribution for delayed photons, selected with  
$2<\tcorr<10$~ns, similar to a typical final analysis requirement. 
The delayed photon 
 requirement shifts the maximum of the
distribution of $\psi$ from $\sim$10\dg to $\sim$25\dg. As the incident angles
of photons from long-lived particles are much larger than for prompt
photons, the standard selection criteria are re-examined and modified
where necessary.  

\label{sec:idvars}

To verify that we can robustly and efficiently identify 
photons from heavy, long-lived particles, 
we examine the efficiencies of the 
photon ID variables as a function of \ag and \bg separately.
 As we will see 
the standard photon identification requirements are slightly modified
for this search; each is listed in Table~\ref{tab:phcuts}.
To study photon showers at a wide variety of angles in the
calorimeter,  we create a number of data
and MC samples of photons and electrons. An electron shower 
in the calorimeter is very similar to that from a photon, but electrons can be selected with high
purity. We create two samples of \Wenu events, one from data, and the
other simulated using the {\sc pythia} MC generator and the standard, {\sc
  geant} based, CDF 
detector simulation~\cite{ref9}. Each must pass   
the requirements listed in Table~\ref{tab:elcuts}. 
 Similarly, two samples of MC photons 
 are generated using  \none$\rightarrow\gamma$\grav decays   
with \mN~=~110~\munit and \tauN~=~0~ns and  \tauN~=~10~ns
  respectively to cover the region $0\leq\psi\leq60\dg$.   The highest \et
 photon in the event is required to be the decay product of a
 \none and to pass the $\et$, $\eta$, and fiducial requirements listed in
  Table~\ref{tab:phcuts}.

\begin{table}[tb]
  \centering
    {\small \addtolength{\tabcolsep}{0.1em}
\centering      
\begin{tabular}{l}        
\hline \hline
      $\et>30$~GeV and $|\eta|\le1.0$ \\ 
      Fiducial: not near the boundary, in $\phi$ or $z$, of a \\
      \ \ \ \ calorimeter tower \\ 
      $E_{\mathrm{Had}}/E_{\mathrm{EM}}<0.125$ \\ 
      Energy in a $\Delta$$R=0.4$ cone around the photon
       \\ \ \ \ \ excluding the photon energy: \\ \ \ \ \ 
      $E^{\mathrm{Iso}}<2.0~\mathrm{GeV}+0.02 \cdot (\et-20~\mathrm{GeV})$   \\    
      No tracks pointing at the cluster or \\
      \ \ \ \ one with $p_{T}<1.0~\punit+0.005\cdot \et$    \\ 
      $\Sigma$\pt of tracks in a 0.4 cone $<2.0~\punit+0.005 \cdot \et$   \\ 
      $E^{\mathrm{2^{nd} cluster}}<2.4~\mathrm{GeV}+0.01 \cdot \et$   \\ 
      ${\rm A_{P}}=\frac{|E_{\mathrm{PMT1}}-E_{\mathrm{PMT2}}|}{E_{\mathrm{PMT1}}+E_{\mathrm{PMT2}}}<0.6$  \\  
\hline \hline

\end {tabular}
}
 \vspace{1em}
    \caption[The sample requirements to select electrons from \Wenu
    events. Note these requirements are described in detail in ref **.] 
{\label{tab:phcuts}\small 
The photon identification and isolation selection requirements.  These are the standard requirements with the $\chi^2_{\rm CES}<20$  requirement removed. 
These variables are described in more detail in~\cite{ggmet,photoncuts} and the Appendix.
}
\end{table}


\begin{table}[tb]
  \begin{center}
    {\small \addtolength{\tabcolsep}{0.1em}
\centering      
\begin{tabular}{l}        
\hline \hline
\multicolumn{1}{c}{
Electron Requirements} \\ \hline
        $\et>30$~GeV and $|\eta|\le1.0$ \\
        Fiducial: not near the boundary, in $\phi$ or $z$, of a \\
      \ \ \ \ calorimeter tower \\ 
        $0.9<E/p<1.1$ or $\pt>50$~\punit  \\
        Track traverses $\ge$3 stereo and $\ge$3 axial COT \\
        \ \ superlayers with 5 hits each \\
        Additional requirements to reject electrons from $\gamma\to ee$ \\ 
         \hline
        \multicolumn{1}{c}{Global Event Requirements} \\ \hline
        \met$>$30~GeV \\
        Exactly 1 vertex with $N_{\mathrm{trks}}\ge4$ and $|z|<60$~cm \\
        Transverse mass of the electron and \met: \\  \ \ \ \ 
        $50<m_{T}<120$~\munit \\ 
        \hline \hline
\end{tabular}
    }
    \vspace{1em}
    \caption[The sample requirements to select electrons from \Wenu
    events. Note these requirements are described in detail in ref **.] 
{\label{tab:elcuts}\small 
      The 
      requirements used to 
      select electrons from \Wenu events to validate
      the ID efficiency of simulated photons. 
      These are topological and global
      event cuts in combination with loose calorimetry but tight track
      quality requirements.   
      This produces a sample that
      contains electrons with  high purity but has a low bias for
      calculating the efficiency of photon ID requirements vs. incident angle.
      The vertex reconstruction algorithm is described in
      Section~\ref{sec:vertexing} and uses tracks passing the
      requirements listed in Table~\ref{tab:trackcuts}. These
      variables are summarized in the Appendix and described in more
      detail in~\cite{CDFII}. } 
  \end{center}
\end{table}

\begin{figure*}[tb]
  \centering 
  \vspace{-35pt}
  \subfigure[]{
    \label{fig:hadem}
    \includegraphics[scale=.42]
{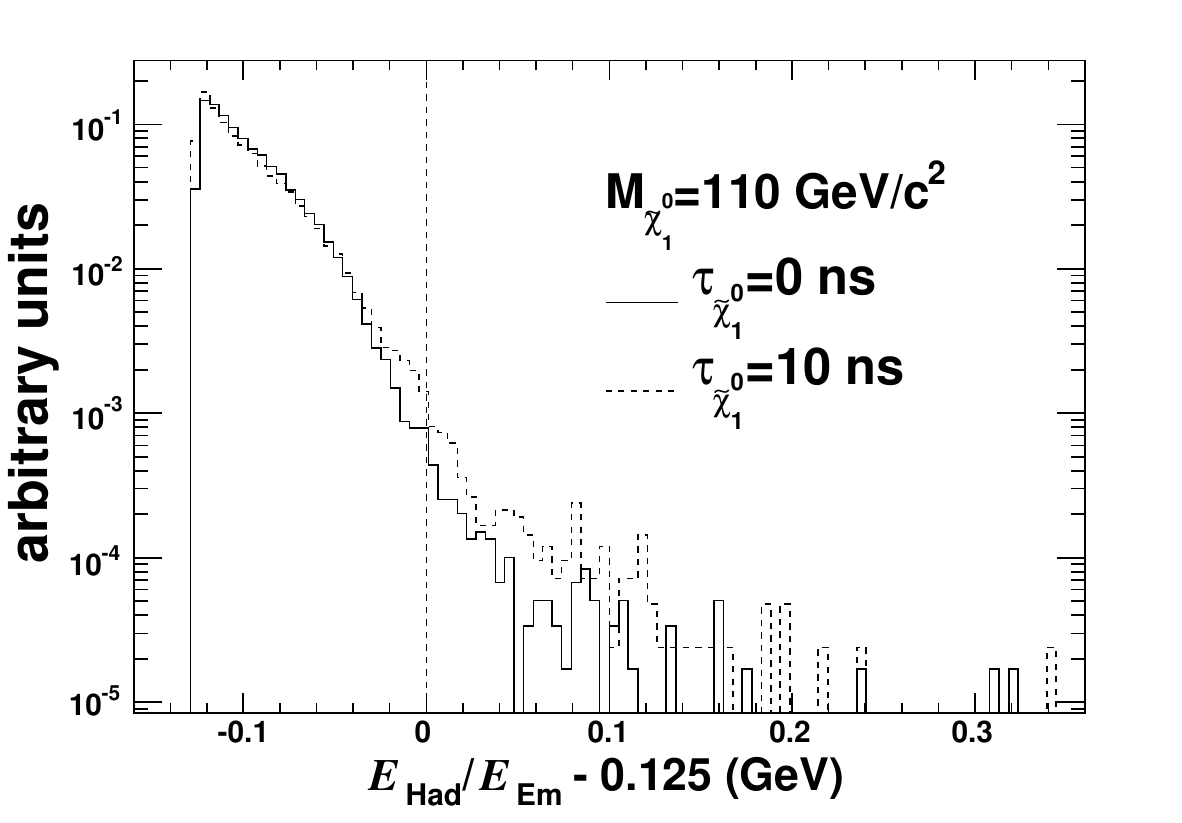}}
  \hspace{15pt}
\subfigure[]{
    \label{fig:eiso4}
    \includegraphics[scale=.42] 
{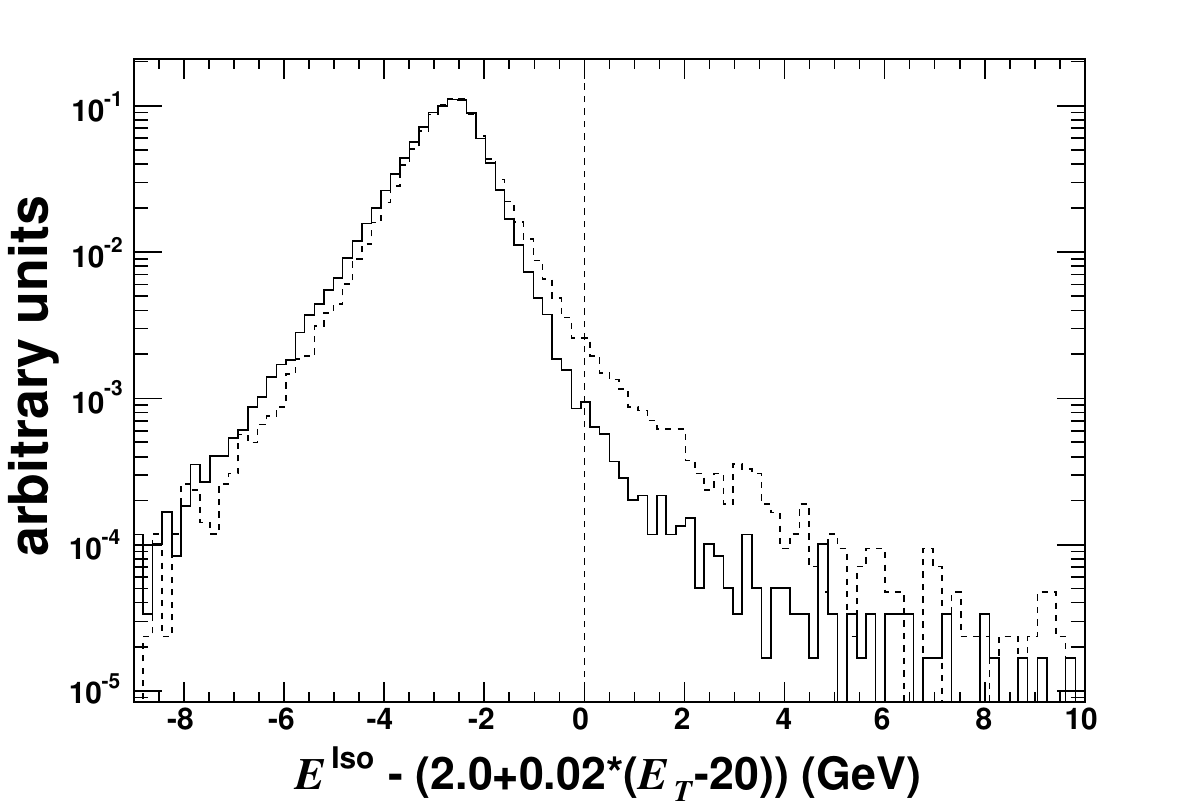}} \\
 \vspace{-19pt}
  \subfigure[]{
    \label{fig:n3d}
    \includegraphics[scale=.42]
{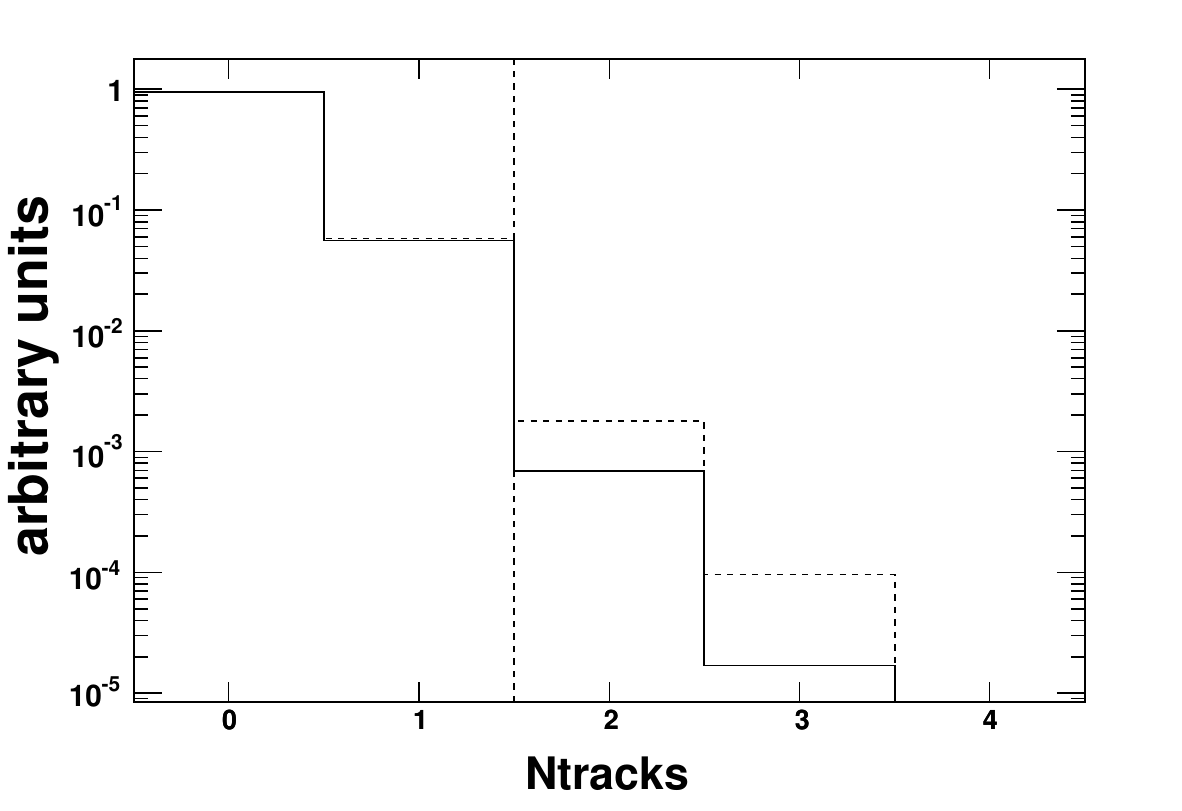}}
  \hspace{15pt}
\subfigure[]{
    \label{fig:pt_et}
    \includegraphics[scale=.42] 
{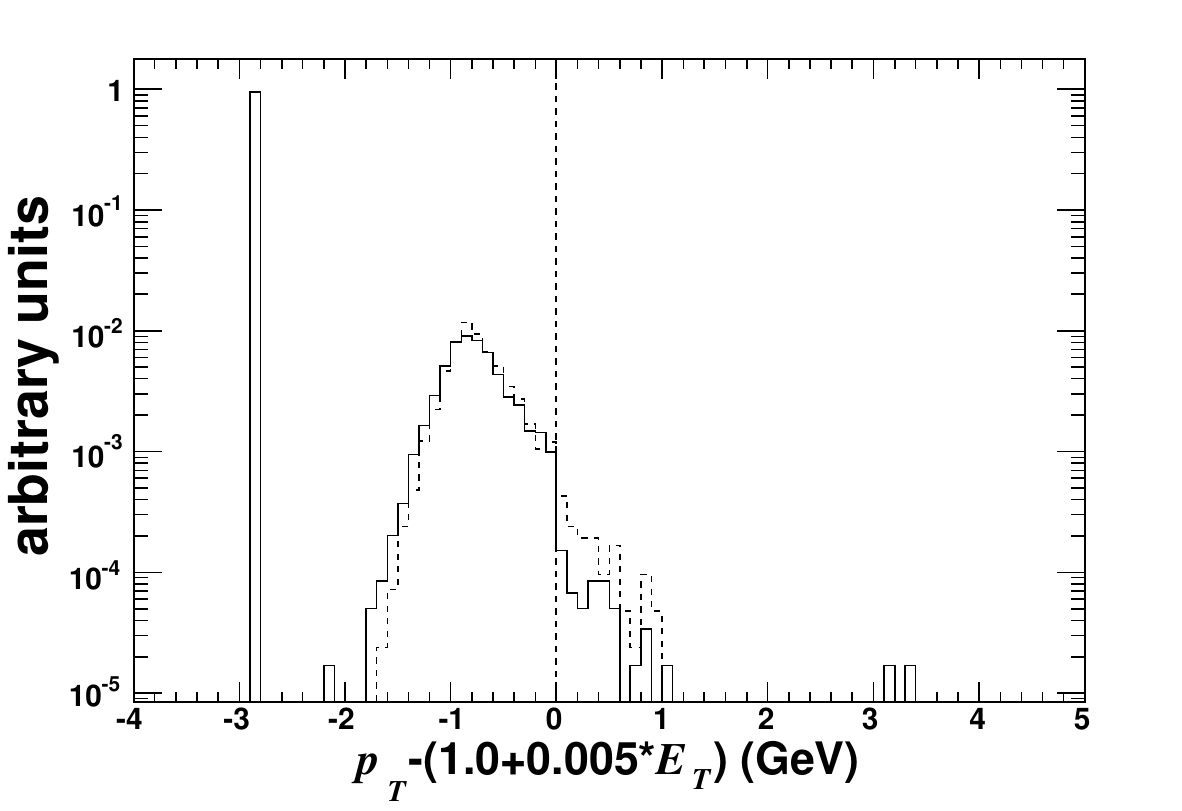}}    \\
 \vspace{-19pt}
  \subfigure[] {
    \label{fig:sumpt4}
    \includegraphics[scale=.42] 
{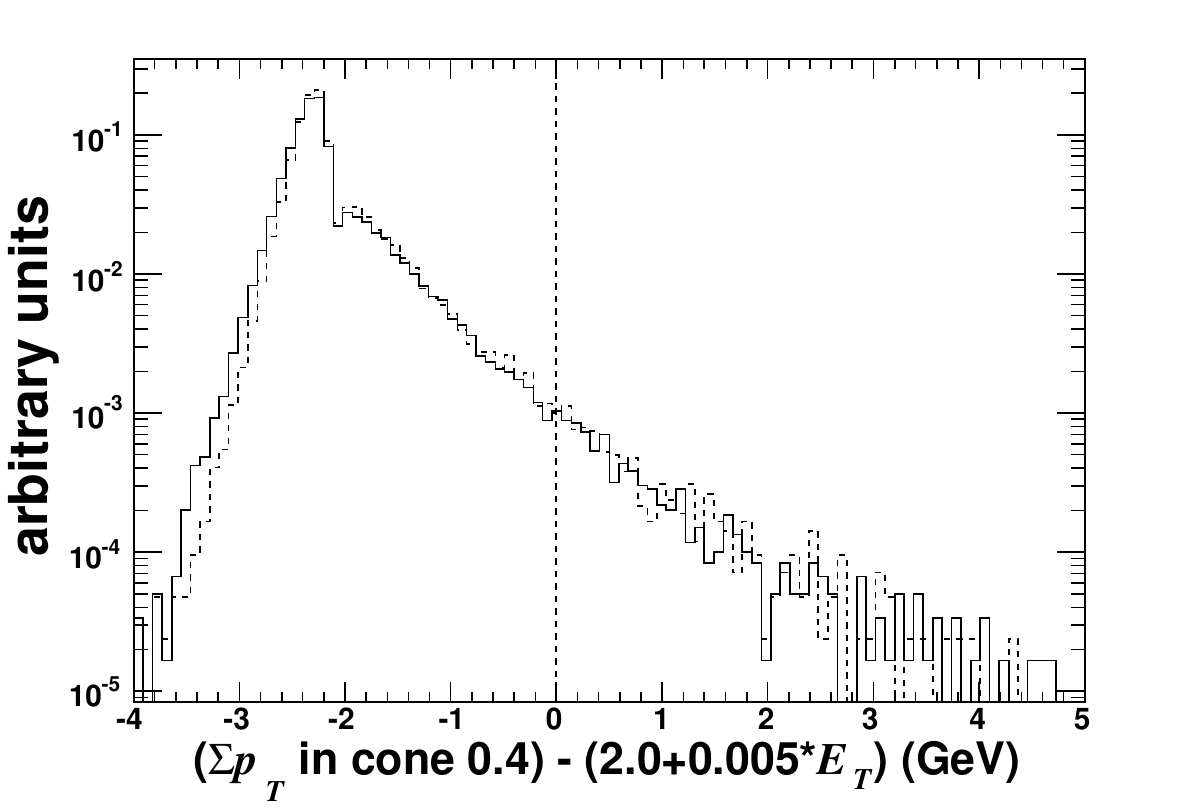}}
  \hspace{15pt}
\subfigure[]{
    \label{fig:cese}
    \includegraphics[scale=.42] 
{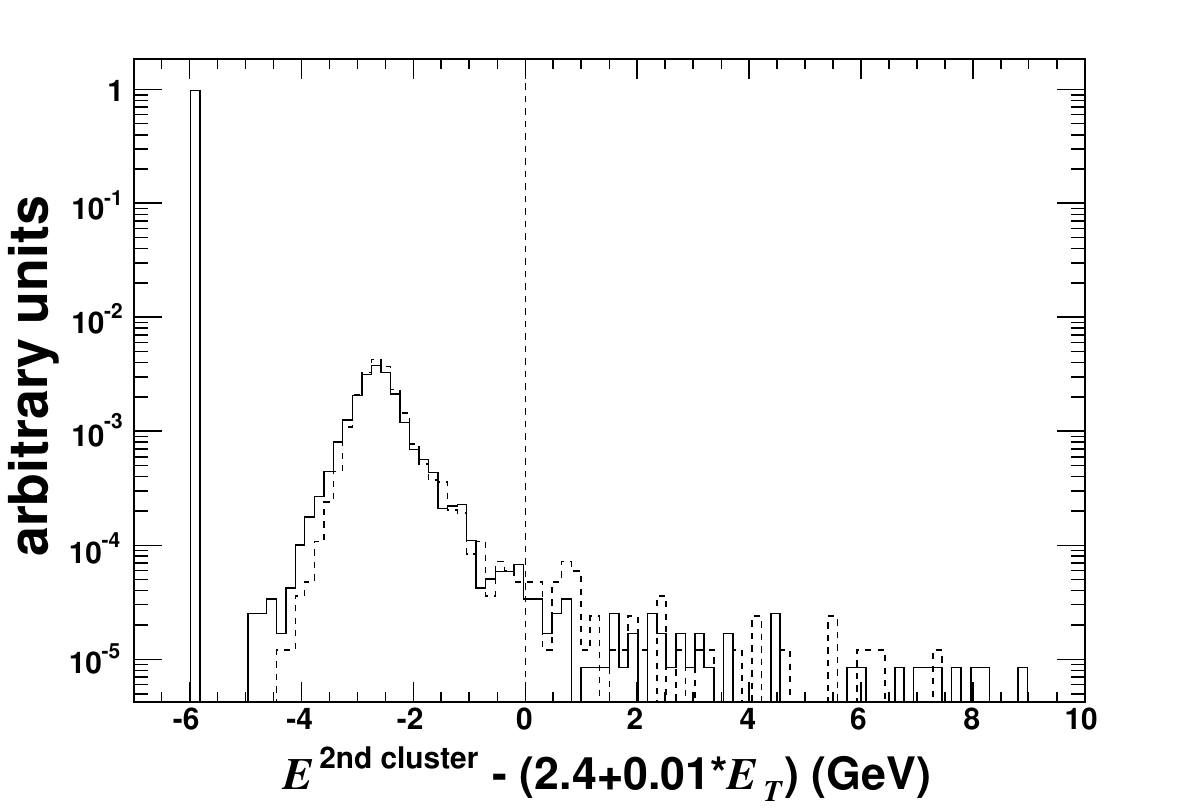}}
 \vspace{-14pt}
  \caption[The ID variable distributions for photons in a GMSB model
  with \mN~=~110~\munit produced with the CDF MC, normalized
  to 1.]{\small A simulation of the ID variable distributions (minus their requirement value) for photons in
    a GMSB model with \mN~=~110~\munit. 
      The solid line is for prompt
    photons, simulated as decay photons from \none's with a lifetime
    of 0~ns and the dashed line is for photons from 
    long-lived \none's with a lifetime of 10~ns.  
    Entries to the left of the dashed vertical line pass the
    corresponding requirement. The bin at $-2.8$ in (d) 
    collects the photons that have no track within the isolation cone. 
    In (f) the bin at $-6$ 
    shows the photons that have no 2nd CES cluster nearby. 
    The distributions for
    all ID variables do not change significantly between the prompt
    and the long-lived case except for slight deviations in the energy
    isolation in (b) as discussed in the text. }
  \label{fig:idvars1}
\end{figure*}

\begin{figure}[tb]
  \centering 
  \vspace{-12pt}
  \subfigure[] {
  \hspace{-7pt}
    \includegraphics[width=.49\textwidth]{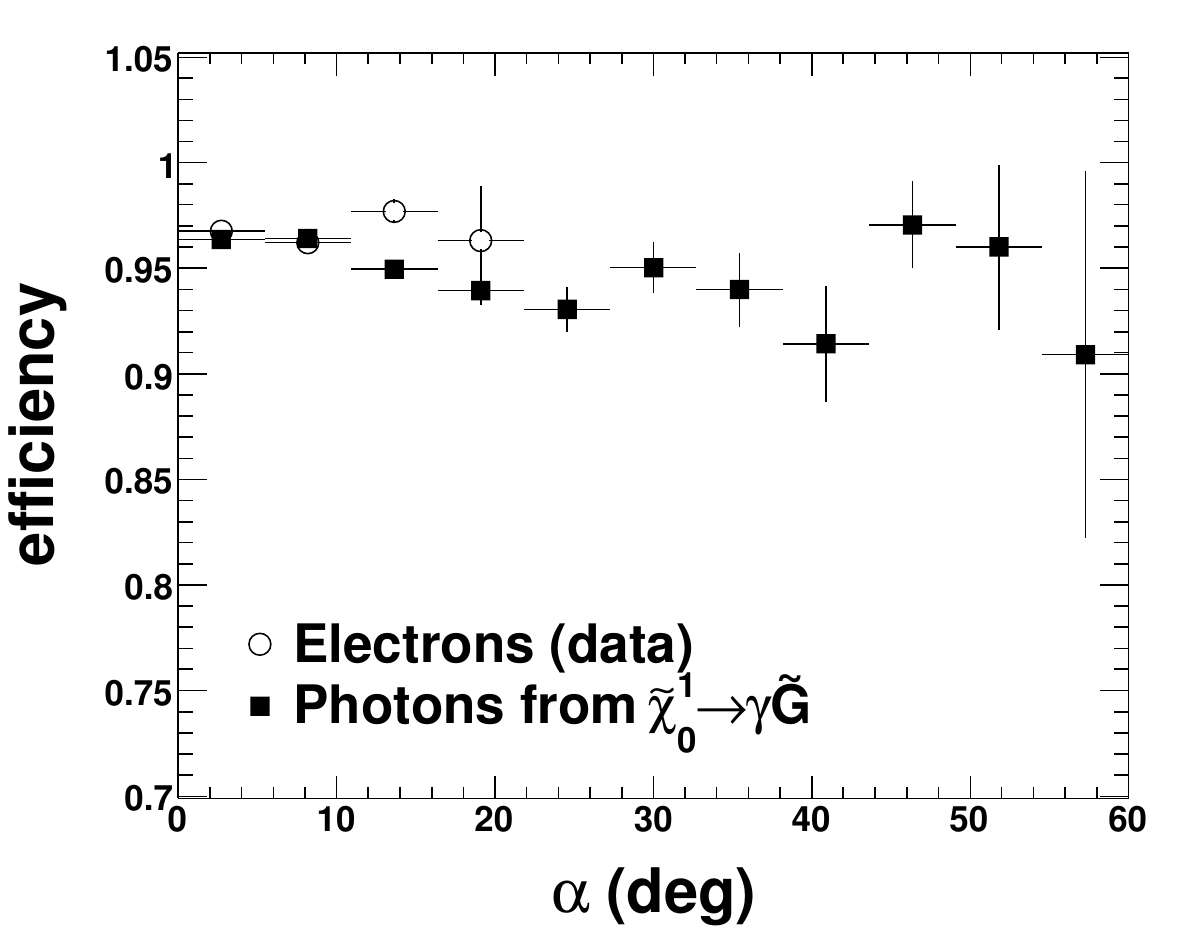}}\\
 \vspace{-19pt}
  \hspace{+22pt}
\subfigure[] {
    \includegraphics[width=.49\textwidth]{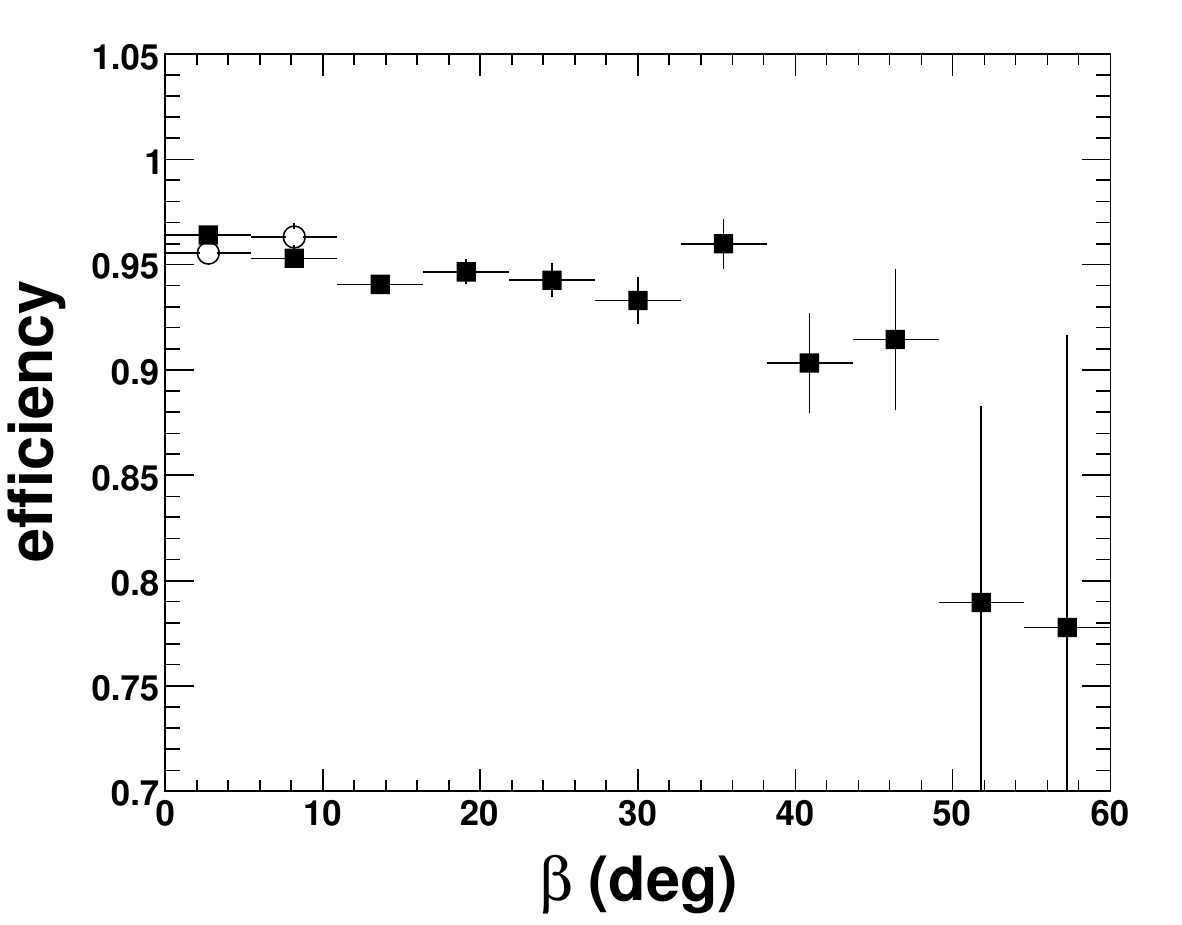}}
  \vspace{-14pt}
  \caption[The total efficiencies for the 
photon ID
  requirements 
(in Table~\ref{tab:phcuts}) for photons in the fiducial
      portion of the CEM vs. incident angles \ag~and \bg. ]{\small The 
     efficiencies for photons and electrons to pass the ID requirements  
    in Table~\ref{tab:phcuts}~vs.~incident angles \ag and \bg. The
      solid squares represent MC 
    photons from \none$\rightarrow\gamma$\grav decays (\mN~=~110~\munit, \tauN~=~10~ns) while 
    the empty circles represent electrons from a $W\rightarrow e\nu$ data sample that pass the 
    requirements in Table~\ref{tab:elcuts}. The efficiency
    falls by $\sim$15\% from 0\dg to 60\dg in \bg.  This effect is mostly 
    due to the energy isolation requirement, as discussed in the text.
  }
  \label{fig:cuteff}
\end{figure}

Figure~\ref{fig:idvars1} compares the distributions of the photon ID
variables for the \tauN~=~0 and \tauN~=~10~ns samples.  A visual comparison shows that
the differences are, on average, very small.
Figure~\ref{fig:cuteff}
shows the 
efficiency of the final set of ID requirements for MC photons and electrons from data as a function of incident
angles \ag and \bg (taking the photon position from the measured center of the calorimeter energy cluster). 
 The efficiencies are very similar and constant except at large values
 of \bg where the efficiency drops, which is where real collision data are not
 available.  The drop in efficiency at large \bg is due to the photon
 shower in the calorimeter traversing into 
the neighboring tower in $\phi$.  Because the photons are identified
and measured as clusters in the calorimeter~\cite{photoncuts}, this
decreases the cluster-energy sum while increasing the isolation
energy. Therefore, the photon 
appears non-isolated and the isolation efficiency falls from   
 $\sim$98\% at \bg=~0\dg to $\sim$90\% at \bg=~50\dg.  This is not a
problem for large \ag as energy leakage into the neighboring tower in
$\eta$ is included in the energy sum.  The total photon identification
efficiency as a function of $\psi$ in this regime falls from
$\sim$93\% to $\sim$80\%.  However, since in our $\psi$ region the
fraction of events with  
 large \bg is small (see Fig.~\ref{fig:delta_exp}), even at large \tauN, 
the ID criteria are only $\sim$1.5\%
less efficient for photons for the 
\tauN~=~10~ns sample than for the prompt sample. 
Thus, the majority of the standard 
requirements are not changed for the search. The efficiency variation as a
function of angle is taken into account 
 by using the detector simulation for the efficiencies and 
 assigning a 5\% systematic
uncertainty to the overall photon ID efficiency measurement. 

The comparison of the photon shower-maximum profile to
test-beam expectations, $\chi^2_{\rm CES}$~\cite{photoncuts}, is removed
from the photon
identification requirements because it becomes inefficient at large angles.
The shower for a photon that hits the shower-maximum detector (CES) at a
large value of  \ag (\bg) angle would spread out and have a
larger-than-expected RMS in the $z$ ($\phi$) 
direction due to the projection. A {\sc
geant} simulation~\cite{ref9}  shows the efficiency of the
$\chi^2_{\rm CES}$ requirement is constant at small angles, but then falls off 
rapidly at large
angles. Thus, the photon $\chi^2_{\rm CES}$ requirement is removed. 

\begin{figure}[tbp]
   \vspace{-6pt}
    \includegraphics[width=.52\textwidth]{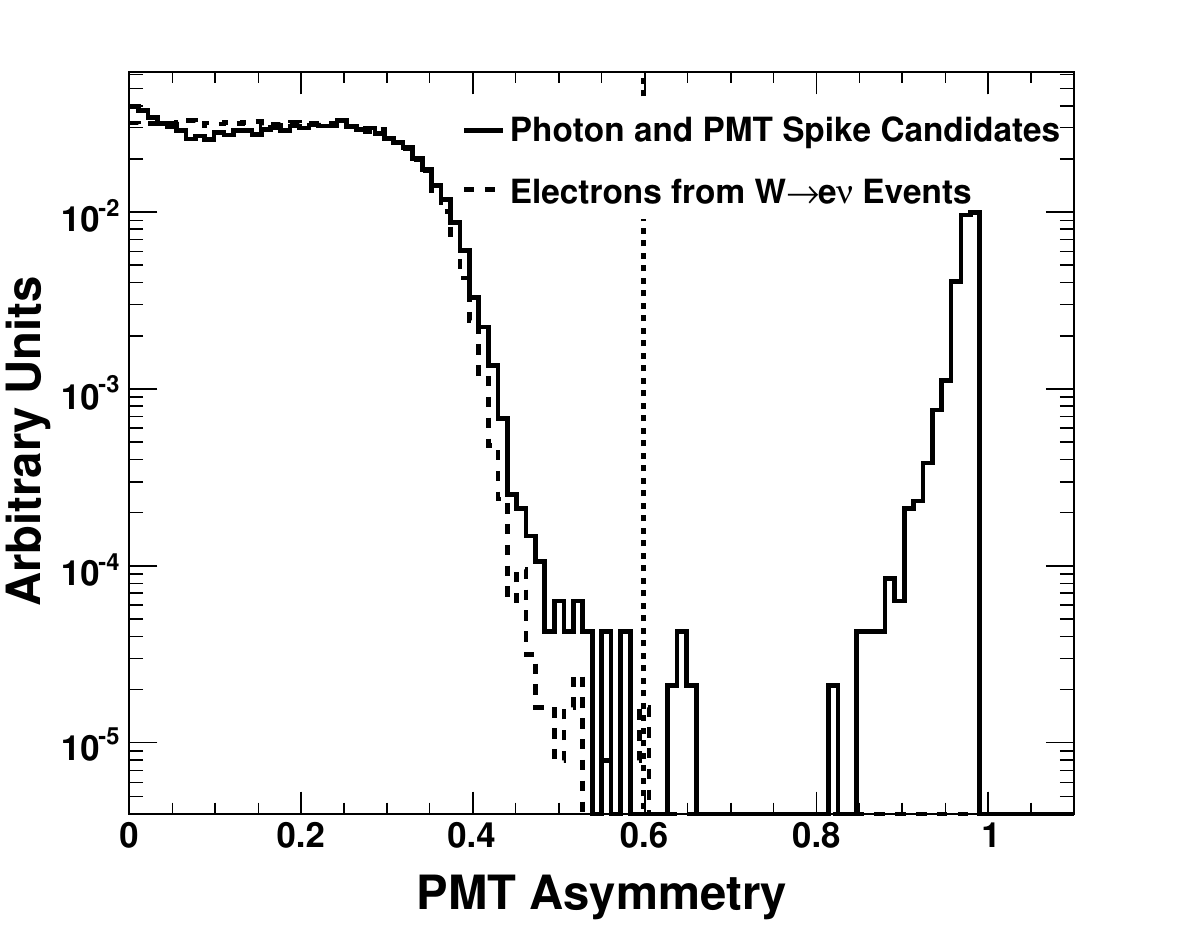}
  \caption[The magnitude of the PMT asymmetry for a the full
  photon+\met sample that contains both PMT spikes and real photons in
  (a), and the asymmetry for electrons from pure \Wenu events in
  (b).]{\label{fig:bkg_spikes_asym}\small A comparison of the PMT
    asymmetry, ${\rm A_{P}}$, for a photon+\met sample that contains both PMT
    spikes and real photons,  
    and  a sample of electrons
    from \Wenu events.  
    PMT spikes can be effectively removed by
    requiring the asymmetry to be less than 0.6.
}
\end{figure} 

A second change to the standard photon ID is to add a requirement to  remove high-energy 
photon candidates that are caused by a high-voltage
breakdown (``spike'') between the PMT photocathode and the surrounding
material. Such an occurrence can produce false photon candidates that
are uncorrelated with the collision and appear delayed in time. 
Spikes are identified by the 
asymmetry of the two energy measurements of the PMTs of a tower: 
\begin{equation}
{\rm A_{P}}=\frac{|E_{\mathrm{PMT1}}-E_{\mathrm{PMT2}}|}{E_{\mathrm{PMT1}}+E_{\mathrm{PMT2}}}
   \label{eq:pmtasym}
\end{equation}
where $E_{\mathrm{PMT1}}$ and $E_{\mathrm{PMT2}}$ are the two PMT energies. 
Figure \ref{fig:bkg_spikes_asym} compares photon candidates from both
real photons and spikes to real electrons from \Wenu events.  The
photon candidates pass all but the ${\rm A_{P}}$ identification
requirements shown in Table~\ref{tab:phcuts} in events with
$\met>30$~GeV, while the electrons selected pass the requirements 
in Table~\ref{tab:bkg_idcuts_el}.
As shown in the figure, a requirement of ${\rm A_{P}}<0.6$
rejects $\sim$100\% of all spikes with a minimal loss in effeciency for real photons.  Thus, this source will be neglected in the background estimate.

\subsection{Measurement of the Collision Time and Position}
\label{sec:vertexing}

The corrected photon time is a combination of 
 the measurements of the photon arrival time and position 
using the EMTiming system and the primary interaction position and time
 using the COT. We begin with a description of a new vertexing
 algorithm that provides this time and  continue with the EMTiming
 measurement and the final \tcorr  calculation.  

The standard vertexing algorithms~\cite{vertalg} reconstruct the
vertex position ($\vec{x}_{i}$) from high quality COT and SVX tracks. 
However, it is important also to measure \tzero and 
to separate tracks from the vertex that produced the photon from any other vertex that lies
close in space but occurs at a different time. This is 
particularly true at high instantaneous luminosities where two or
more collisions can occur in one event and can lie close to each other 
in $z$.  Misassigned vertex events are a dominant
contribution to the background estimate. 

\begin{table}[tbp]
  \begin{center}
    {\small \addtolength{\tabcolsep}{0.1em}
    \begin{tabular}{l}
      \hline\hline
      $\pt>0.3$~\punit \\ 
      $\pt>1.4$~\punit or passes the slow proton rejection cuts \\  \ \ \ \  if charge$>0$ \\ 
      $|\eta|<1.6$ \\ 
      $|z_{0}|<70$~cm \\  
      $\mathrm{Err}(z_{0})<1$~cm \\ 
      $|t_{0}|<40$~ns \\
      $0.05<\mathrm{Err}(t_{0})<0.8$~ns \\
      Traverses $\ge$3 stereo and $\ge$3 axial COT \\
      \ \ superlayers with 5 hits each \\  \hline \hline
    \end{tabular}
    }
    \vspace{1em}
    \caption[The set of requirements for tracks to be included in the
    vertex reconstruction.]{\label{tab:trackcuts}\small The set of
      requirements for tracks to be included in the vertex
      reconstruction. These are the standard tracking 
      requirements~\cite{CDFII}, 
but with additional quality
      requirements on the \tzero measurement and a 
      slow proton rejection requirement~\cite{slow_proton} 
      to remove tracks that likely have a mis-measured track
      \tzero. These variables are described in the Appendix.}
  \end{center}
\end{table}
\begin{table}[tbp]
  \begin{center}
    {\small \addtolength{\tabcolsep}{0.1em}
      \begin{tabular}{l}
        \hline \hline
        $\et>20$~GeV and $|\eta|\le1.0$ \\
	Fiducial: not near the boundary, in $\phi$ or $z$, of a \\
      \ \ \ \ calorimeter tower \\ 
        $E_{\mathrm{Had}}/E_{\mathrm{Em}}<0.055+0.00045 \cdot E$ \\ 
        $\chi^{2}_{\mathrm{Strip}}<10$    \\ 
        $Lshr<0.2$ \\
        $\pt>10$~\punit  \\ 
        $E^{\mathrm{Iso}}<0.1 \cdot \et$ \\
        $-3 < \Delta x \cdot q < 1.5$ and $|\Delta z|<3$~cm \\
        $|z_{0}|<60$~cm \\
        $\pt>50$~\punit or $0.5<E/p<2.0$  \\
        Track traverses $\ge$3 stereo and $\ge$3 axial COT \\
        \ \ \ \ \ superlayers with 5 hits each 
        \\ \hline \hline
      \end{tabular}
    }
    \vspace{1em}
    \caption[
The electron ID requirements and global
    event requirements for the data and MC samples. For use in creating 
the \Wenu timing sample. ``$q$'' is the
    charge of the electron.]{\label{tab:bkg_idcuts_el}\small The identification requirements for use in selecting electrons with high purity to study the vertexing performance. Note that ``$q$'' is the
      charge of the electron. The identification requirements are
      summarized in the Appendix and described in 
    more detail in~\cite{CDFII}. }
  \end{center}
\end{table}

To solve this problem, we have developed a new vertex reconstruction
algorithm based on track clustering.  The procedure~\cite{vertex} uses tracks with a
well measured  \tzero and \zzero  that pass the requirements in
Table~\ref{tab:trackcuts} and groups those  
that are close to each other in both space and time. The algorithm can
be separated into  three phases: (1) the initial  
assignment of tracks that are nearby in \tzero and \zzero into
clusters, (2) the determination of the \tzero, \zzero, and \sumpt of
the vertex, 
 and (3) the adjustment of the 
  number of clusters by merging clusters that are close
 to each other.  

A simple algorithm is used to make a preliminary  assignment of all tracks into clusters.  It is designed to overestimate, initially, the number of vertices in the event to obviate the need for dividing a single cluster into two separate clusters, called splitting. The highest-\pt track is designated as the ``seed'' of the first cluster, and 
any lower-\pt tracks that lie within three times
the typical cluster RMS 
 (0.6~ns 
and 1.0~cm for \tzero and \zzero, respectively) are also assigned to it. The highest-\pt track
from the remaining set of tracks is then picked as a second seed and
tracks are assigned to it, and so forth until no tracks are left.
The mean position and time for each cluster,  $z_{\rm vertex}$
and $t_{\rm vertex}$  respectively,
 is then  calculated. 

The vertexing algorithm is essentially a likelihood fit
using an iterative procedure to get a best estimate of the true number
of vertices and their parameters~\cite{emax}.  We allow the cluster parameters to float in the fit and maximize the probability that each track is a member of a vertex with a track density that is Gaussian in both space and time. 
All clusters are fitted simultaneously. 
If during the procedure the means of two clusters are within
both 3~cm in \zzero and 1.8~ns in \tzero or if two clusters share the same set
of tracks, then the clusters are merged. 
No splitting is done because 
 the initial seeding  is designed to  
overestimate the number of clusters.  Splitting a cluster 
with a too-large RMS can result in two clusters that both do not
pass the final requirements and would reduce the clustering
efficiency. Having two clusters merged that are close in both space
and time does not substantially affect the \tcorr measurement. We choose the primary
vertex for an event to be the highest \sumpt cluster that has at least 4
tracks.

\subsubsection{Vertexing Resolution and Efficiency}

\begin{figure}[tbp]
  \centering 
  \vspace{-20pt}
  \subfigure[] {
   \hspace{-10pt}
    \includegraphics[scale=.46]{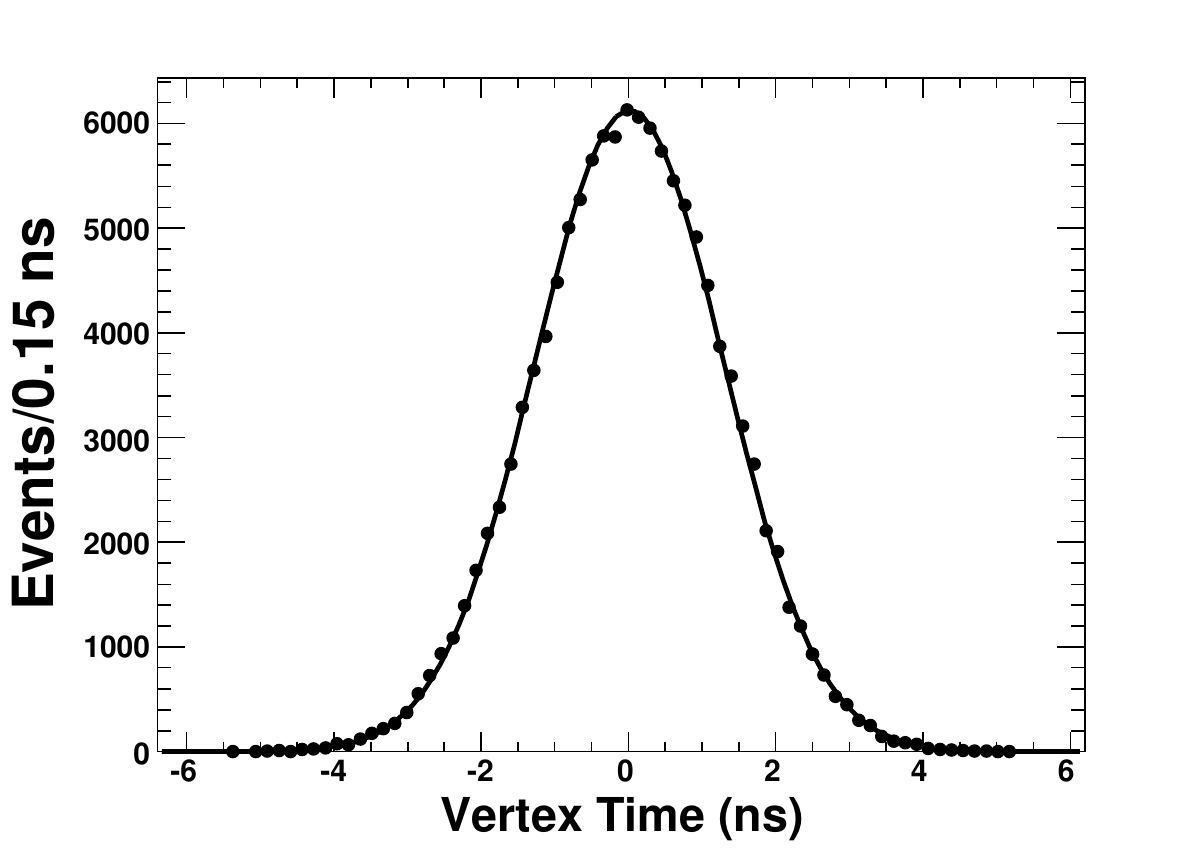}}\\
   \vspace{-19pt}
  \subfigure[]{
 \hspace{-10pt}
    \includegraphics[scale=.46]{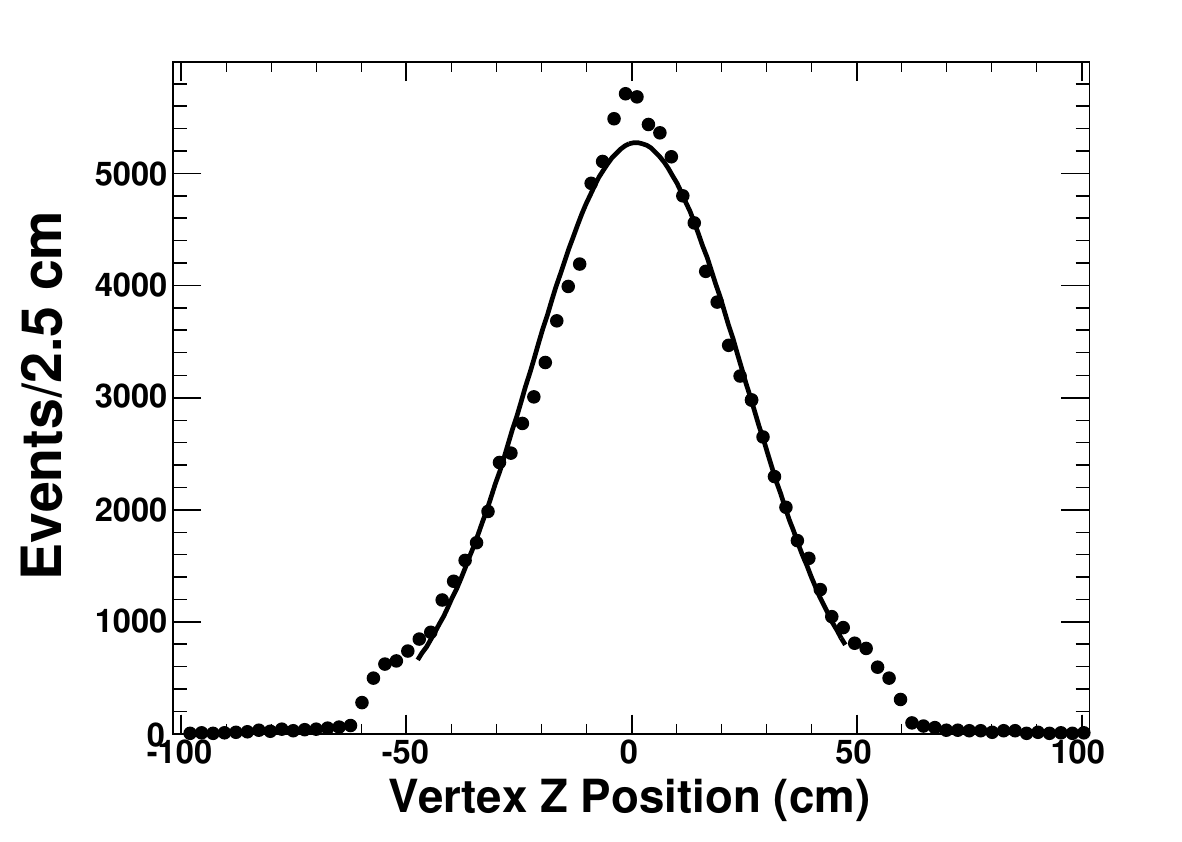}}\\
  \vspace{-19pt}
\subfigure[]{
\hspace{-10pt}
    \includegraphics[scale=.46]{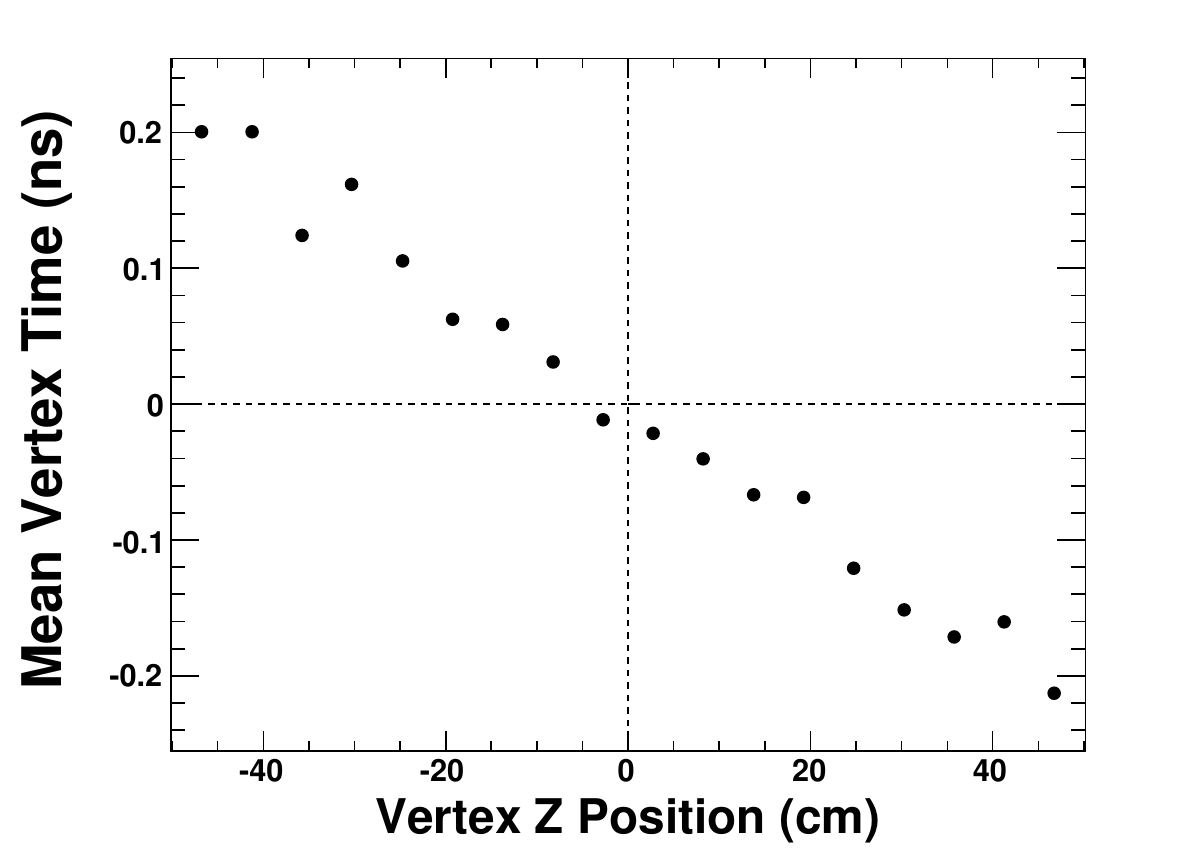}}
 \vspace{-19pt}
  \caption[The \tzero and \zzero of the reconstructed highest \sumpt
  vertex in \Wenu events without allowing the electron track to
  participate in the vertexing.]{\label{fig:vertdist}\small Plots (a), (b), and (c) show the \tzero,
     \zzero, and their correlation respectively for the reconstructed highest \sumpt vertex in \Wenu
    events.  The fits in (a) and (b) are both a single Gaussian.
     The falloff in the (b) at $|z|\simeq 60$~cm is due to the
     requirement that all tracks have $|z|<70$~cm.   
     In the search the vertex is 
     required to have $|z|<60$~cm.
  }
\end{figure}

The cluster resolution, the reconstruction
efficiency, and beam properties are measured using a high purity \Wenu data
sample, selected using the cuts in Table~\ref{tab:bkg_idcuts_el}.
To measure the performance for possible events with photons, 
the electron track is removed from the vertexing and is used to measure
the vertexing performance as it identifies the correct event vertex.  
Figure~\ref{fig:vertdist} shows the   
  \zzero and \tzero distributions as well as their correlation for the vertices in this sample. Both are roughly Gaussian and
centered at zero with an RMS of 25~cm and 1.28~ns, respectively,
reflecting the accelerator performance. There is a non-Gaussian excess around
zero in the \zzero distribution that comes from events that contain more than one vertex. In this
case the clustering has merged two vertices that are close to each
other, which most likely happens at $z=0$~cm. The correlation between the collision position and time distributions is caused by the differences in the proton and anti-proton bunch structure within the accelerator 
 ($\sigma_p\approx 50$~cm and
 $\sigma_{\bar{p}}\approx 70$~cm~\cite{pwt}).


\begin{figure}[tbp]
 \vspace{-18pt}
 \centering
        \subfigure[] {
   \hspace{-15pt}
    \includegraphics[scale=.47]{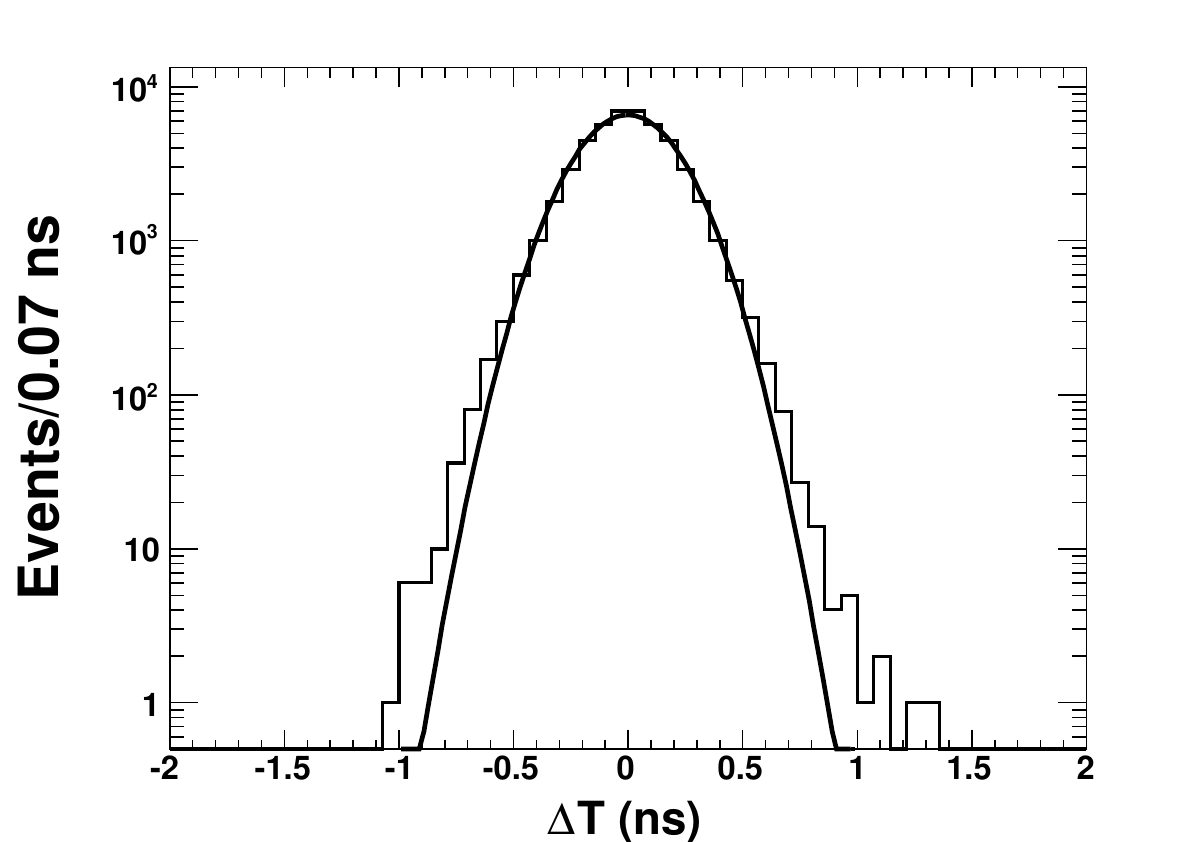}}\\
 \vspace{-19pt}
   \subfigure[] {
   \hspace{-15pt}
    \includegraphics[scale=.47]{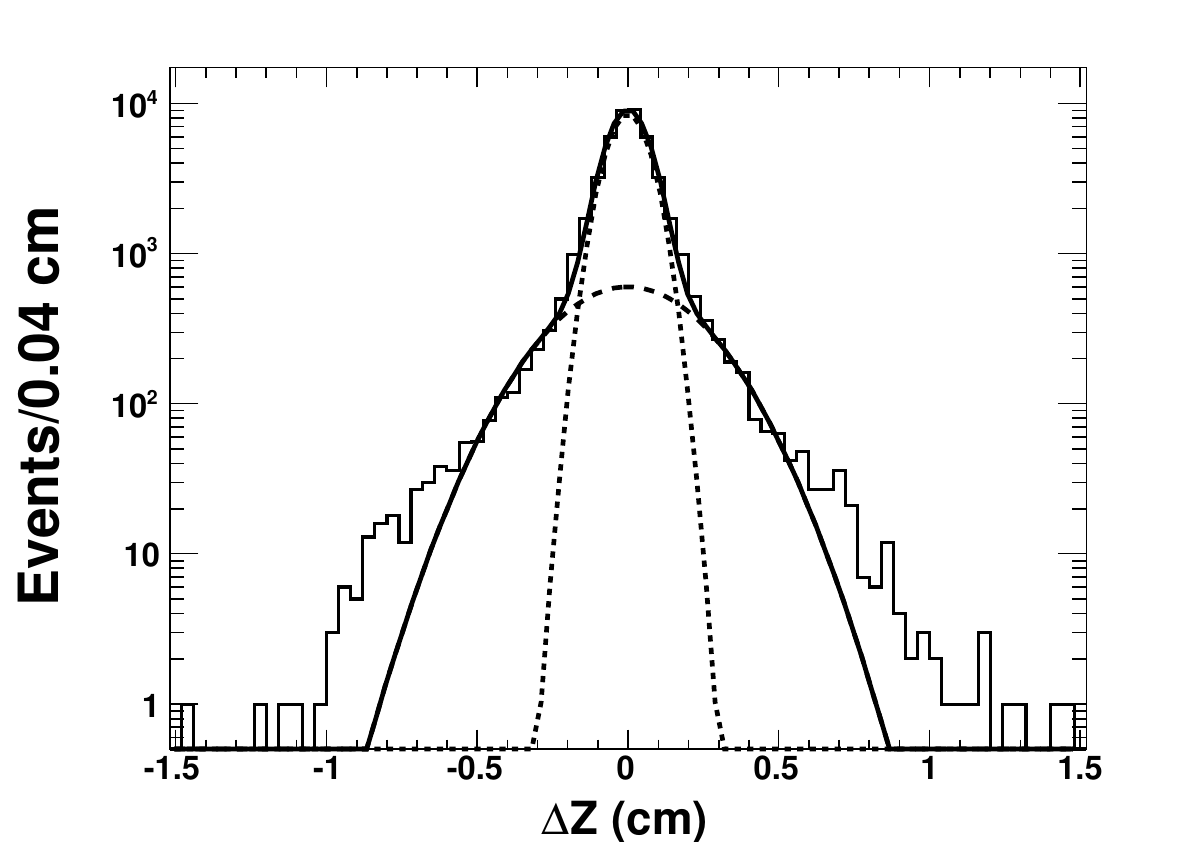}}\
 \vspace{-14pt}
\caption[The difference in $t$ and $z$ between two arbitrarily
  selected sets of tracks from the same reconstructed vertex in a
  \Wenu dataset with the electron track removed from the vertexing.
  ]{\label{fig:vertres}\small The difference in $t$ and $z$ between
    two arbitrarily selected sets of tracks from the same
    reconstructed vertex in a \Wenu dataset with the electron track
    removed from the vertexing. This is a measure of the vertex
    resolution. (a) is fit with one Gaussian while (b) is fit with two.  Note that the factor of $\sqrt{2}$ is already taken
    out.} 
\end{figure}

\begin{figure}[tbp]
\vspace{-20pt}
  \centering 
  \subfigure[]{
    \includegraphics[scale=.45]{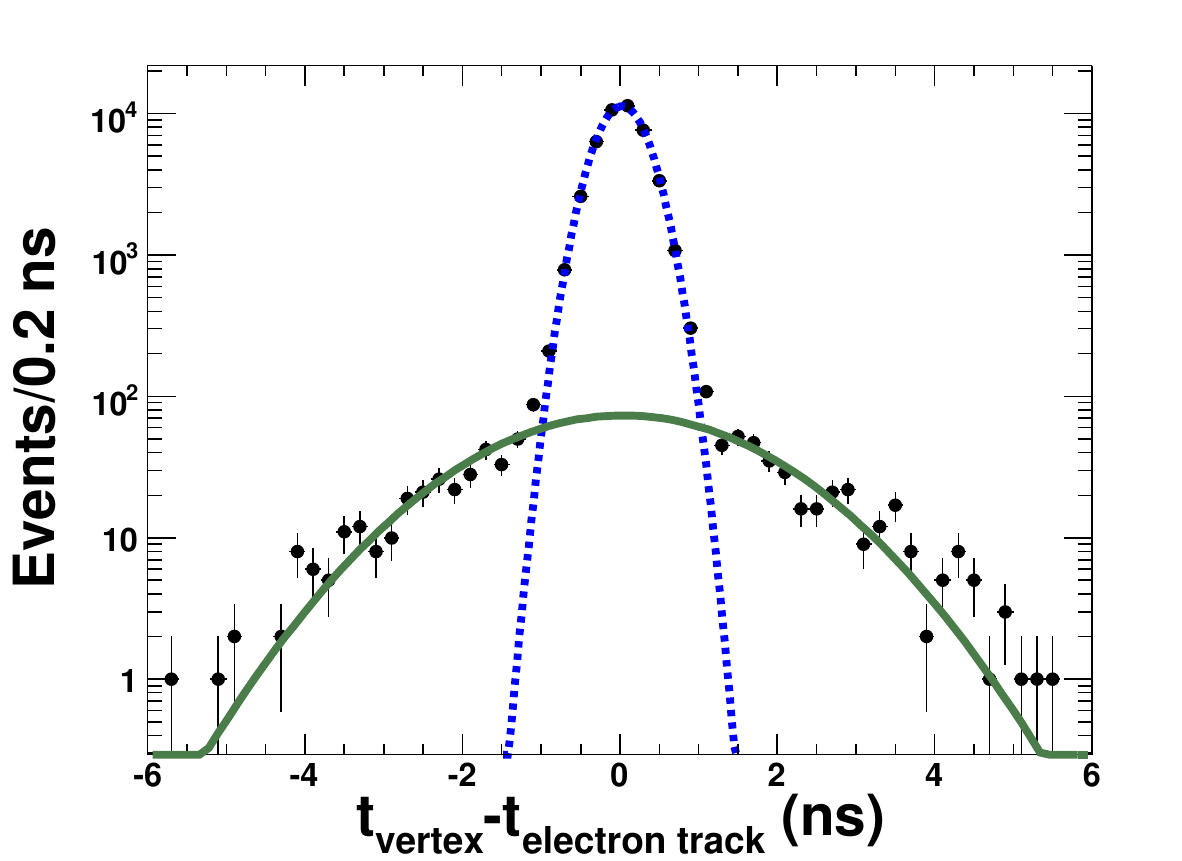}} \\
  \vspace{-19pt}
  \subfigure[] {
    \includegraphics[scale=.45]{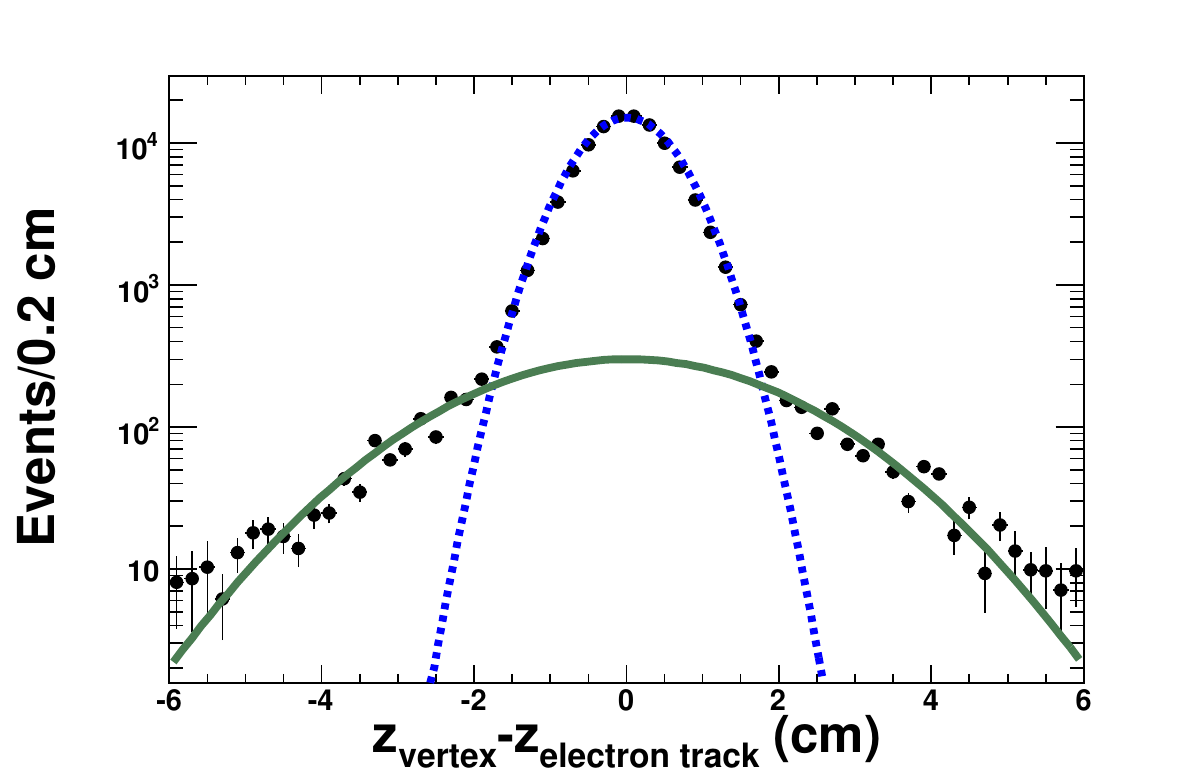}}
  \vspace{-14pt}
  \caption[The difference in \tzero (a)  and in \zzero (b)  between the electron
  track and the highest \sumpt reconstructed vertex without the
  electron track participating in the vertexing.
  ]{\label{fig:vertbias}\small The difference in $t$ (a) and in $z$
    (b) between the electron track and the highest \sumpt reconstructed
    vertex (without the electron track participating in the vertexing) in \Wenu events. 
    The distributions are centered at zero and fit with double Gaussians, indicating that there is
    no bias in the clustering procedure. The secondary Gaussian
    contains events where the electron does not originate from the highest 
    \sumpt vertex in the event.}
\end{figure}

The vertexing resolution 
is estimated using a subsample of the events with only one
reconstructed  vertex. For each
event the tracks in the vertex are randomly divided into two 
groups that are then separately put through the vertexing algorithm.
Figure~\ref{fig:vertres} shows the distance between the two clusters,
divided by $\sqrt{2}$ to take into account the two measurements, 
giving a resolution measurement of $\sigma_{t}=0.22~$ns and
$\sigma_{z}=0.24~$cm.  
 The secondary Gaussian in Fig.~\ref{fig:vertres}(b) indicates
cases where two different vertices have been combined into one cluster. 
Figure~\ref{fig:vertbias}
shows the difference in time and position between the reconstructed
cluster and the electron track (not included in the vertexing)  
for the full sample. The distributions are well described by two Gaussians that are both
symmetric and centered at zero, indicating no measurement bias. The
primary Gaussian distribution contains events where the reconstructed
cluster is the vertex that produced the electron.  Its RMS is dominated
by the resolution of the electron track position and time. The
secondary Gaussian distribution contains events where the electron
 does not originate from the highest \sumpt vertex in the event. 

\begin{figure}[tbp]
  \centering 
  \vspace{-28pt}
  \subfigure[] {
  \hspace{-10pt}
    \includegraphics[scale=.46]{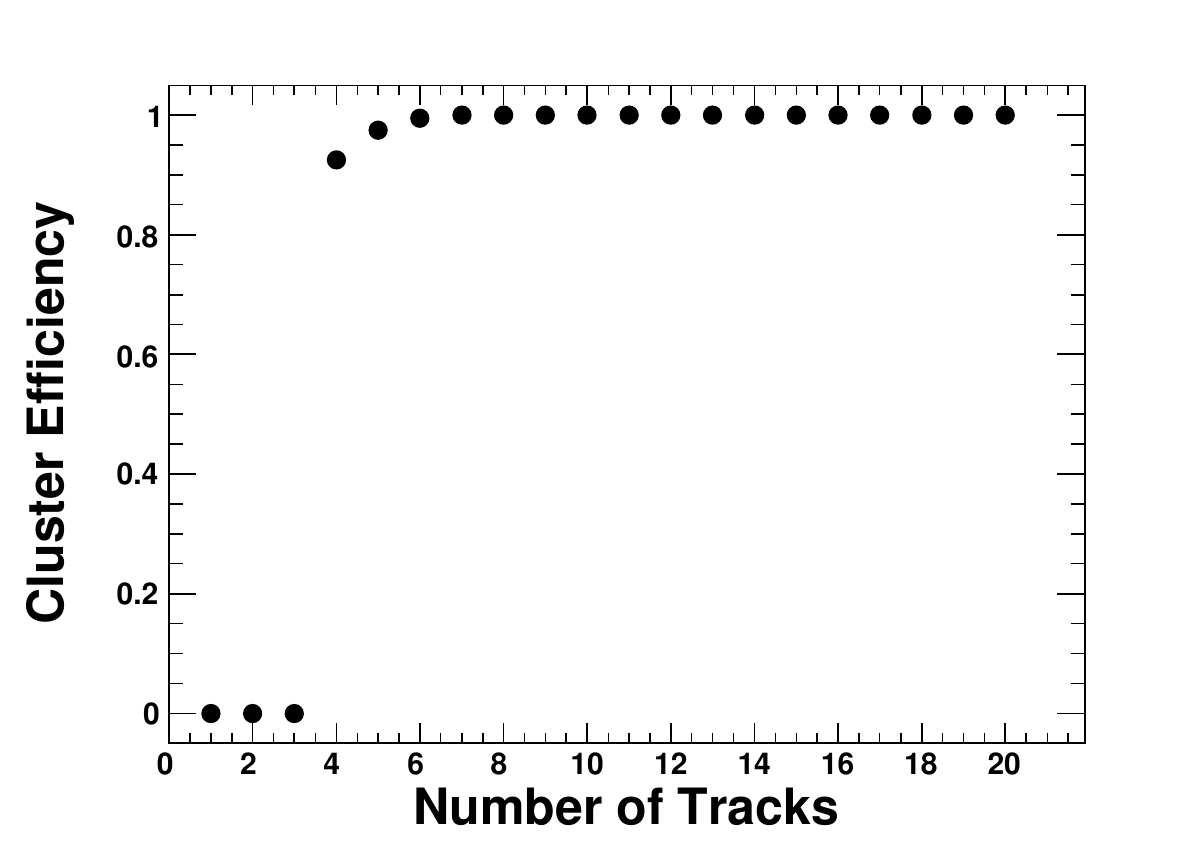}}\\
\vspace{-19pt}
 \subfigure[]{
  \hspace{-10pt} 
    \includegraphics[scale=.46]{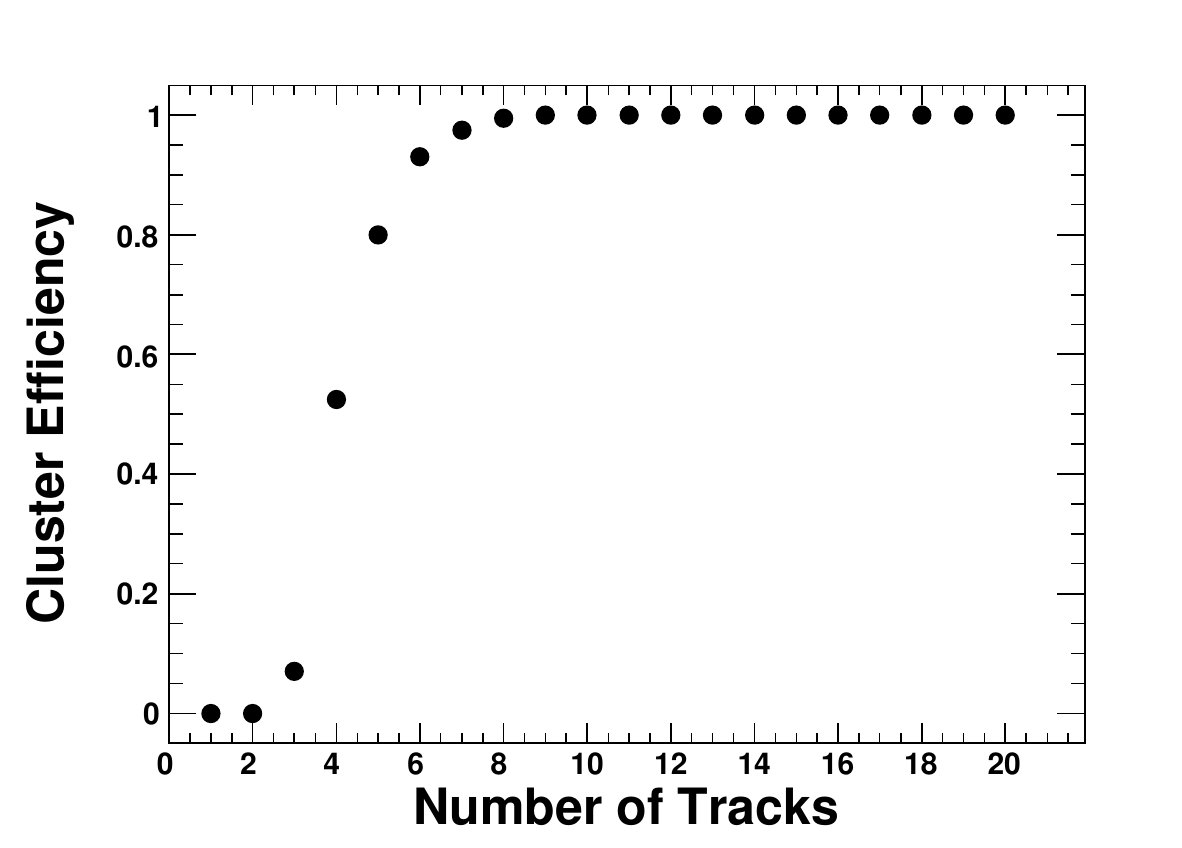}}\\
\vspace{-19pt}
  \subfigure[] {
  \hspace{-10pt}
    \includegraphics[scale=.46]{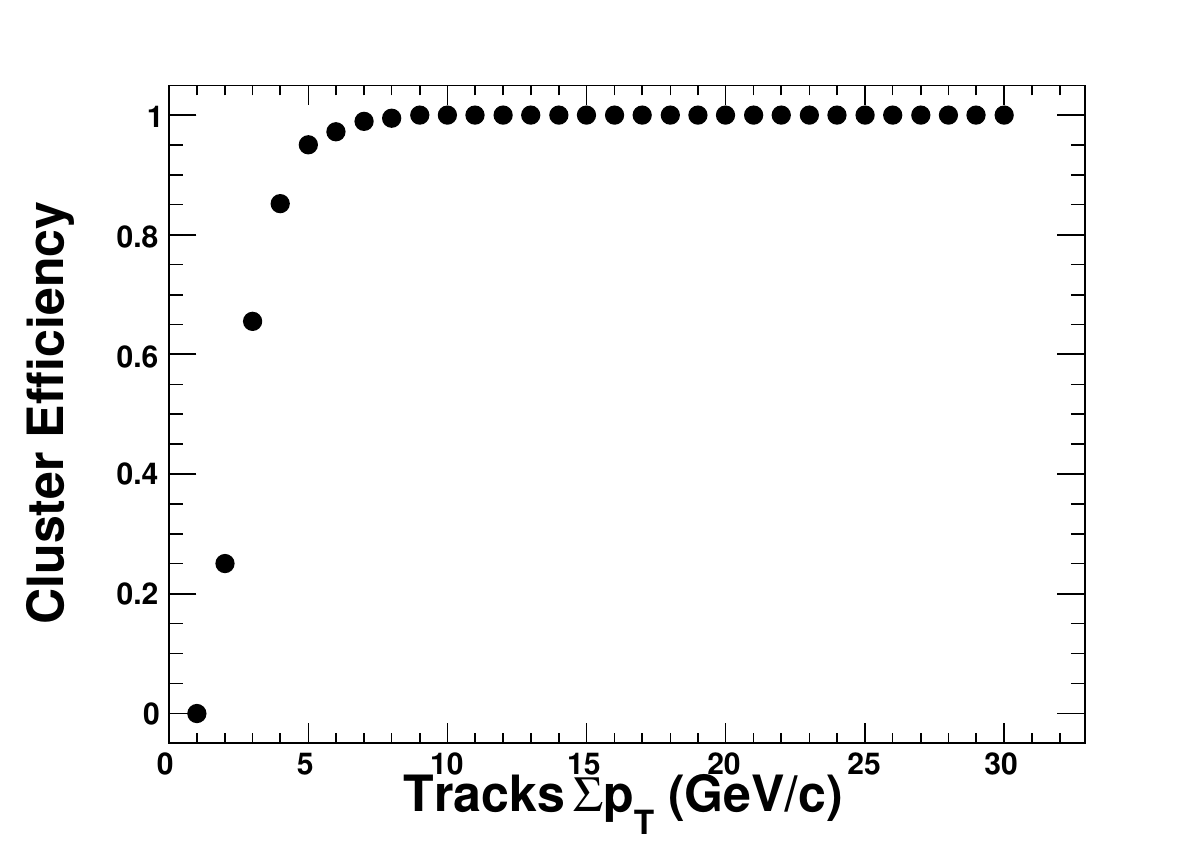}}
\vspace{-14pt}
  \caption[The clustering efficiency as a function of the number of
  tracks using (a) the subset method and (b) the window method, and
  (c) as a function of the \sumpt of the tracks using the window
  method.]{\label{fig:verteff}\small The clustering efficiency as a
    function of the number of tracks using (a) the subset method and
    (b) the window method, and (c) as a function of the \sumpt of the
    tracks using the window method. 
    Note that a cluster is required to have at least 4 tracks and the
    efficiency is 100\% for $\sumpt>15$~\punit in this search.}
\end{figure}

The efficiency of the vertex reconstruction algorithm is investigated
using two separate methods. The efficiency as a function of the number
of tracks is determined by selecting events that contain a cluster 
with a high track multiplicity. 
 Next, various random subsets of the tracks are taken that belong to
this cluster to see if they alone could produce a cluster.  
Figure~\ref{fig:verteff} shows the ratio of
subset samples in which a cluster is reconstructed to
all cases tried for a given set of tracks as a function of the number
of tracks in the various subsets. 
The algorithm
is over 90\% efficient if 4 tracks are present, where the inefficiency
is usually caused by the algorithm reconstructing two separate clusters each with 
$<$4 tracks, and 100\% efficient with 6 tracks (the 
final analysis requires at least 4 tracks). 
A second method that also allows for a measurement of the efficiency 
as a function of the \sumpt is to consider tracks in a $2\ {\rm 
  cm}\times 2\ {\rm ns}$ window around the electron track 
($\sim$5$\sigma$ in each direction) and search for clusters.
 Only events with at most one reconstructed vertex are considered.  
While this result is not biased by 
selecting cases with a known vertex, the disadvantage is that 
 for resolution reasons not 
all tracks are in the window, resulting in a 
small under-counting of the number of tracks. Figure~\ref{fig:verteff} 
shows that the efficiency as a function of the number of tracks in the 
vertex yields a similar result for the  two very different methods. 
 This gives confidence in the results as a function of
\sumpt. The efficiency plateaus at $ \sumpt=7$~\punit, as higher \pt tracks have a better \tzero resolution measurement.
It is important to note that the efficiency measurements 
 as a function of the number of tracks are sample-dependent. 
 For instance, if a sample is chosen that is biased towards a higher
average track \pt then the efficiency might be higher for a smaller
 number of tracks, or if a sample contains many high-\pt tracks,
the efficiency as function of \sumpt might plateau earlier. Since the
 search requires  $\sumpt>15$~\punit, as
 stated later, we take the efficiency for the vertex selection
 requirements to be 100\%.

\subsection{The Corrected Photon Time}
\label{sec:timing}

\begin{figure}[tbp]
  \centering
  \vspace{-14pt}
    \includegraphics 
[scale=0.46]{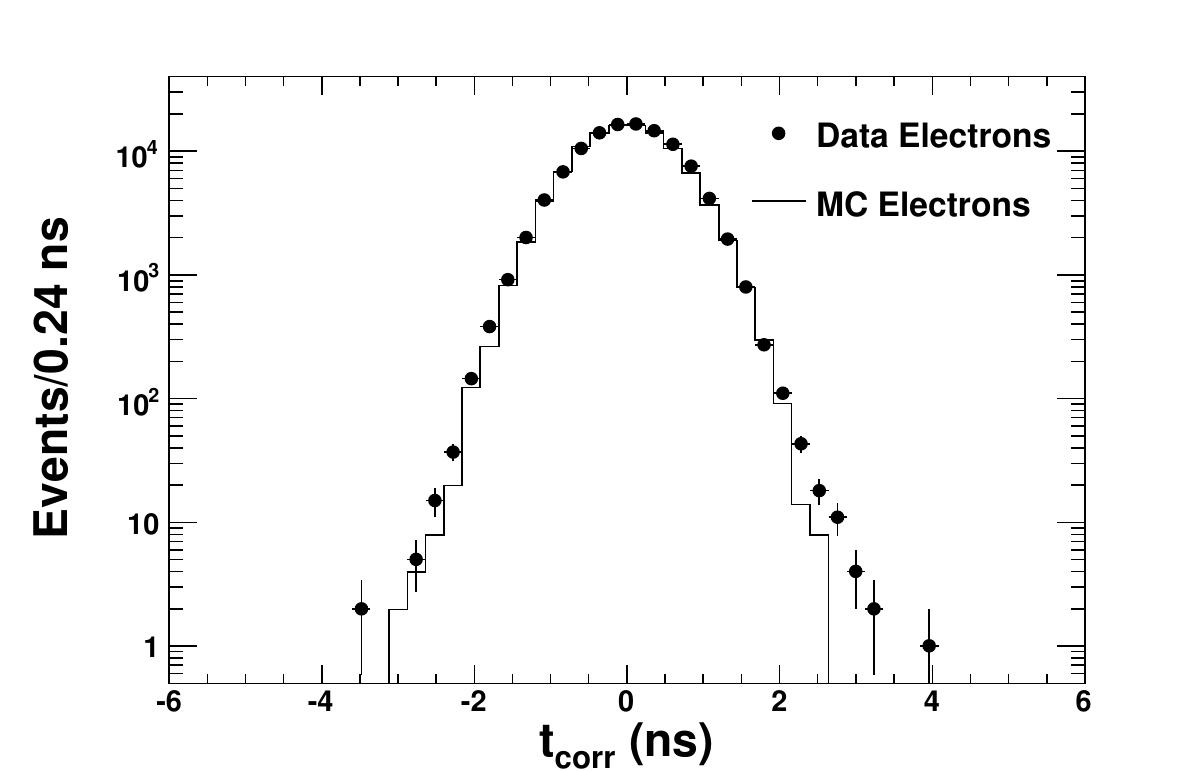}
  \caption[A comparison between MC (solid) and data (points) for the
\tcorr for electrons from a $W\rightarrow e\nu$ sample
  after cumulatively applying the various corrections (left side) and
  a comparison of the corrections themselves (right side).]{\small A
    comparison between MC (solid) and collision data (points) for  
\tcorr for electrons from a $W\rightarrow e\nu$ sample.  
The distributions are well
    centered around 0 and the resolutions of collision data and MC fit well with
    a fully corrected RMS of 0.64~ns. \label{fig:dataMC_cem}}
   
\end{figure}

With the vertex time and position in hand, we move to a full  measurement 
of \tcorr by incorporating the EMTiming information. 
The time of arrival recorded by the EMTiming system TDCs 
is corrected using calibrations 
that take into account 
 channel to channel variations and  an
energy-dependent (``slewing'') effect due to the fixed-threshold discriminators. A full description of the hardware as well as the correction and calibration 
procedure is described in Ref.~\cite{nim}.
The \tcorr resolution
for electrons from \Wenu events is 0.64~ns (0.63~ns) for collision data (MC),
dominated by the intrinsic resolution (0.5~ns), the precision of the
TDC output (0.29~ns) and the vertex \tzero resolution (0.22~ns). 
A comparison of detector simulation to collision data for \Wenu events is shown in Fig.~\ref{fig:dataMC_cem}. There are no non-gaussian tails out to $\sim5\sigma$.

\section{Triggers, Datasets and Event Preselection}
\label{sec:selection}

The event selection is a three stage process. The stages are (1) an
online  sample is selected (during data taking), (2) a \gmetjet
``preselection sample'' 
 is selected offline, and (3) the event selection uses optimized 
 final event selection requirements.
The full set of requirements that determine the preselection sample 
for the search are summarized in Table~\ref{tab:idcuts}. The
optimization and final event requirements are described in
Section~\ref{sec:optimization}. 

The analysis begins by selecting events online using a single set of
3-level trigger requirements that 
 require a photon candidate and \met. 
The Level 1 trigger requires   
a single tower in the calorimeter with $|\eta|<1.1$, $\et>8$~GeV,  
$E_{\mathrm{Had}}/E_{\mathrm{EM}}<0.125$~GeV,  and $\met>15$~GeV. 
For a description of the ID variables, see the Appendix.  
The Level 2 trigger requires the event to have an  EM cluster with 
 $\et\ge20$~GeV and \met$\geq15$~GeV. 
At Level 3 the requirements are tightened with $\et>25$ GeV,
 $E_{\mathrm{Had}}/E_{\mathrm{EM}}<0.125$, and  \met$>$25~GeV. 
The data consist of events from the data-taking
period from December 2004, when the EMTiming system became fully
functional, until November 2005. The data  
 correspond to an integrated luminosity of
(570$\pm$34)~\invpb. 

The sample of \gmet candidate events that pass the trigger requirements is 
processed offline where the event
characteristics are refined to increase the signal purity and 
further reduce the backgrounds. The offline preselection requirements
include photon ID and \met requirements as well as jet, vertex, and
cosmic ray rejection requirements.  To ensure that all 
signal events would have passed the trigger with 100\% efficiency
 each event is required to have $\met>30$~GeV, a photon with 
$\et>30$~GeV, and pass
the identification criteria shown in Table \ref {tab:phcuts}. 
 
We require the presence of at least one jet and a high \sumpt
vertex in each event for the preselection sample. 
This preserves the acceptance of $\ntwo\conepm$ and
$\conep\conem$ production while maintaining a search strategy that is as
model-independent as possible. 
While the term ``jet'' typically refers to the hadronization of a high
energy quark or gluon that is produced in the collision, at CDF  
jets are identified as clusters of energy in the calorimeter. Hence,  
the hadronic decays of taus and/or the energy deposits from electrons or photons are also efficiently reconstructed as jets. 
Requiring at least a single jet with $\et>30$~GeV and
$|\eta|<2.1$~\cite{jets} retains high efficiency and significantly
reduces  non-collision backgrounds which typically only produce a
single photon candidate.  
As previously mentioned, each event must also have a good space-time vertex with at least 4 good tracks and 
a \sumpt of at least 15~\punit. 
This allows for a good \tcorr measurement and further helps reduce the non-collision backgrounds.
 We also require $|z|<60$~cm and $|t_{0}|<5$~ns for tracks to be included in the vertexing 
 so that both the COT tracking and the calorimeter are able to produce
 high quality measurements.

A cosmic ray that traverses the detector can create hits in the muon system that are not associated with tracks in the COT  and deposit a photon candidate nearby in the calorimeter. 
An event is rejected from the preselection sample if there are potential cosmic--ray hits in the muon chamber within 30 degrees in $\phi$ of the photon that are not matched to any track.
Table~\ref{tab:evtsel} lists the cumulative
number of events that pass each of the successive requirements to create our
preselection sample. 

\begin{table}[tb]
  \begin{center}
    {\small \addtolength{\tabcolsep}{0.1em}
    \begin{tabular}{l}
      \hline \hline 
     
\multicolumn{1}{c}{Photon} \\ \hline

      $\et>30$~GeV and $|\eta|\le1.0$ \\ 
      Fiducial: not near the boundary, in $\phi$ or $z$, of a \\
      \ \ \ \ calorimeter tower \\ 
      $E_{\mathrm{Had}}/E_{\mathrm{EM}}<0.125$ \\ 
      Energy in a $\Delta$$R=0.4$ cone around the photon
       \\ \ \ \ \ excluding the photon energy: \\
      \ \ \ \ $E^{\mathrm{Iso}}<2.0~\mathrm{GeV}+0.02 \cdot (\et-20~\mathrm{GeV})$   \\  
      No tracks pointing at the cluster or \\
      \ \ \ \ one track with $p_{T}<1.0~\punit+0.005\cdot \et$    \\ 
      \sumpt of tracks in the \\ \ \ \ \  $\Delta$$R=0.4$ cone $<2.0~\punit+0.005 \cdot \et$   \\ 
      No second cluster in the shower maximum detector \\ \ \ \ \ or
      $E^{\mathrm{2^{nd} cluster}}<2.4~\mathrm{GeV}+0.01 \cdot \et$   \\ 
      ${\rm A_{P}}=\frac{|E_{\mathrm{PMT1}}-E_{\mathrm{PMT2}}|}{E_{\mathrm{PMT1}}+E_{\mathrm{PMT2}}}<0.6$  \\  
   \hline
\multicolumn{1}{c}{Jet} \\ \hline
      $\et^{\mathrm{jet}}>30$~GeV \\
      $|\eta^{\mathrm{jet}}|<2.0$ \\ 
       \hline
      \multicolumn{1}{c}{Highest $\Sigma p_{T}$ Space-Time Vertex
      with:} \\ \hline
      $N_{\mathrm{trks}}\ge 4$ \\
      $\sumpt\gt15$~\punit \\
      $|z|<60$~cm \\
      $|t_{0}|<5$~ns \\
       \hline
      \multicolumn{1}{c}{Global Event Cuts} \\ \hline
      $\met>30$~GeV \\
      Passes cosmic ray rejection requirements \\
       \hline \hline
    \end{tabular}
    }
    \vspace{1em}
    \caption[The jet, vertex and global event requirements
    used to obtain the preselection sample of \gmetjet
    events.]{\label{tab:idcuts}\small The 
      requirements used to obtain the preselection sample
      of \gmetjet events.  
       The cosmic ray rejection cut is described in
      more detail in~\cite{onyisi}. The number of
      events in the data that pass each cut are shown in
      Table~\ref{tab:evtsel}.  For more detail on the ID variables, see the Appendix. 
}
 \end{center}
\end{table}

\begin{table}[tb]
  \begin{center}
    {\small \addtolength{\tabcolsep}{0.1em}
    \begin{tabular}{lcc}
      \hline \hline Selection & No. of Observed \\&Events \\ \hline
      $\et>30$~GeV, $\met>30$~GeV, & \\
      \ \ photon ID and fiducial requirements & 119944 \\
      Vertex with $\sumpt>15$~\punit, $\geq$4 tracks  & 19574 \\
      $\ge$1 jet with $\et>30$~GeV and $|\eta|<2.0$  & 13097 \\
      Cosmics rejection  & 12855 \\
      \hline \hline
    \end{tabular}
    }
    \vspace{1em}
    \caption[Event reduction for the preselection \gmetjet
    dataset.]{\label{tab:evtsel}\small Event reduction for the
      preselection \gmetjet sample. 
      For the
      individual requirements see Table~\ref{tab:idcuts}.
    } 
  \end{center}
\end{table}

\section{Backgrounds}
\label{sec:bkgs}


Backgrounds to the \gmetjets signature 
 can be categorized into two different
 classes: collision and non-collision events.  
The rate that each type of background contributes to the final signal
 time-window is estimated solely from collision data using control samples of
 events that pass all of the final requirements excluding
 timing.  We define the ``kinematic sample'' as the events
 that pass the final event requirements (summarized in
 Section~\ref{sec:optimization},  Table~\ref{tab:prl}) 
 except the timing requirement.
The \tcorr distributions outside the timing signal region are used to
 normalize  each background,  which is then extrapolated into the
 signal time region.  In this section each of the backgrounds is
 described, and the signal estimation techniques are outlined.

\subsection{Standard Model Backgrounds: Prompt Photons}
\label{sec:bkg_prompt}

Prompt collision events dominate the sample and populate the region
around $\tcorr=0$~ns. As shown later, it is not important for this
search to distinguish further 
 between the various prompt photon sources. Most events are from 
 $\gamma j$ and $jj$ events with 
one jet reconstructed as a photon and with 
 \met from the  mis-measurement of the photon and/or jet
 in the calorimeter.
 A smaller source is from 
 SM \Wenuj events where the electron is misidentified as
a photon  
 and the $\nu$ leaves undetected to cause the \met. 
 In both cases these events can 
fall into the large \tcorr 
signal time window due to either
Gaussian fluctuations of the timing  measurement  
or a  wrong
collision vertex selection. The latter case dominates the SM background
estimate and is more likely at high
instantaneous luminosity when there are multiple collision vertices
reconstructed.

\begin{figure}[tbp!]
  \centering
  \vspace{-10pt}
\subfigure[]{
  \hspace{-8pt}
    \includegraphics[scale=.46]{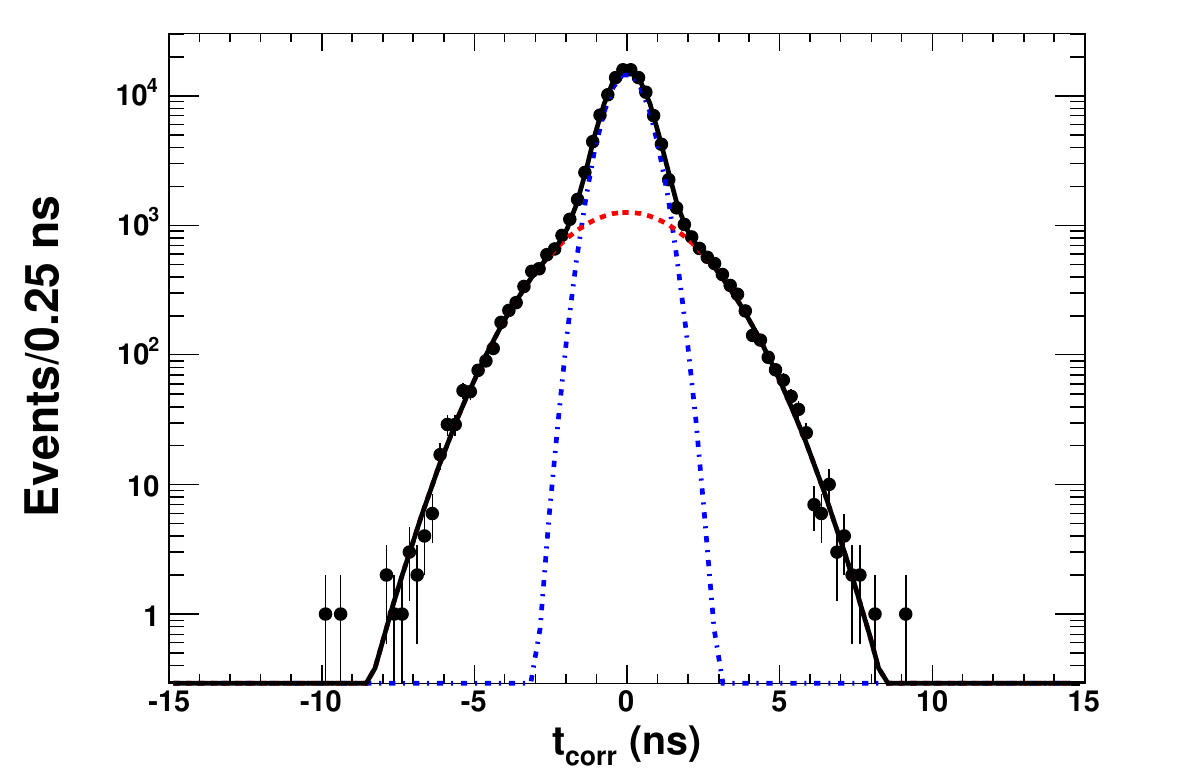}} \\
  \vspace{-19pt}
\subfigure[]{
  \hspace{-8pt}
 \includegraphics[scale=0.46]{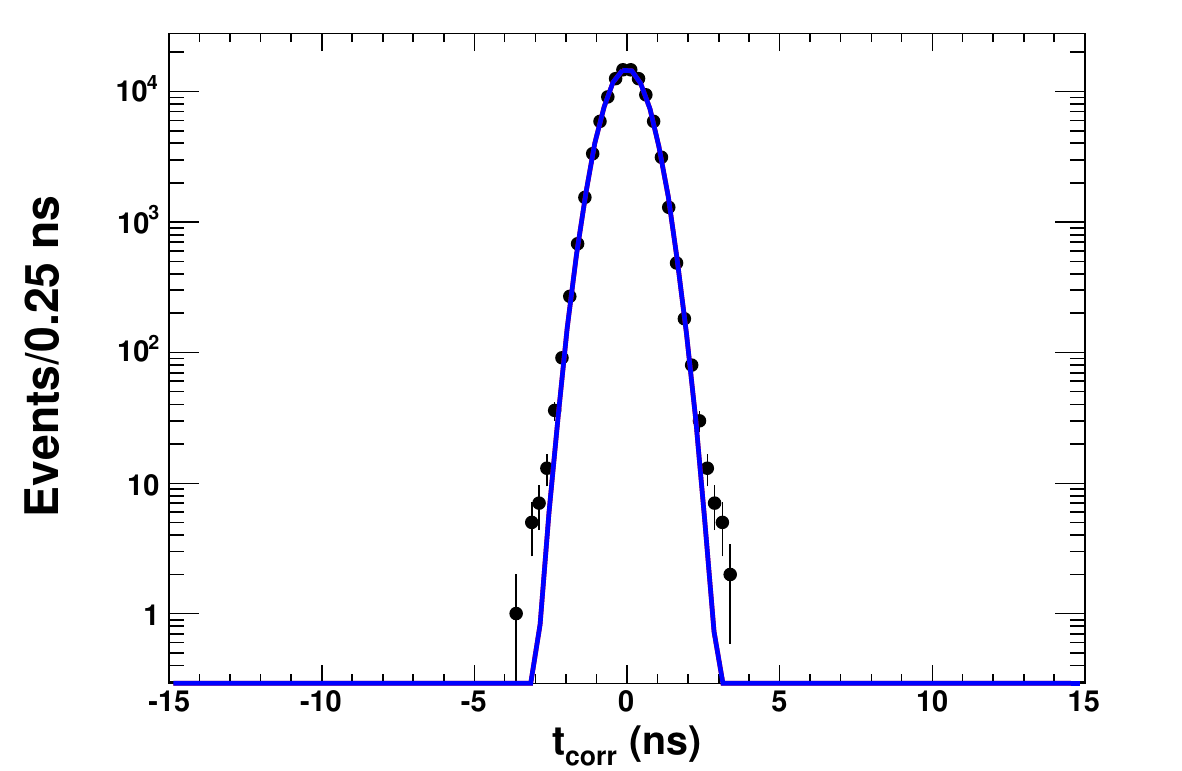}} \\
  \vspace{-19pt}
\subfigure[]{
  \hspace{-8pt}
 \includegraphics[scale=.46]{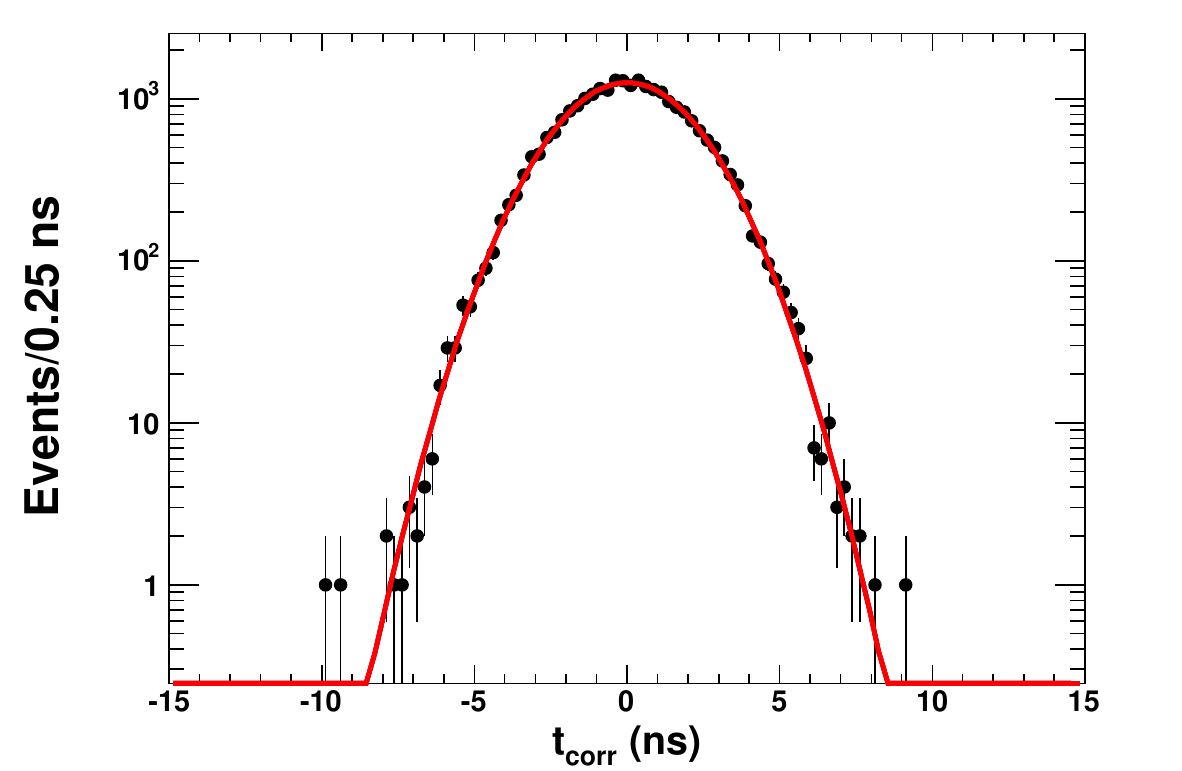}}
  \vspace{-14pt}
  
\caption[The electron \tcorr in a sample of \Wenu
  events.]{\label{fig:w_time}\small The \tcorr distribution for 
    electrons in a sample of
    \Wenu events. In plot (a) the two Gaussians correspond to the cases when the
    highest-\sumpt vertex is associated to the electron track and when
    it is not. These cases can be separated by requiring a match (b) and an
    anti-match (c), respectively, between the vertex and the electron track in
    both space and time.}
\end{figure}

To study the \tcorr distribution for promptly produced photons, a sample of 
\Wenu events is selected  using the requirements described in
Table~\ref{tab:elcuts}. This sample is used for the reasons described
in Section~\ref{sec:photonid}, and has the additional advantage that the 
 electron track in the COT allows for a determination of the
correct vertex.  To mimic closely the vertexing for
events with photons, the electron track is dropped from the vertex
clustering.  The highest-\sumpt vertex is chosen as the most likely
to have produced the EM cluster (the ``photon'').
Figure~\ref{fig:w_time} shows the
resulting \tcorr distribution and has a double-Gaussian shape.  One
Gaussian comes from events where the vertex choice is correct and the
other Gaussian comes from events where the vertex choice is
 incorrect~\cite{note}.

Figure~\ref{fig:w_time} also shows 
these events 
separated into right and wrong-vertex subsamples, selected by requiring a tight
match ($|z_{\rm track}-z_{\rm vertex}|<2$~cm and $|t_{\rm
  track}-t_{\rm vertex}|<2$~ns) and anti-match between the electron
track and the vertex, respectively. In both cases the distributions
are Gaussian and centered at zero. The right vertex selection has an
RMS of 0.64~ns, reflecting the system resolution, 
and the wrong-vertex selection has an RMS of $\sim$2.0~ns. 
 The wrong-vertex time distribution  can be understood by combining
 the RMS of the 
  time distribution without the vertex \tzero and \zzero corrections 
 (RMS$=1.6$~ns)  
 with the
RMS of the collision \tzero distribution (RMS$=1.28$~ns as shown in
Fig.~\ref{fig:vertdist}): ${\rm RMS}_{\rm wrong~vertex}=\sqrt{1.6^2+1.28^2}=2.05$~ns. The number of
events in the \tcorr signal region ($\tcorr>0$) for prompt, SM sources 
 can thus be estimated by simple extrapolation from a measurement of
 the timing distribution for $\tcorr<1.2$~ns.

\begin{figure}[tbp]
\vspace{-16pt}
\centering
\subfigure[]{
    \includegraphics [scale=0.47] {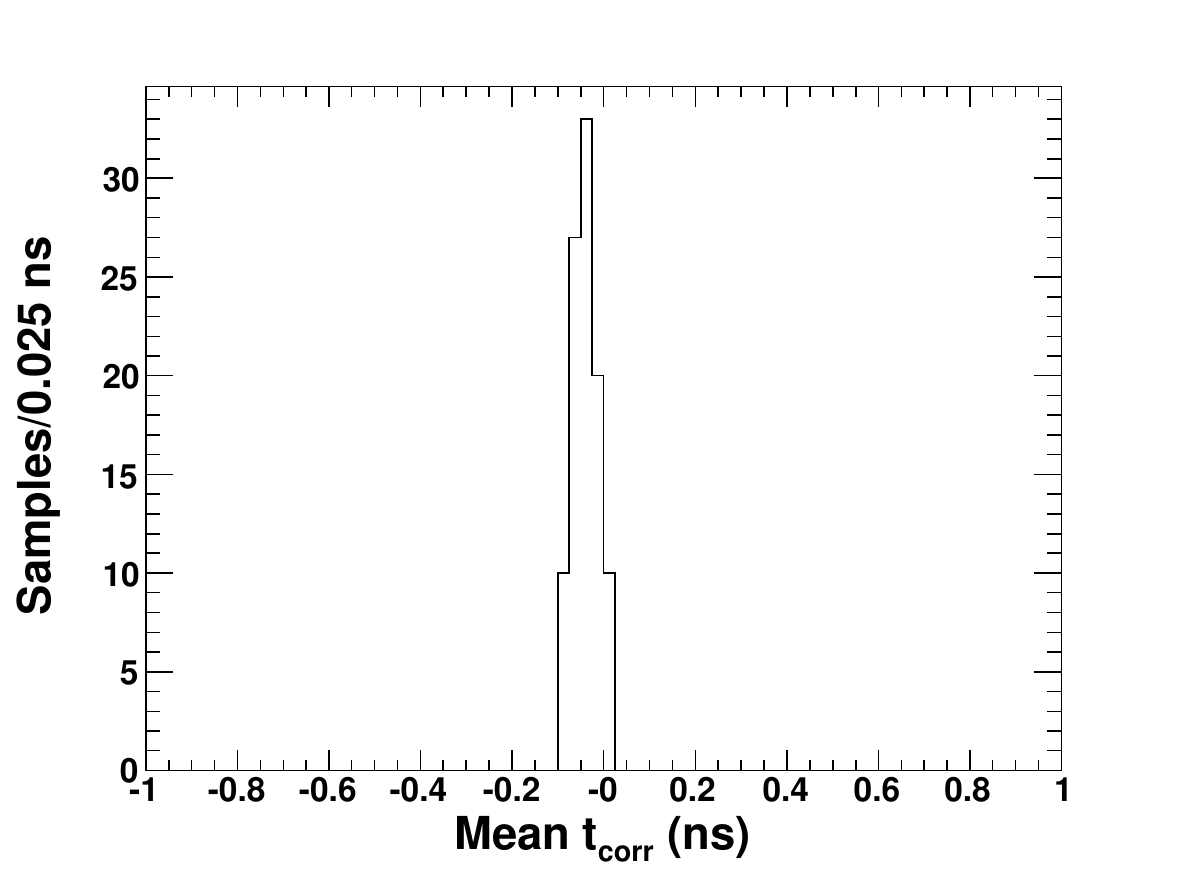}}\\
\vspace{-19pt}
 \subfigure[]{
     \includegraphics [scale=0.47] {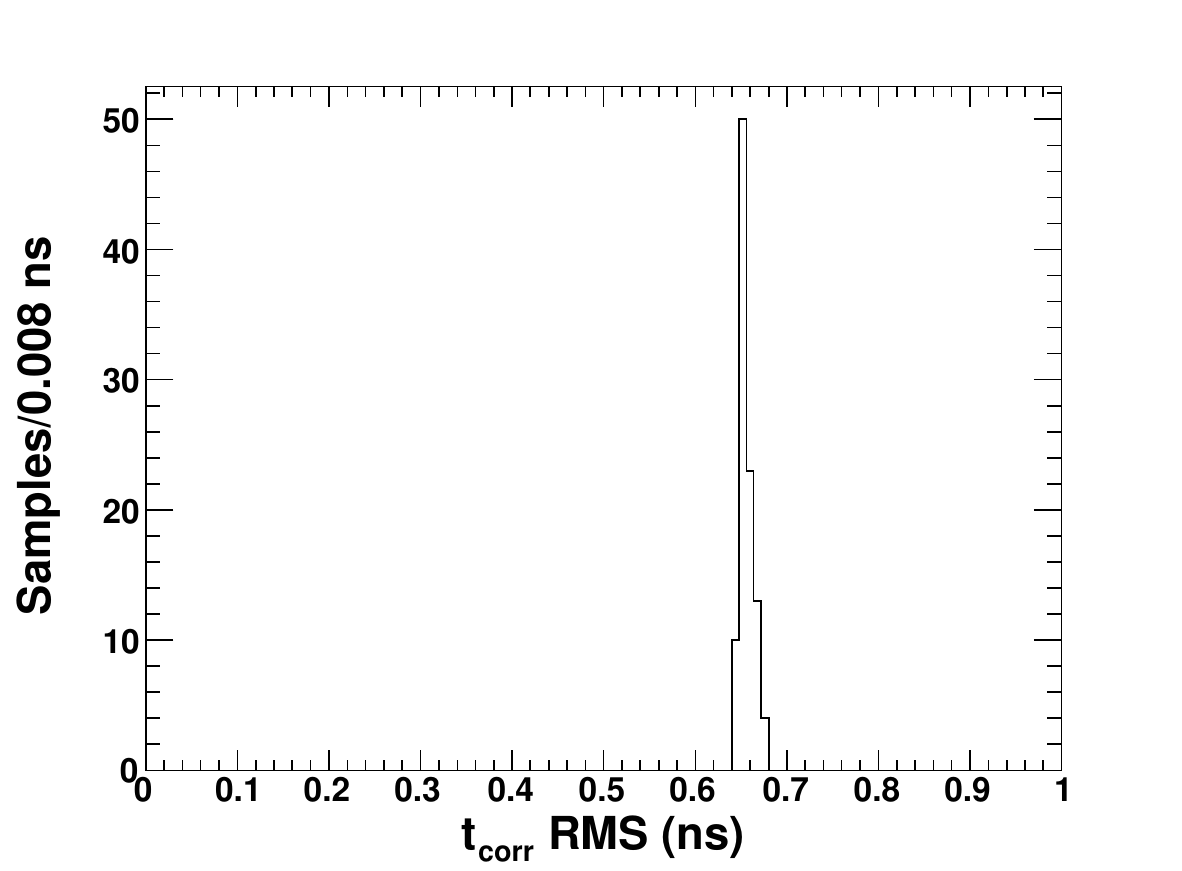}}
\vspace{-14pt}
\caption[The mean of the \tcorr distribution for electrons from various
    subsamples of \Wenuj events where each entry reflects a different
    combination of the electron \et, jet \et and \met event
    requirements.]{\label{fig:meant}\small The mean and RMS 
    of the \tcorr distribution for electrons from various 
subsamples of \Wenuj events where each
      entry reflects a different combination of the electron \et, jet
      \et and \met event requirements. There are slight shifts as the
      these requirements vary. 
       While the mean of the distribution is close to
      zero, the systematic variation on the mean of the primary
      Gaussian of the prompt time distribution is conservatively
      taken to be 0.2~ns in the background estimates.}
\end{figure}

The systematic uncertainty on the number of prompt events in
the signal region is  
dominated by the observed variation in the mean and RMS of the \tcorr
distribution as a function of the \met, jet $\et$, and photon
$\et$ requirements. To estimate the variation, we study the \tcorr
distribution  for samples of electrons in \Wjets events for various
electron \et, jet \et and \met event requirements ($20\leq\et^{\rm \ ele}\leq40~\munit$, $25\leq\et^{\rm \ jet}\leq40~\munit$, and $30\leq\met\leq50$~GeV). 
The results are shown in Fig.~\ref{fig:meant}.
The variation in the mean is up to 0.1~ns and is conservatively
rounded up to 0.2~ns.  Similarly, the systematic uncertainty on the
RMS of \tcorr is conservatively overestimated from a fit to 
Fig.~\ref{fig:meant}   to be 0.02~ns and is only a small addition.
 
\begin{figure}[tb]
  \centering
  \vspace{-15pt}
  \subfigure[]{
    \includegraphics[width=.51\textwidth]{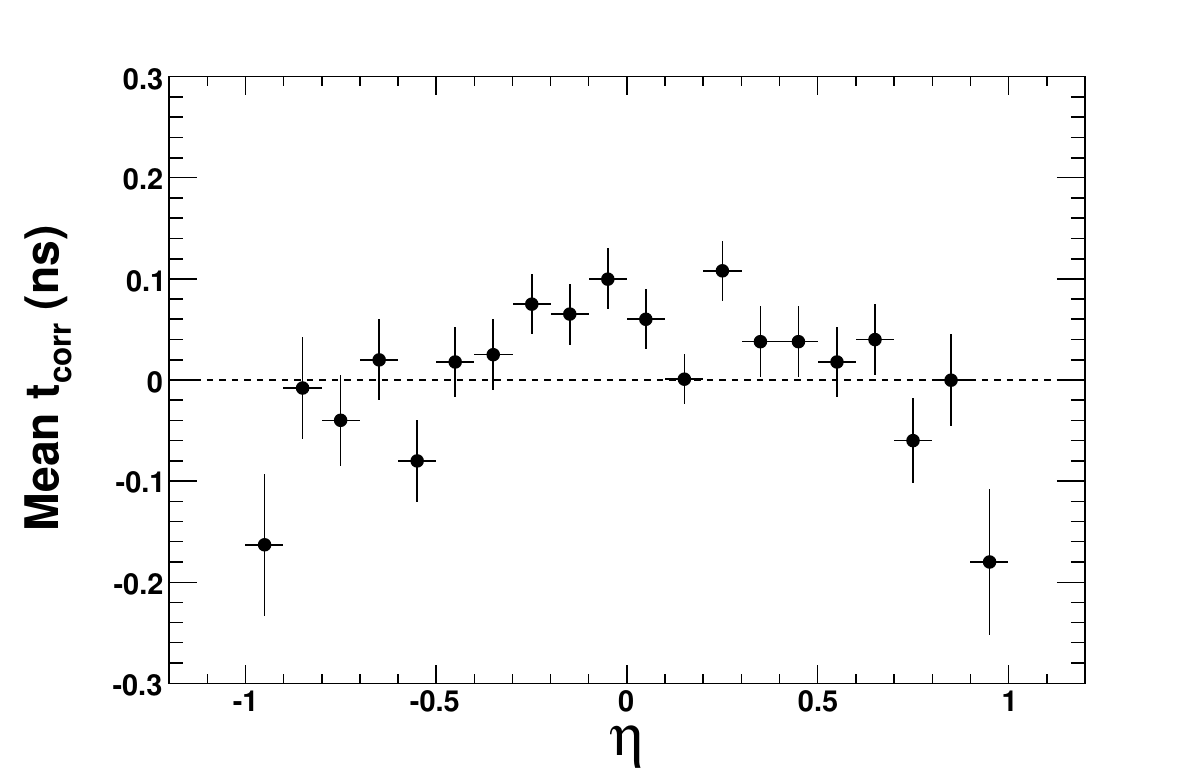}}\\
\vspace{-19pt}
  \subfigure[]{
    \includegraphics[width=.51\textwidth]{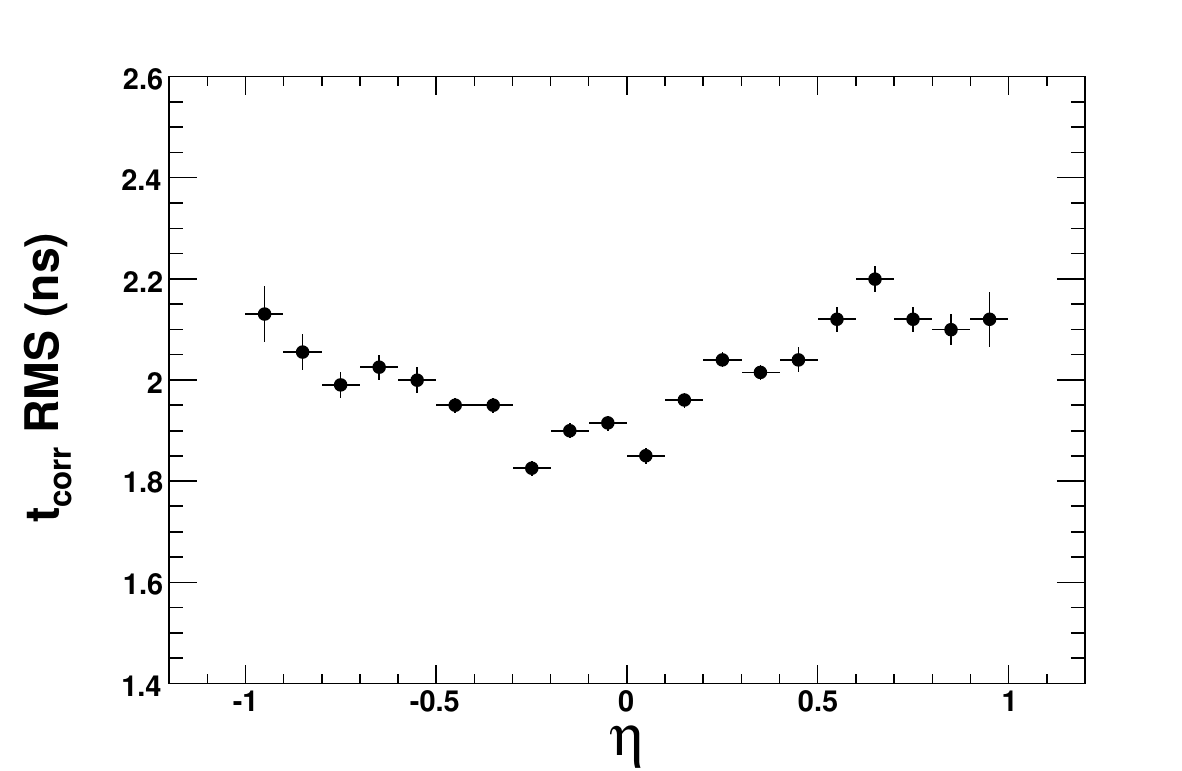}} \vspace{-14pt}
  \caption[The \tcorr distribution mean and RMS for electrons from
  \Wenu events, where the wrong-vertex is picked, as a function of the
  tower-$\eta$ where -9 (+9) corresponds to $\eta=-1$
  ($\eta=+1$).]{\label{fig:time_meanrms_wrong}\small The 
    mean and RMS of the \tcorr
    distribution for electrons from \Wenu events, where
    the wrong vertex is picked, as a function of $\eta$.
  }
\end{figure}

For wrong-vertex assignments, there is an additional variation in the
 \tcorr distribution as a function of photon $\eta$ due to the
 incorrect time-of-flight calculation.
Figure~\ref{fig:time_meanrms_wrong} shows the mean and the RMS
of the \tcorr distribution for electrons from \Wenu events where the
wrong-vertex is selected for the timing correction, as a function of
tower-$\eta$. 
  We take a systematic uncertainty 
 on the mean and the RMS of the wrong-vertex contribution to the \tcorr 
 distribution to be equal to the full variation.  We assign values of 0.33~ns and  0.28~ns,
 respectively, to these  systematic 
 uncertainties, the latter arising from the largest variations in
 Fig.~\ref{fig:time_meanrms_wrong}. 

\subsection{Non-Collision Backgrounds}
\label{sec:bkg_ncoll}

\begin{figure} [tpb]
\centering
\vspace{-9pt}
  \subfigure[]{
\includegraphics[scale=0.32]{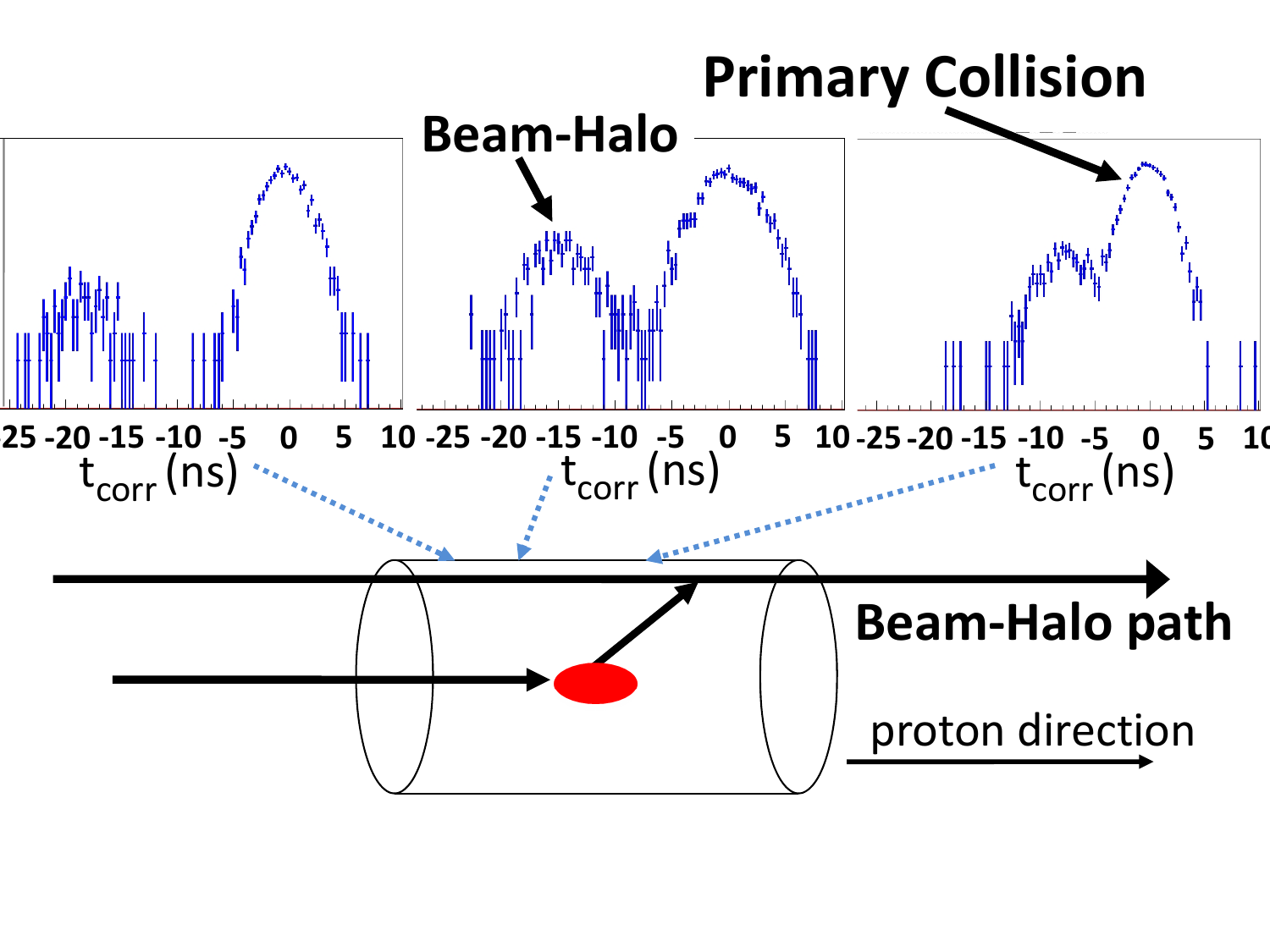}} \\
\vspace{-17pt}
\subfigure[]{
\includegraphics[scale=0.32]{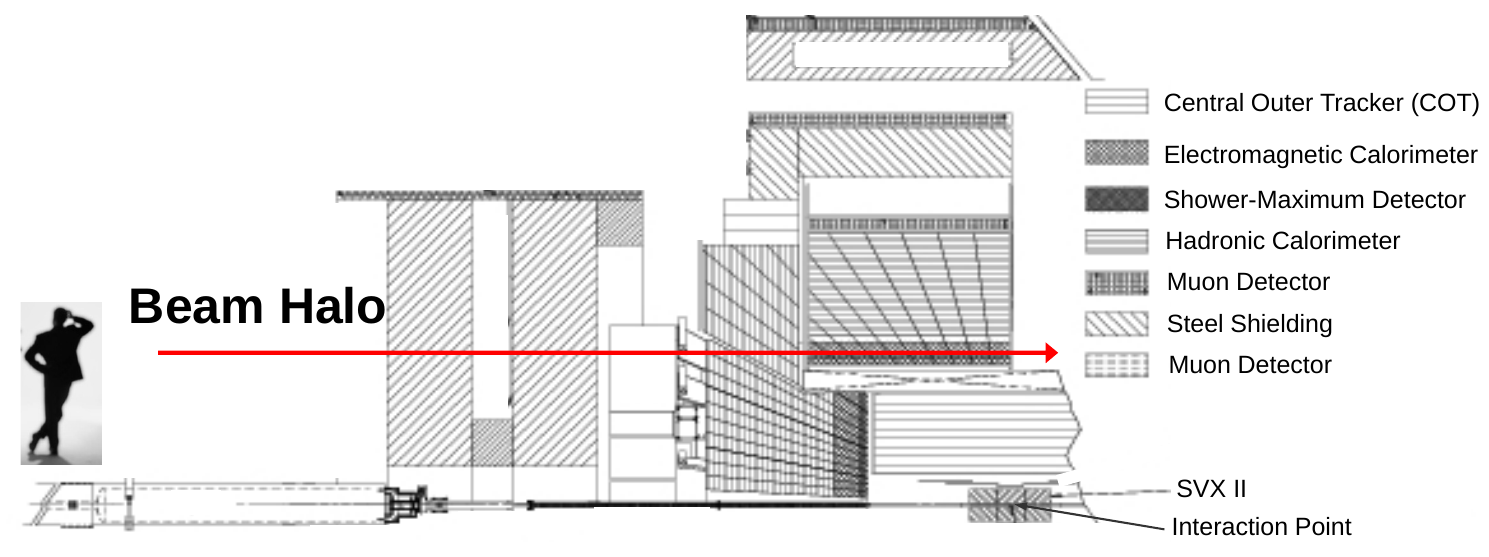}}
\vspace{-10pt}
\caption[ ] {\label{fig:bh_evd} \small  Illustrations of a beam
    halo event interacting with the detector. In both figures 
    the muon path is indicated with an arrow. (a) A
    comparison of the time distributions of prompt collision events
    with beam halo photon candidates for three example towers in
    the calorimeter shows that the mean time changes  
    as a function of tower $\eta$ and is always less than zero. The
    $y$-axes are in arbitrary units.   
    (b) An illustration of how the beam halo interacts with the 
    calorimeter. The muon travels through multiple towers  
    in the hadronic calorimeter at high $\eta$ before hitting the 
    electromagnetic calorimeter.
     } \end {figure}

The fraction of non-collision backgrounds in the kinematic sample 
 that fall in the timing signal window is significant.
 To study these backgrounds, we divide them into two separate sources,  
 cosmic ray muons and beam related backgrounds.
Cosmic ray events (cosmics) come from cosmic ray 
muons that emit photons via bremsstrahlung as they traverse the detector 
 or produce significant ionization in a large $q^{2}$ interaction with the EM calorimeter.
 Beam halo events (beam halo) are caused by beam particles (mostly
 from the more intense proton beam) that 
 hit the beam pipe upstream of the detector and produce muons. 
These muons travel almost parallel to the proton beam direction and 
 shower into
the EM calorimeter to create a photon candidate; see
Fig.~\ref{fig:bh_evd}.  In both cases the event has significant
\met that is highly correlated with the photon \et 
and is uncorrelated with any collision that might occur
coincidentally  at high luminosity. As cosmic ray muons 
interact with the detector and produce a photon randomly in time,
their time distribution is roughly constant over the entire calorimeter  
energy integration window range of 132~ns. Beam halo
``photons'' typically arrive a few ns earlier than prompt photons for geometric reasons as  shown in Fig.~\ref{fig:bh_evd}. 
However in this case, while the rate is lower, the photon candidate
can also  have a \tcorr of $\sim$18~ns (and multiples later and
earlier)  if the muon was created in one of the bunched beam interactions that
can occur every 18~ns in the accelerator.

\begin{figure}[tbp]
  \centering
  \vspace{-15pt}
   \hspace{6pt}
    \includegraphics[width=.495\textwidth]{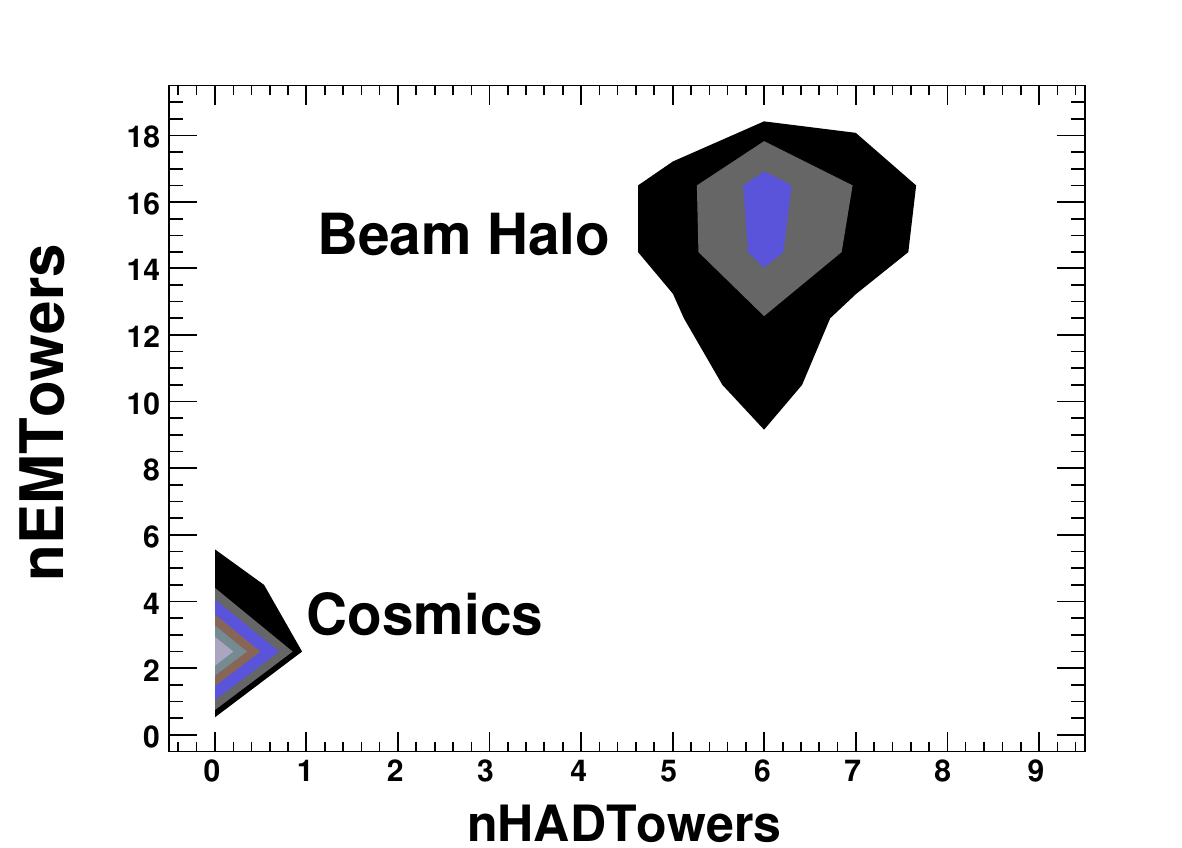}
  \caption[The separation of cosmic and beam halo backgrounds in the
  \gmet sample without tracks.]{\label{fig:ncoll}\small The variables used to 
    separate cosmic and beam halo backgrounds in the \gmet sample without
    a vertex. Beam halo muons deposit energy in many HAD towers as
    they interact with the detector at high $\eta$ 
    and many EM towers as they traverse the central portion of the
    calorimeter along the  beam halo direction.
  }
\end{figure}

\begin{figure}[tbp]
  \centering
  \vspace{-17pt}
\subfigure[]{
  \includegraphics[width=.50\textwidth]{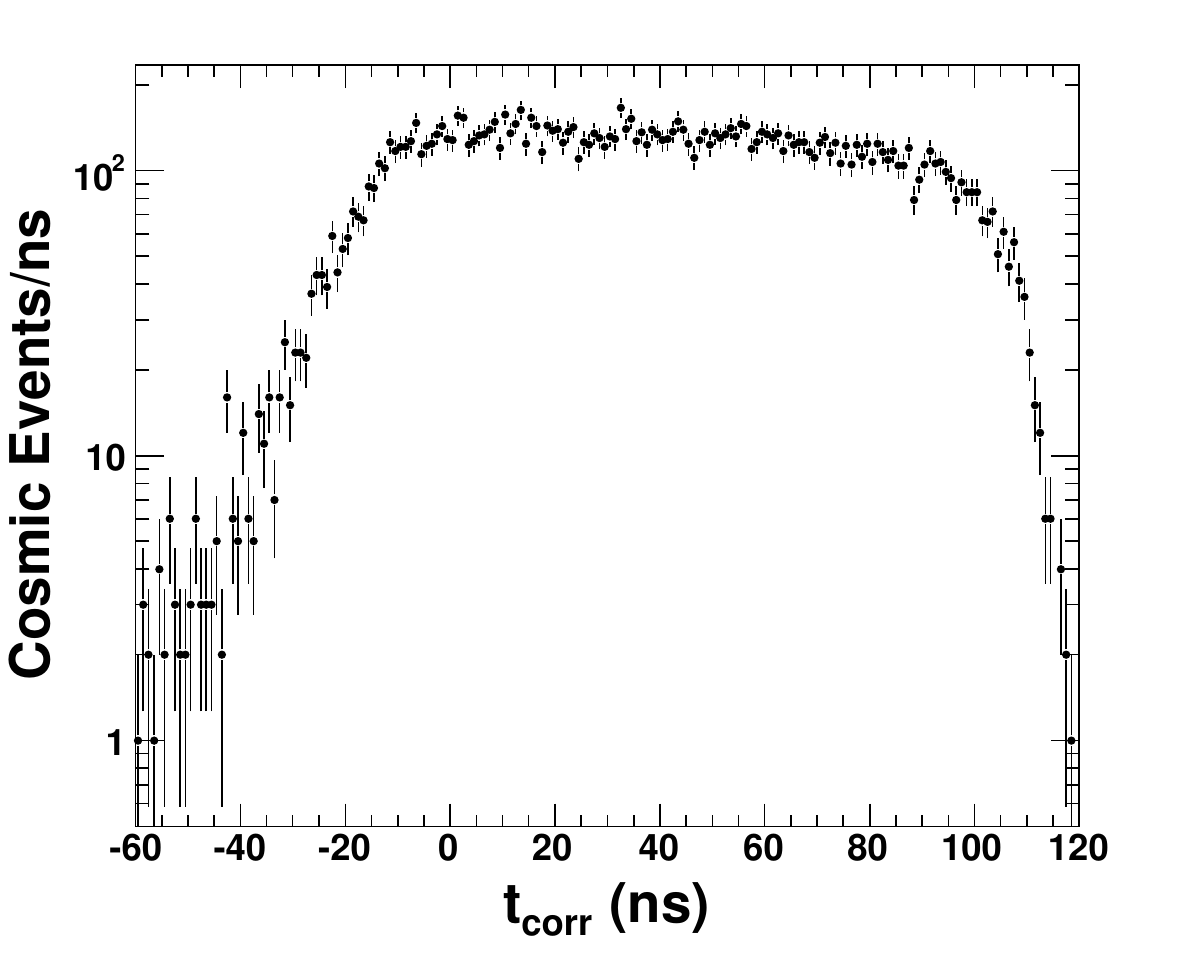}}\\
\vspace{-19pt}
\subfigure[]{
  \includegraphics[width=.50\textwidth]{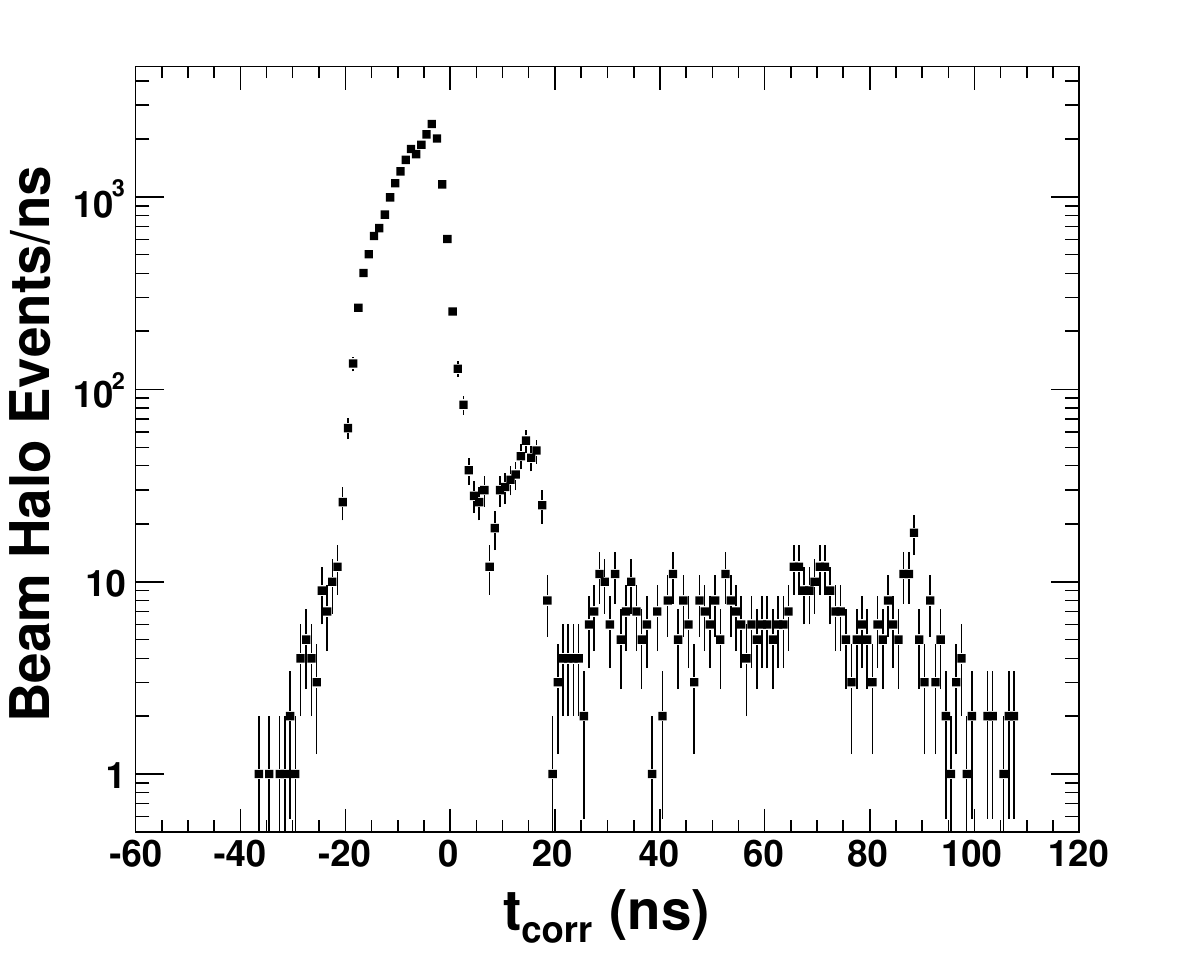}}
\vspace{-14pt}
  \caption[More separation of cosmic and beam halo backgrounds in the
  \gmet sample without tracks.]
{\label{fig:ncolltwo}
\small The \tcorr distributions for the cosmic ray (a) and beam halo  (b) backgrounds in the \gmet sample without a collision. 
}
\end{figure}

\begin{figure}[tbp]
  \centering
  \vspace{-11pt}
   \hspace{6pt}
    \includegraphics[width=.49\textwidth]{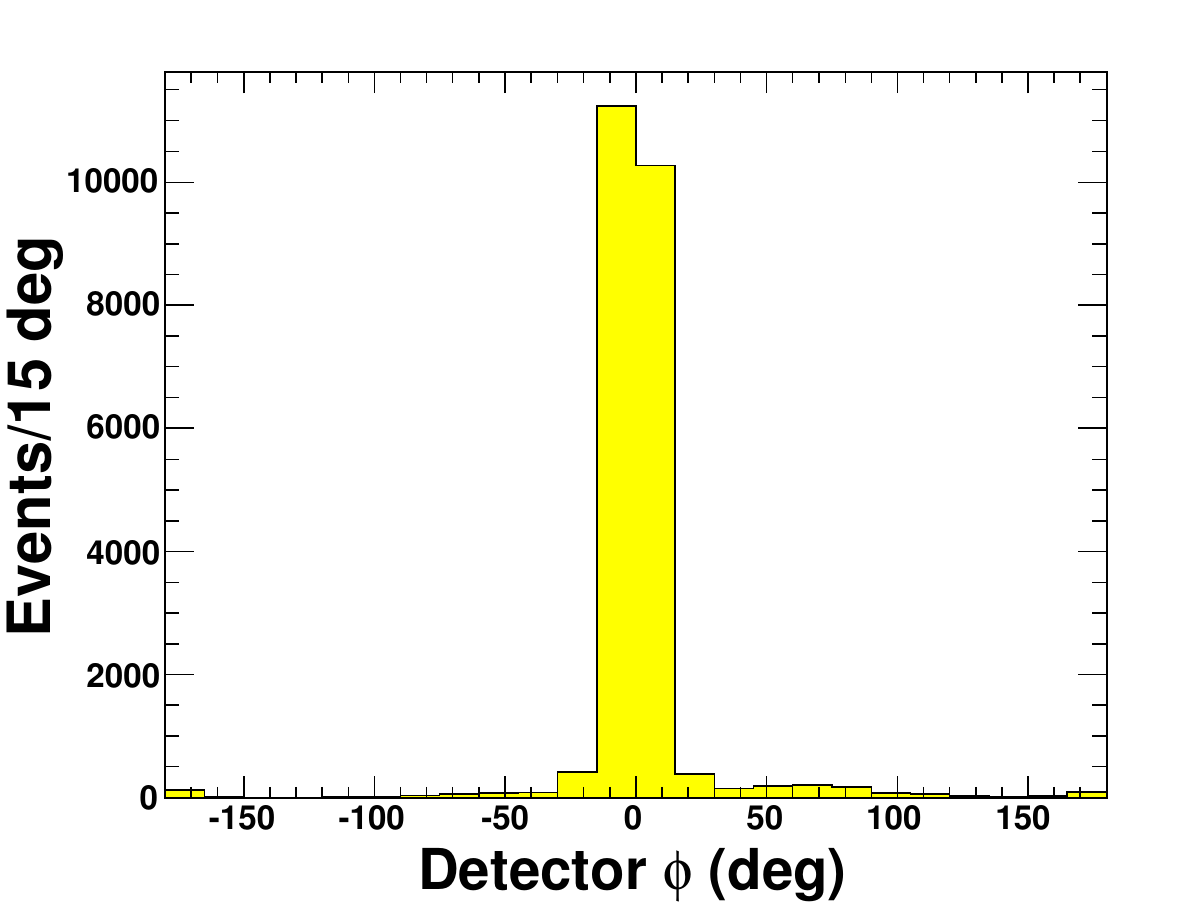}
  \caption[
  ]{\label{fig:bh_time}\small 
    The  number of beam halo photon candidates as a function of 
     detector $\phi$.  
    Most photons arrive 
    at $\phi\approx0$. 
  }
\end{figure}

The rate at which both non-collision backgrounds populate the signal
region is estimated from collision data using events with no identified
collision. The non-collision sample consists 
 of events with a photon that passes the photon ID criteria listed in
Table~\ref{tab:phcuts}, $\met>30$~GeV, and  no
reconstructed vertex.  This sample is used to make timing distribution
templates from pure samples of each type of non-collision background. 
 Beam halo events are identified   
by the energy deposition of the muon as it passes through the high
$\eta$ towers of the plug hadronic calorimeter ($|\eta|\geq~1.1$) and
 the central EM calorimeter ($|\eta|\leq~1.1$) towers at the same
 $\phi$ as the photon candidate; see Fig.~\ref{fig:bh_evd}. 
 The muon deposits a small amount of energy in  
 most towers along its path. Hence, 
 we count the number of towers in the hadronic calorimeter 
 with $\ge0.1$~GeV and $|\eta|\geq~1.1$ (nHADTowers) 
and the number of  towers in the EM calorimeter  with $\ge0.1$~GeV and
$|\eta|\leq~1.1$ 
 (nEMTowers). 
The results are shown in
Fig.~\ref{fig:ncoll} for the full non-collision sample.  
 Cosmic ray candidates are easily separated from beam halo
 candidates.  This is because cosmics do not deposit energy in the
 hadronic calorimeter with $|\eta|\geq~1.1$ and typically only deposit 
 significant energy in a single EM tower. An event is identified as a
 cosmic if it has nHADTowers~=~0 and nEMTowers~$<5$. (Note that we also ignore
 all photon candidates with $-15\dg<\phi<15\dg$ as beam halo dominates
 there.)  Conversely, beam halo events are identified if they have no
 muon stubs and have both  nHADTowers~$>1$ and nEMTowers~$>4$.  
The \tcorr distribution for each is shown in 
Fig.~\ref{fig:ncolltwo} 
for the entire calorimeter energy integration window and indicates that the real collision contamination is negligible. 
As these events lack a vertex, the photon
arrival time is corrected assuming \zzero=~0 and \tzero=~0 in Eq.~\ref{eq:delta_s}.
 To create the \tcorr distribution for use in extrapolating the number of non-collision events in the signal time window from the control regions, we  
convolute  
 the distributions in Fig.~\ref{fig:ncolltwo} with the RMS of the interaction
time of 1.3~ns as the collision time is
uncorrelated.  As will be seen, the 
uncertainty on the rate of the number of events in the signal time region 
is dominated by the statistical uncertainty 
on  the number of non-collision events in the control regions. 
We note that because of the accelerator geometry there are $\sim$40
times more beam halo events that occur  
around the region $\phi\simeq0\dg$ as can be seen in Fig.~\ref{fig:bh_time}. 
 This explains the $-15\dg<\phi<15\dg$ separation requirement and will
 be  further used in the final background estimate procedure.


\subsection{Background Estimation Methods}
\label{sec:bkg_est}


The number of background events in the signal region is estimated from collision data 
by fitting a set of control regions with background timing shapes and extrapolating 
into the signal time window. The \tcorr distribution shape ``templates'' for 
each background source are given in Figs.~\ref{fig:w_time} 
and~\ref{fig:ncolltwo}. Since a sample is defined by kinematic cuts alone 
we can estimate the number of background events in any potential signal time 
window using sensibly chosen control regions. Thus, we can predict the 
background rate for a large variety of final kinematic and timing cuts and use 
these estimates as part of our optimization procedure. 

The background prediction for the signal timing region for each subsample of 
\gmetjet events after the kinematic sample requirements is done as a two-step 
process with multiple control regions. There are a number of reasons for this. 
Multiple control regions are used to get a robust estimate of each of the 
background event contributions that are hard to separate; for example
 the timing region \{$-15,0$\}~ns is populated by both 
the wrong-vertex backgrounds and   
beam-halo backgrounds.  Secondly,   
for many of the potential kinematics-only samples, low statistics can
bias the fit results. We define a set of control regions chosen such
that each is largely dominated by a single background source, and use
an iterative fitting procedure  
to ensure that each background is well estimated for each kinematic
requirement  choice during optimization.



The control regions are designed to allow for a good estimation of each background separately. 
 Since the cosmics rate is essentially constant in time,  the time 
control region is defined to be \{$25,90$\}~ns and is chosen such that
(a) it is well above the beam halo secondary peak at  $\sim$18~ns
and (b) it does not include the region close to the end of the 
calorimeter energy integration window where the event
rate falls sharply. The beam halo control region is defined to be 
\{$-20,-6$\}~ns and is
chosen such that (a) it contains most of the beam halo events but (b)
stays well away from the region dominated by the prompt photon
production.  The standard model control region is defined to be
\{$-10,1.2$\}~ns.  An additional requirement on this region is that the photon must have
$|\phi|\geq15\dg$.  This allows us (a) to include as much of the
collision data as possible to get good precision on the 
ratio of right to wrong-vertex events, 
(b) allows for a potential signal
region above 1.2~ns, and (c) removes most of the beam halo
contamination.  We note that while the $\phi$ restriction is useful for
estimating backgrounds, it is not an effective tool in improving the
 sensitivity. While the upper time limit of the signal
region at 10~ns is not quantitatively motivated, it contains most of a
long-lived signal on the order of nanosecond lifetimes as the time
distribution falls exponentially (Fig.~\ref{fig:delta_s}).

The background prediction for the signal timing region for each subsample of
 \gmetjet events after the
kinematic sample requirements is done as a two-step
process. 
In step 1, the wrong-vertex fraction and the overall prompt photon
 rate are measured using the collision data control sample after a
 correction for non-collision contamination. 
 The non-collision contamination to the control sample is estimated by
 fitting the beam halo template and cosmics, 
simultaneously in the control regions \{$-20,-6$\}~ns (beam halo
dominated) and \{$25,90$\}~ns (cosmics dominated).  The fit is extrapolated to
the collision control region \{$-10,1.2$\}~ns, where the non-collision
contamination is subtracted
off. The remaining data are then fit using the two single Gaussian functions as
shown in Fig.~\ref{fig:w_time}. While the mean and RMS of both 
functions are fixed, the normalizations are allowed to float. After
fitting, the final normalization is scaled by a factor of 12/11 to account for the removed data in the $30\dg$ slice around $\phi=0$.
The statistical error on the prediction in the
signal region is determined by the fit. 
The  uncertainty on the number of
events in the signal time window is estimated by varying the 
collision background fractions, means, and RMS's according to their systematic and statistical uncertainties. We note that if we choose the kinematic sample to be the dataset that passes the preselection sample criteria (see Table~\ref{tab:idcuts}), the
resulting fraction of wrong-vertex events 
 is (3$\pm$1)\%.

In step 2, the rate of the non-collision backgrounds is estimated using the entire $\phi$ region. In the full data sample a 
simultaneous fit is performed for the normalization   
of the beam halo and cosmic ray backgrounds  using the control regions, 
after subtracting off the expected contamination from collision
sources in both regions obtained in step 1. The
uncertainties on this estimate are dominated by the statistical
error on the number of events in the control regions and the  
uncertainty on the extrapolation from the prompt background. With this
technique, the background estimation for all sources is robust enough to be
applied to a variety of kinematic requirements. This feature
will be used along with the simulated acceptance of GMSB events
for the optimization.

\section{Acceptances For GMSB Events and Their Systematic Uncertainties}
\label{sec:acc_eff}

We use  MC techniques to  estimate the acceptance and overall
 sensitivity to GMSB models.
 The sparticle properties (mass, branching fractions etc.) are
 calculated  with {\sc  isasugra}~\cite{isajet}. 
Samples of events of GMSB processes are simulated according to their
 production cross sections using {\sc pythia}~\cite{pythia}, a full
 detector simulation, as well as parton distribution functions
 (PDFs)~\cite{WZcross}.   All sparticle production mechanisms are
 simulated as this maximizes the sensitivity to the
 model~\cite{simeon}.   
To map out the sensitivity for GMSB models as a function of 
\none mass and lifetime, MC samples are generated 
 for $65\leq\mN\leq150$~\munit and  $0\leq\tauN\leq40$~ns. 
As $\sim$5\% of the simulated 
events pass all the selection requirements,  
the size of the MC samples is chosen to be 120,000 events so that
their statistical uncertainty is 
$\sim$1\% and negligible compared to the combined systematic uncertainty.

The total event acceptance 
is 
\begin{equation}
A\cdot\epsilon=(A\cdot\epsilon)_{\mathrm{Signal\ MC}}\times C_{\mathrm{MC}},
\label{eq:aeps}
\end{equation}
where the MC program is used to estimate $A$, the fraction of events
that pass the kinematic sample requirements and to
estimate $\epsilon$, the fraction of
these events that remain after the \tcorr requirement.   
  $C_{\mathrm{MC}}$  is a correction factor 
 for efficiency loss due to the cosmic ray rejection requirement and  
 is not simulated.  Table~\ref{tab:sig_pass} shows the breakdown of
 the number of MC events after each of the preselection sample  
requirements in Table~\ref{tab:idcuts} for an example GMSB point at
\mN~=~100~\munit and \tauN~=~5~ns, near the expected sensitivity limit.

The loss of signal events due to the cosmic ray rejection requirement is
chiefly caused  by real cosmic rays overlapping the
signal events and causing the requirement to fail.
 This efficiency is estimated simply to be equal to the efficiency of the  
requirement as measured from the preselection sample but 
 additionally requiring the
photons to be within $|\tcorr|<10$~ns to select collision events with
high purity. There are 12,583 events in this sample.  
12,360 events remain after the cosmic ray rejection requirement,
giving an efficiency of 
$C_{\mathrm{MC}}=\frac{12,360}{12,583}=(98\pm1)$\%, with the error
conservatively overestimated. 

\begin{table}[tp]
\begin{center}
    {\small \addtolength{\tabcolsep}{0.1em}
    \begin{tabular}{lcc}
      \hline \hline Requirement & Events passed & $(A \cdot \epsilon)_{\mathrm{Signal\ MC}}$ \\ & & (\%) 
      \\ \hline
      Sample events &  120000 & 100.00  \\
      Central photon with &
\\ \ \ \ \ $\et>30$~GeV, and \\ \ \ \ \  $\met>30$~GeV &  64303 & 53.6 \\
      Photon fiducial and ID cuts &  46730 & 38.9 \\
      Good vertex & 37077 & 30.9  \\
      $\ge$1 jet with $\et>30$~GeV &
\\ \ \ \ \ and $|\eta|<2.0$ &  28693 & 23.9 \\ 
      Cosmic ray rejection ($\times C_{\mathrm{MC}}$) &  N/A & 23.5  \\ 
      \hline \hline
    \end{tabular}
    }
    \vspace{1em}
    \caption[Summary of the event reduction for a GMSB example point
    at \mN~=~100~GeV and \tauN~=~5~ns as it passes the preselection
    cuts of Table~\ref{tab:idcuts}.]{\label{tab:sig_pass}\small
      Summary of the MC event reduction for a GMSB example point at
      \mN~=~100~\munit and \tauN~=~5~ns as a function of the preselection
      sample cuts of Table~\ref{tab:idcuts}. Note that the efficiency
      loss caused by the cosmic ray rejection 
      requirement is implemented as an MC correction
      factor, $C_{\mathrm{MC}}$. 
    }
 \end{center}
\end{table}

The systematic uncertainty that enters the limit 
 calculation (and thus a proper optimization) is dominated by the 
potential shift of the \tcorr measurement 
 for the kinematic sample requirements. 
 This along with the remaining systematic effects on the acceptance, 
 luminosity, and production cross section are summarized in
 Table~\ref{tab:sysunc}.  The uncertainty is evaluated at
 $\mN=95$~\munit  and $\tauN=10$~ns.
 The effect of varying \mN and \tauN is negligible when compared
 to the other systematic effects.
  We next describe the estimation of these important effects.

\begin{table}[bp]
\centering
   {\small \addtolength{\tabcolsep}{0.1em}
    \begin{tabular}{lc}
      \hline \hline & Relative \\ Factor & Systematic \\ & Uncertainty (\%) \\ \hline
      Acceptance: & \\
     \hspace{0.1cm} \tcorr measurement and vertex selection & 6.7 \\
      \hspace{0.1cm} Photon ID efficiency & 5.0 \\
      \hspace{0.1cm} Jet energy scale & 1.0 \\
      \hspace{0.1cm} Initial and Final State Radiation & 2.5 \\
      \hspace{0.1cm} Parton Distribution Functions & 0.7 \\
      Total  & 8.8 \\ \hline
      Cross section: & \\
      \hspace{0.1cm} Parton Distribution Functions & 5.9 \\
      \hspace{0.1cm} Renormalization scale & 2.4 \\
      Total  & 6.4 \\ \hline 
      Luminosity &  6.0 \\
      \hline \hline
    \end{tabular}
}    
    \vspace{0.1em}
    \caption[Summary of the systematic uncertainties on the acceptance
    and production cross section for an example GMSB point at
    \mN~=~94~\munit and \tauN~=~10~ns.]{
\label{tab:sysunc} 
\small Summary
      of the systematic uncertainties on the acceptance and the total
      production cross section. 
    }
\end{table}

\begin {itemize}
\item {\it Time measurement:} 
There is an uncertainty on the acceptance due to the systematic variations
in the \tcorr measurement shown in Fig.~\ref{fig:w_time}. Three types of uncertainties are considered
simultaneously: (1) a shift in the mean 
of \tcorr measurement, (2)
a change in the RMS variation of the \tcorr measurement, and (3) a change in the
fraction of events that have an incorrectly chosen vertex. 
The variation of the mean of  
the right (wrong) vertex \tcorr measurement has been conservatively
overestimated to be 0.2~ns (0.33~ns) and  can shift events into and
out of the signal region. The fractional variation in acceptance due
to this effect is estimated to be 6.7\%. 
The fractional change in acceptance due to 
changing the RMS of the \tcorr measurement  
is estimated to be 0.03\%. 
The variation due to fluctuations in the number of additional vertices, 
 is $\sim$1.5\%~\cite{multiple_collisions}. 
Taken in quadrature the total uncertainty is 6.7\% and 
forms the dominant contribution to the systematic uncertainty on the
acceptance.

\item {\it Photon ID efficiency:} 
As described in Section~\ref{sec:idvars}, the systematic uncertainty
on the photon ID efficiency is estimated to be 5\%.

\item {\it Jet energy:} 
As the event selection requires a jet with 
$\et>30$~GeV 
a systematically mismeasured jet can contribute to the acceptance
uncertainty.  
We use the standard CDF procedure~\cite{jets} of varying the jet energy 
 by $\pm1\sigma$ of the estimated energy systematic uncertainty and find  the resulting variation in the acceptance to be 1.0\%.

\item {\it Initial and final state radiation:} 
The uncertainty in the MC simulation of ISR and FSR effects can cause
the photon, the jet, or 
the \met to be systematically more likely to pass or fail the kinematic
 sample requirements and affect the acceptance. This  is estimated using the
standard CDF procedure of varying the ISR/FSR parameters as
described in~\cite{vertalg}. 
The systematic variation in the acceptance is estimated to be 2.5\%.

\item {\it Parton distribution functions (PDFs):}
The production cross section and the acceptance  have uncertainties
due to uncertainty in the PDFs. The uncertainty is estimated using the 
standard CDF procedure of varying the PDFs within the uncertainties 
  provided by {\sc cteq-6m} as described in~\cite{WZcross}. 
 We find a relative uncertainty of $0.7$\% on the
acceptance and $5.9$\% on the cross section. 

\item {\it Renormalization scale:}
 There is a systematic uncertainty of the NLO
production cross section which is estimated using the standard technique
of varying the renormalization scale between 0.25$\cdot q^{2}$
and 4$\cdot q^{2}$ using \prospino~\cite{prospino}. The variation of
the cross section is estimated to be 2.4\%. 
\end {itemize}


\section{Optimization and Expected Search Sensitivity}
\label{sec:optimization}

The sensitivity to sparticle production is estimated in the
form of the expected 95\% C.L. upper cross section limits
($\sigma_{95}^{\mathrm{exp}}$) for various points in parameter space.
Before unblinding the signal region in the data we optimize the
search sensitivity and determine the best event selection requirements for a
prospective GMSB signal. This is done using the background rates 
and the signal acceptances for all sparticle production, with uncertainties, 
available for different sets of selection requirements.
 The procedure is to consider the number of events  ``observed'' in a
pseudo-experiment, $N_{\rm obs}$, assuming no GMSB signal 
exists, and to calculate  $\sigma_{95}(N_{\rm obs})$  
 using a Bayesian method with a
constant cross section prior~\cite{limitcalc_cdf}. 
 The uncertainties on the signal efficiencies,
backgrounds, and luminosity are treated as nuisance parameters with
Gaussian probability distributions. 
We write  $\sigma_{95}(N_{\rm obs},\mathrm{cuts})$ 
 since the limit is also a function of the number of predicted
background events and  $A \cdot \epsilon$, 
where both factors depend on the set of requirements (cuts) used. 

\begin{figure}[tbp]
  \centering
  \vspace{-15pt}
   \hspace{13pt}
    \includegraphics[width=.51\textwidth]{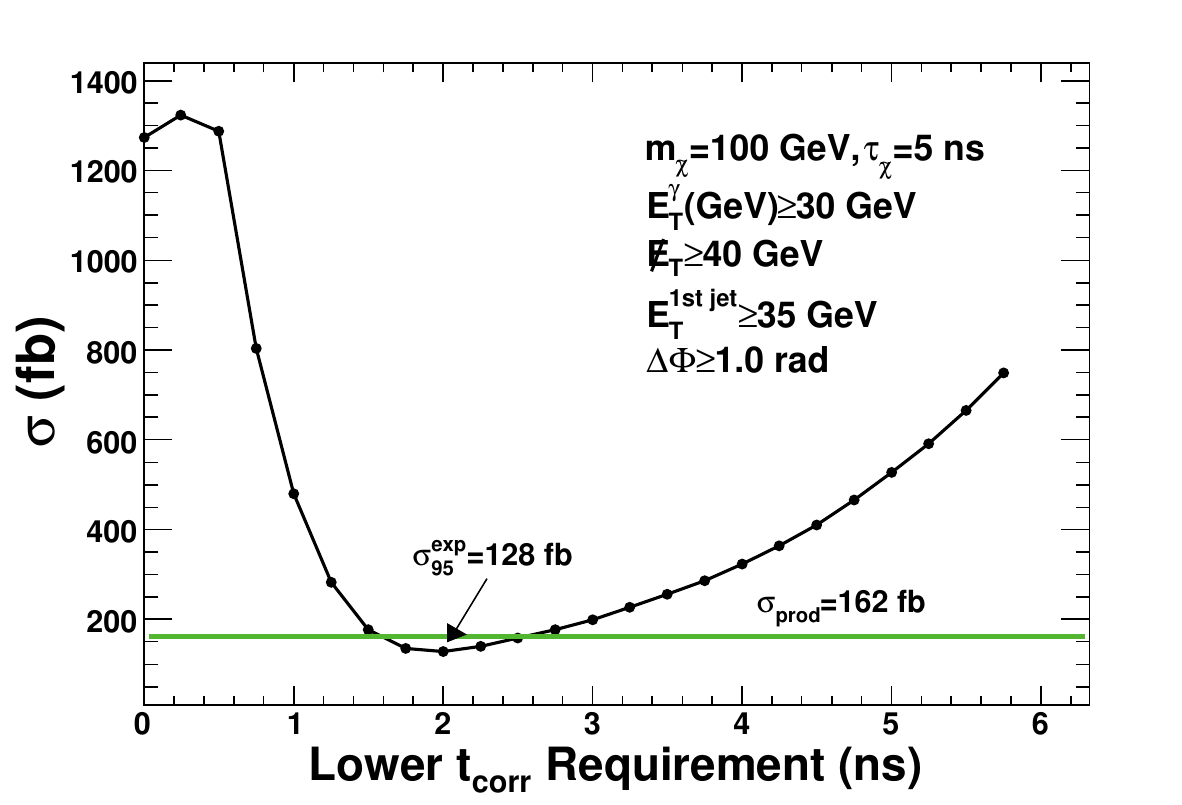}\begin {comment}
  \hspace{-13pt}\subfigure[]{
    \includegraphics[width=.49\textwidth]{Pdf/ana/sigma95vsSumPt_1005_newcut.pdf}
}\end {comment}
  \caption[The expected 95\% C.L. cross section limit as a function of
  the \tcorr and the vertex \sumpt requirement, which is shown for
  completeness, for a GMSB example point (\mN~=~100~GeV and
  \tauN~=~5~ns).]{\small The expected 95\% C.L. cross section limit as
    a function of the lower value of the \tcorr requirement
    for a GMSB example point with \mN~=~100~\munit
    and \tauN~=~5~ns. The values of the kinematic sample requirements
    are held at their optimized values. } 
  \label{fig:xsection_cuts2}
\end{figure}

The expected cross section limit is calculated from
$\sigma_{95}(N_{\rm obs},\mathrm{cuts})$ and takes into account the
 outcomes of the pseudo-experiments determined by their
relative Poisson probability~\cite{limitcalc}, $\cal{P}$.
 The expected cross section limit and its RMS are given
by:
\begin{eqnarray} 
\sigma_{95}^{\mathrm{exp}} (\mathrm{cuts}) = \sum_{N_{\rm obs}=0}^{\infty}\sigma_{95}(N_{\rm obs},\mathrm{cuts}) \nonumber      
   \\ \cdot \cal{P}(N_{\rm obs}, 
N_{\rm back}(\mathrm{cuts}))  \label{eq:opt1} \\
  \mathrm{RMS}^{2} (\mathrm{cuts}) = \sum_{N_{\rm obs}=0}^{\infty} (\sigma_{95}(N_{\rm obs},\mathrm{cuts})-\sigma_{95}^{\mathrm{exp}}(\mathrm{cuts}))^{2} \label{eq:opt} \nonumber
\\\cdot\ \cal{P}(N_{\rm obs},N_{\rm back}(\mathrm{cuts})), 
\end{eqnarray}
where $N_{\rm back}({\rm cuts})$  is the number of expected background
 for a given set of cuts and $\cal{P}$($N_{\rm obs}$,$N_{\rm back}({\rm
 cuts})$) is  the normalized Poisson distribution of $N_{\rm obs}$
 with a mean $N_{\rm back}({\rm cuts})$.  
The expected maximal sensitivity for each GMSB parameter choice is
 found when the  set of requirements 
minimizes $\sigma_{95}^{\mathrm{exp}}(\mathrm{cuts})$. 
To find the minimal  $\sigma_{95}^{\mathrm{exp}}$ 
 we simultaneously vary the photon \et, \met, and jet
\et thresholds,  \dphi, and the lower limit on \tcorr.  Here \dphi is
 the azimuthal angle between \met and the 
highest-\et jet.  This angle cut helps reject events where the \met is
overestimated because of a poorly measured jet.  
The upper limit on \tcorr is kept constant at 10~ns.
As an illustration of the optimization,
Fig.~\ref{fig:xsection_cuts2} shows the
expected cross section limit for a GMSB example point~\cite{snowmass} at
\mN~=~100~\munit and \tauN~=~5~ns  as a function of the lower  
 \tcorr requirement. All other requirements are kept fixed at
 their optimized values. This point is close to the boundary of the
 exclusion region.

In the region $65<\mN<150$~\munit, 
$0<\tauN<40$~ns the optimal cut values have negligible variation except for a small variation in the optimal jet \et
requirements and the lower limit on \tcorr.  
A single fixed set of final requirement values is chosen since, far
from the expected exclusion boundaries, this results in at most a
4\% loss of sensitivity. 
 The final values are Photon $\et\gt30$~GeV, jet $\et\gt35$~GeV,
$\dphi\gt1.0$, $\met\gt40$~GeV, and $2<\tcorr<10$~ns. 
For $\mN=100~\munit$ and $\tauN=5$~ns we find an acceptance of (6.3$\pm$0.6)\%.
 Table~\ref{tab:prl} gives more details on the acceptance reduction 
 as a function of the requirements.
Our fit to the data outside the signal region predicts total backgrounds of
6.2$\pm$3.5 from cosmic rays, 6.8$\pm$4.9 from beam halo background
sources, and the rest from the standard model with a measured wrong-vertex fraction of (0.5$\pm$0.2)\%. Inside the signal region, $2<\tcorr<10$~ns, we predict 1.25$\pm$0.66 events: 0.71$\pm$0.60 from
standard model, 0.46$\pm$0.26 from cosmic rays, and 0.07$\pm$0.05 from
beam halo.
 Table \ref {tab:xsecdetail} shows 
 the various possible number of ``observed'' events and their probability in the no-signal hypothesis. 
  We find for this point in parameter space  $\sigma_{95}^{\rm
    exp}=128$~fb with an RMS of 42~fb.  
  The total sparticle production cross sections, $\sigma_{\rm
    prod}$, 
 are calculated at next to leading order (NLO) by multiplying the 
 LO production cross section from {\sc pythia}~\cite{pythia} by the 
 theoretical K-factors from~\cite{kfactors} ($\sim$1.2 for this mass range). 
 A total sparticle production cross section of 162~fb is predicted for this
 point, and thus we expect to  exclude it. A total of 5.7$\pm$0.7
  signal events is expected for this mass/lifetime combination. 

\begin{table}[tbp]
  \begin{center}
    {\small \addtolength{\tabcolsep}{0.1em}
    \begin{tabular}{l@{\hspace{-5.5pt}}cc} \hline \hline
      \bf{Preselection Sample} & \bf{Individual}  & \bf{Cumulative} \\ \bf{Requirements} & \bf{Efficiency} & \bf{Efficiency} \\ 
      & \bf{(\%)}  &  \bf{(\%)} \\
      \ $\et^{\gamma} > 30$~GeV, $\met>30$~GeV & 54 & 54 \\
      \ Photon ID and fiducial, $|\eta|<1.0$~~~ & 74 & 39 \\
      \ Good vertex, & & \\ \ \ \ \ $\sum_{\rm tracks}\pt>15$~\punit\ \ & 79 & 31 \\
      \ $|\eta^{\mathrm{jet}}|<2.0$, $E^{\rm jet}_{T}>30$~GeV & 77 & 24 \\
      \ Cosmic ray rejection  & 98 & 23 \\
      \bf{Requirements after} & & \\ 
      \bf{Optimization} & & \\
      \ $\met>40$~GeV, $E^{\rm jet}_{T}>35$~GeV & 92 & 21 \\
      \ $\Delta\phi$(\mett, jet) $>$ 1.0 rad & 86 & 18 \\
      \ 2~ns $< \tcorr <$ 10~ns & 33 & 6 \\
      \hline \hline
    \end{tabular}    
    }
    \vspace{1em}
    \caption[The data selection criteria and the total, cumulative
    event efficiency for an example GMSB model point at
    \mN~=~100~\munit and \tauN~=~5~ns, shown for completeness as it is
    presented in Ref.~\cite{prl}.]{\label{tab:prl}\small The data
      selection criteria and the total, cumulative event efficiency
      for an example GMSB model point at \mN~=~100~\munit and
      \tauN~=~5~ns. The listed requirement efficiencies are in
      general model-dependent. The good vertex requirement (95\% 
      efficient) includes the $|z_0|<60$~cm cut.  The efficiency of
      this cut, as well as that of 
      the photon fiducial and cosmic ray rejection cuts, is 
      model-independent and estimated from data. }
  \end{center}  
\end{table}
\begin{table}[tbp]
  \begin{center}
    {\small \addtolength{\tabcolsep}{0.1em}
    \begin{tabular}{ccc}
      \hline \hline $N_{obs}$  &  $\sigma_{95}(N_{obs})$ (fb)  &  Probability (\%)  \\ \hline
      0 & 79.9 & 28.7  \\
      1 & 120 & 35.8  \\
      2 & 153 & 22.4  \\
      3 & 196 & 9.32  \\
      4 & 239 & 2.91  \\
      5 & 280 & 0.729  \\
      \hline \hline
    \end{tabular}
    }
    \vspace{1em}
    \caption[The 95\% C.L. cross section limit as a function of the
    hypothetically observed number of events and the Poisson
    probability for this number of events based on the no-signal
    hypothesis.]{\label{tab:xsecdetail}\small The 95\% C.L.  cross
      section limit as a function of the hypothetically observed
      number of events, the Poisson probability for the number of
      events based on the no-signal hypothesis (1.3 events expected) at an example GMSB point of \mN=100~\munit and \tauN=5~ns, and the requirements listed in Table~\ref{tab:prl}. 
      We find for this point in parameter space  $\sigma_{95}^{\rm
    exp}=128$~fb with an RMS of 42~fb. A total sparticle production
    cross section of 162~fb is predicted for this  
    point, and thus on average we expect to  exclude it.  
      }
  \end{center}  
\end{table}

\begin{table}[tbp]
  \begin{center}
    {\small \addtolength{\tabcolsep}{0.1em}
    \begin{tabular}{l@{\hspace{-10pt}}cc}
      \hline \hline  
      Control Region & Dominant Background & Observed Events \\ \hline
      $-20\le\tcorr\le-6$~ns & Beam halo & 4 \\
      $-10\le\tcorr\le1.2$~ns & SM & 493 \\
      $25\le\tcorr\le90$~ns & Cosmics & 4 \\
      \hline \hline
    \end{tabular}
    }
    \vspace{1em}
    \caption[The observed number of events in each control region
    after all optimized requirements, except the timing
    cut.]{\label{tab:control}\small The observed number of events in
      each control region after all the optimized kinematic sample requirements.
     }
  \end{center}  
\end{table}

\section{Data, Cross Section Limits and Final Results}
\label{sec:results}

\begin{figure}[tbp]
  \centering
  \vspace{-18pt}
  \subfigure[]{
    \includegraphics [scale=0.45]
{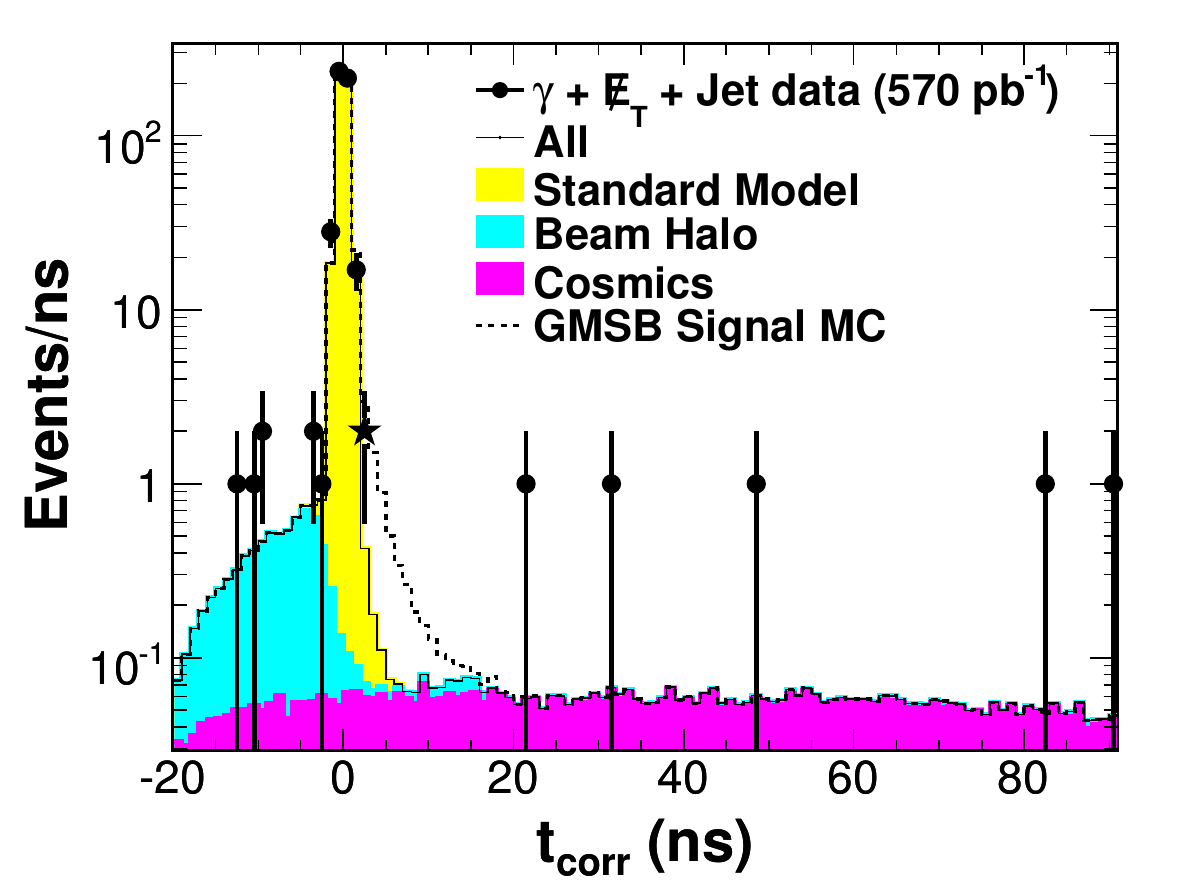}}\\
\vspace{-19pt}
  \subfigure[]{
 \includegraphics[scale=0.45]
{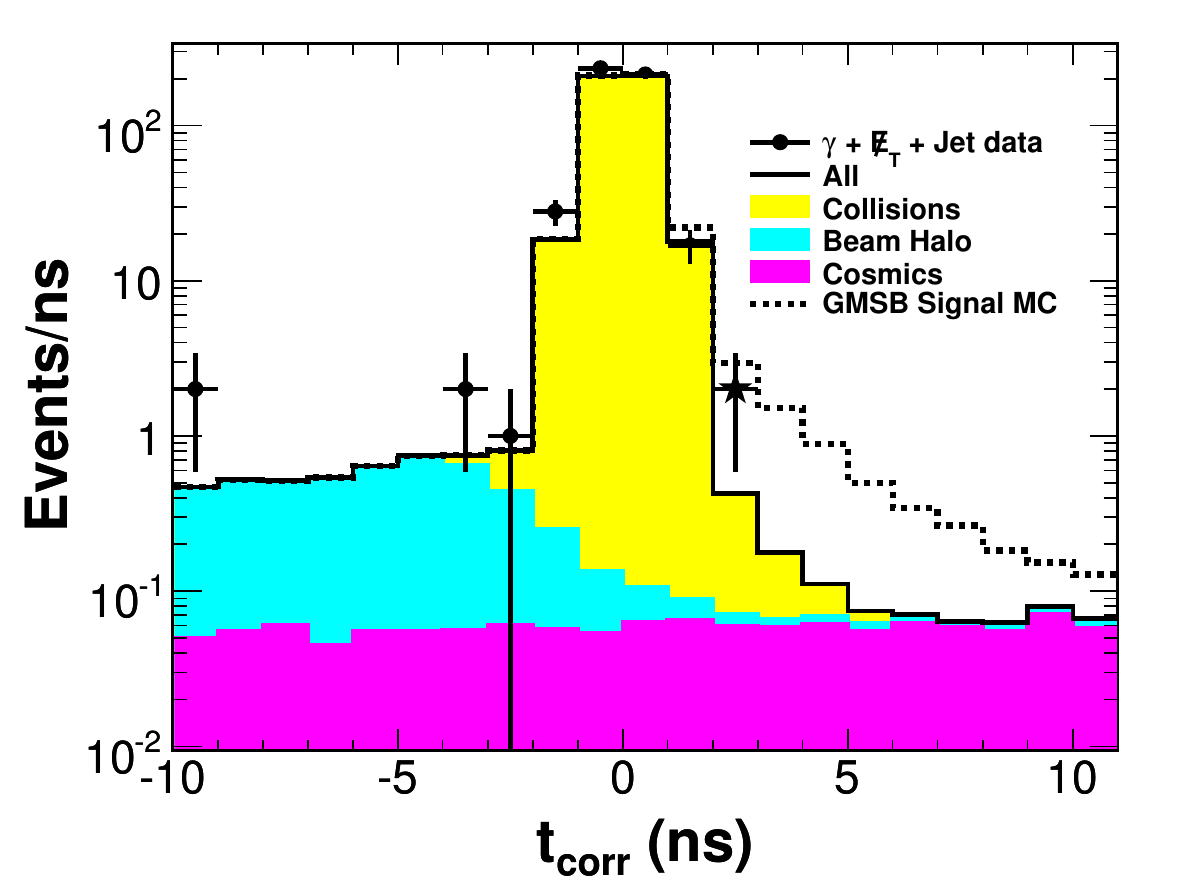}}
\vspace{-14pt}
  \caption[The \tcorr distribution including control and signal
  regions, after all but the timing cut for all backgrounds, the
  expected signal and the observed data.]{\small The \tcorr
    distribution including the control and signal regions, after all but
    the timing cut for all backgrounds, for the expected signal and the
    observed data. A total of 508 events is observed in the full time
    window.  The two observed events in the signal region,
  $2<\tcorr<10$~ns, are in the first signal time  bin (marked with a
  star).  This is consistent with the background
    expectation of 1.3$\pm$0.7 events.  
  }
  \label{fig:sig_time_window}
\end{figure}

After the kinematic requirements (Table~\ref{tab:prl}) 508
events remain in the 
data sample.  Table~\ref{tab:control}  
lists the number of events observed in
the three control regions.  Figure~\ref{fig:sig_time_window} shows the
\tcorr distribution 
from data along with the signal expectations and the background
shapes, normalized using the control regions.

Since the number of events in the timing window $1.2\le\tcorr\le10$~ns is predicted by the background estimation techniques we can compare the number of predicted and observed events. Table~\ref{tab:expevts} shows the results as each
of the optimized requirements is applied sequentially along with the expectations for a GMSB example point.  
The large fractional errors on the backgrounds 
are due to the systematic uncertainty on the mean and RMS of the SM
distributions as discussed in Section~\ref{sec:bkgs}. 
The large fractional errors on the beam halo and
cosmic ray estimates are primarily due to the small 
number of events in the
control regions. Neither is a problem in the final analysis as the absolute number of background events is small in the signal region.
After each requirement, sparticle production would have increased
the number of events observed in the signal region above the
background levels. However, there is good
agreement between the background prediction and the number of events
observed in all cases.  The bulk of the beam halo and cosmics background are rejected by the timing
requirement. 

\begin{figure*}[tbp]
  \centering
 \vspace{-25pt}
  \subfigure[]{
 \hspace{-15pt}
    \includegraphics[scale=.44]{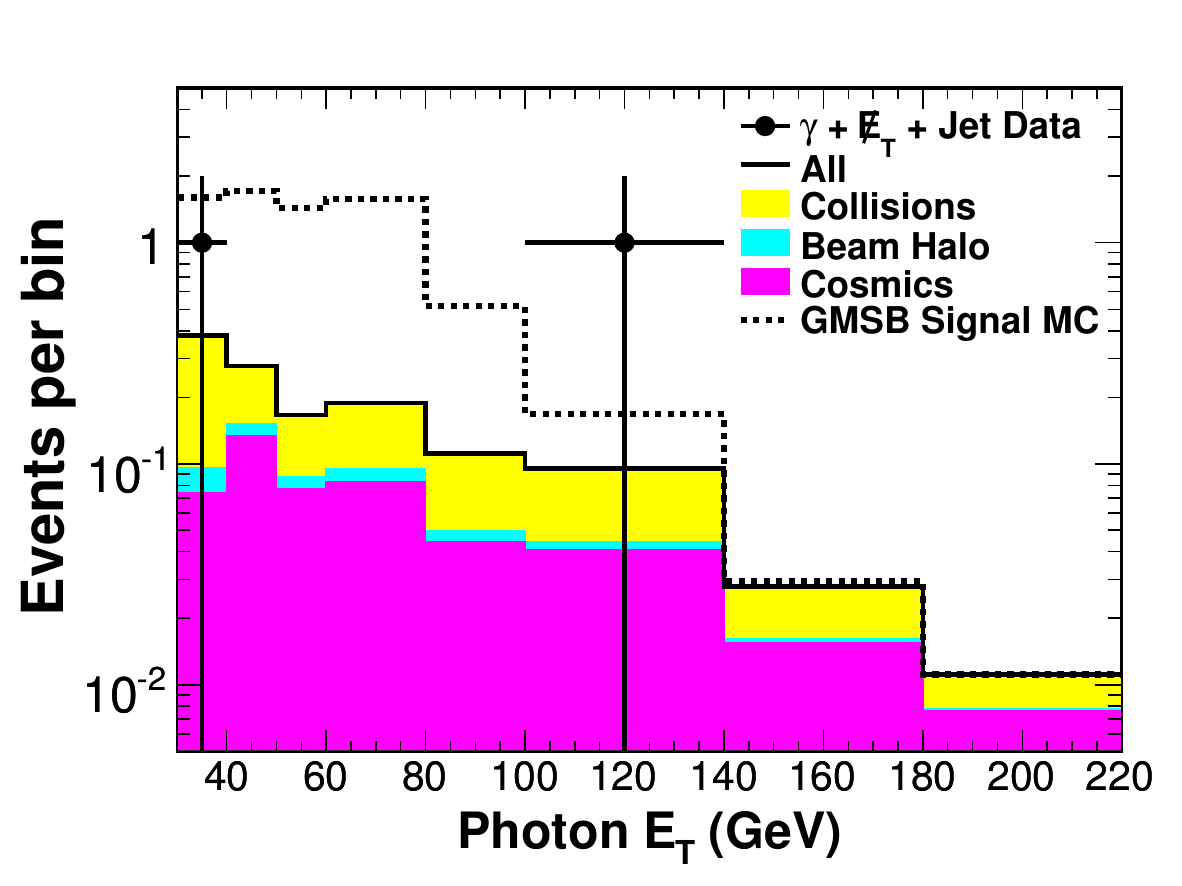}}
  \subfigure[]{    \hspace{8pt}
    \includegraphics[scale=.44]{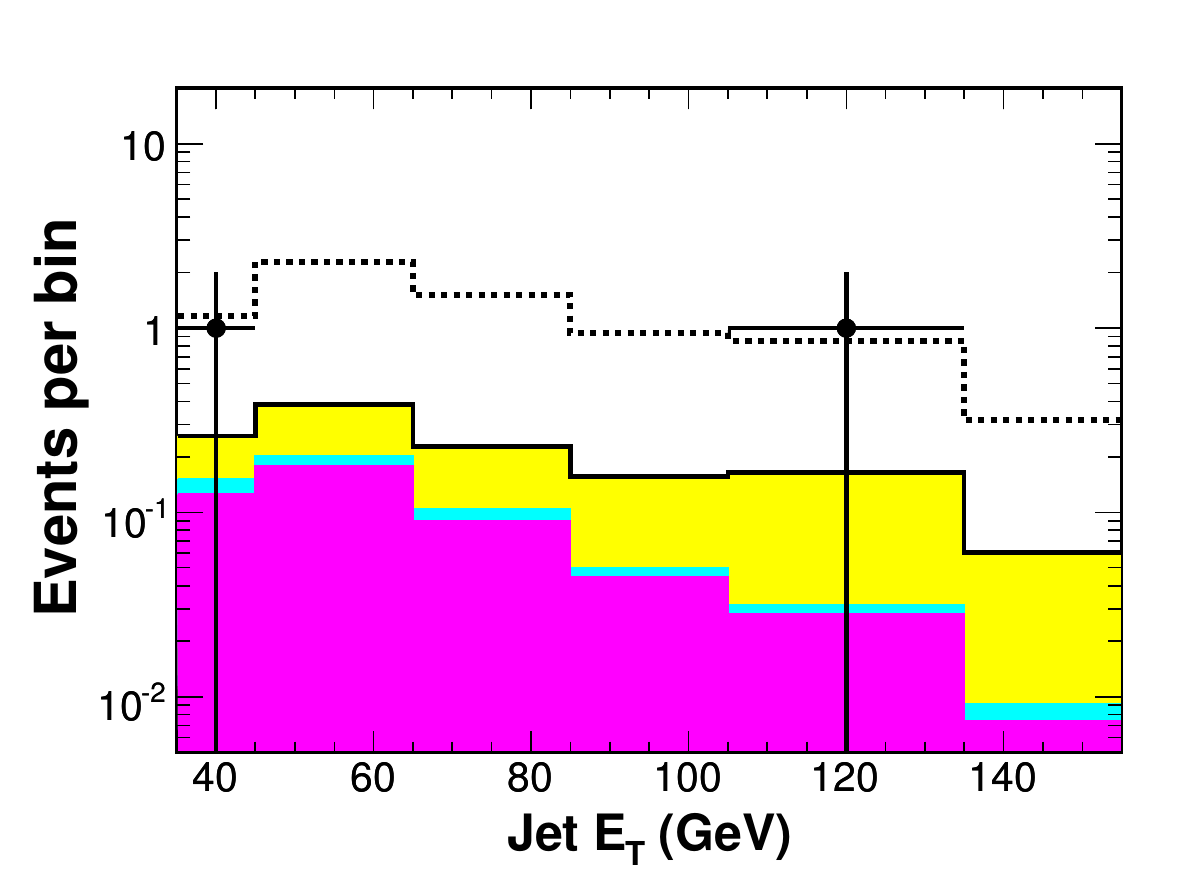}} \newline \\ 
    \vspace{-19pt}
  \subfigure[]{     \hspace{-15pt}
    \includegraphics[scale=.44]{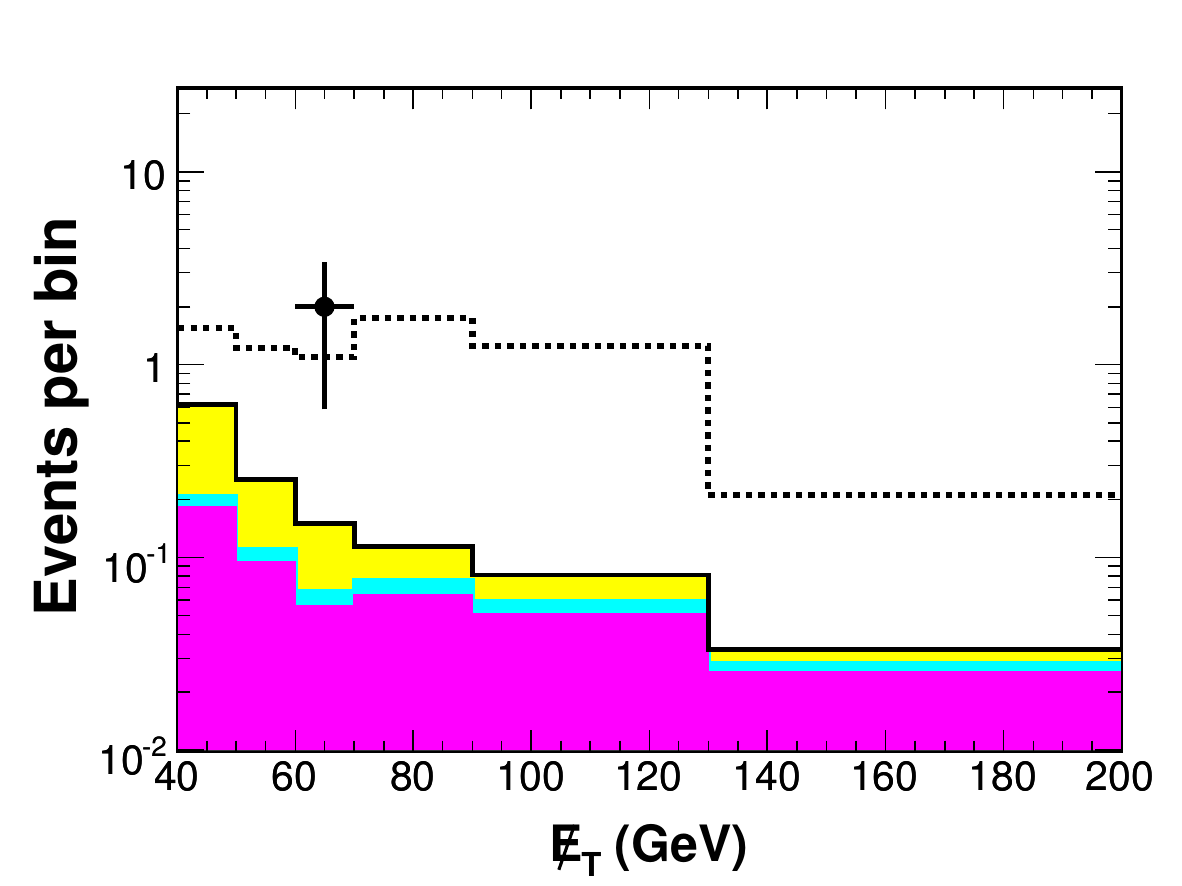}}
  \subfigure[]{      \hspace{8pt}
    \includegraphics[scale=.44]{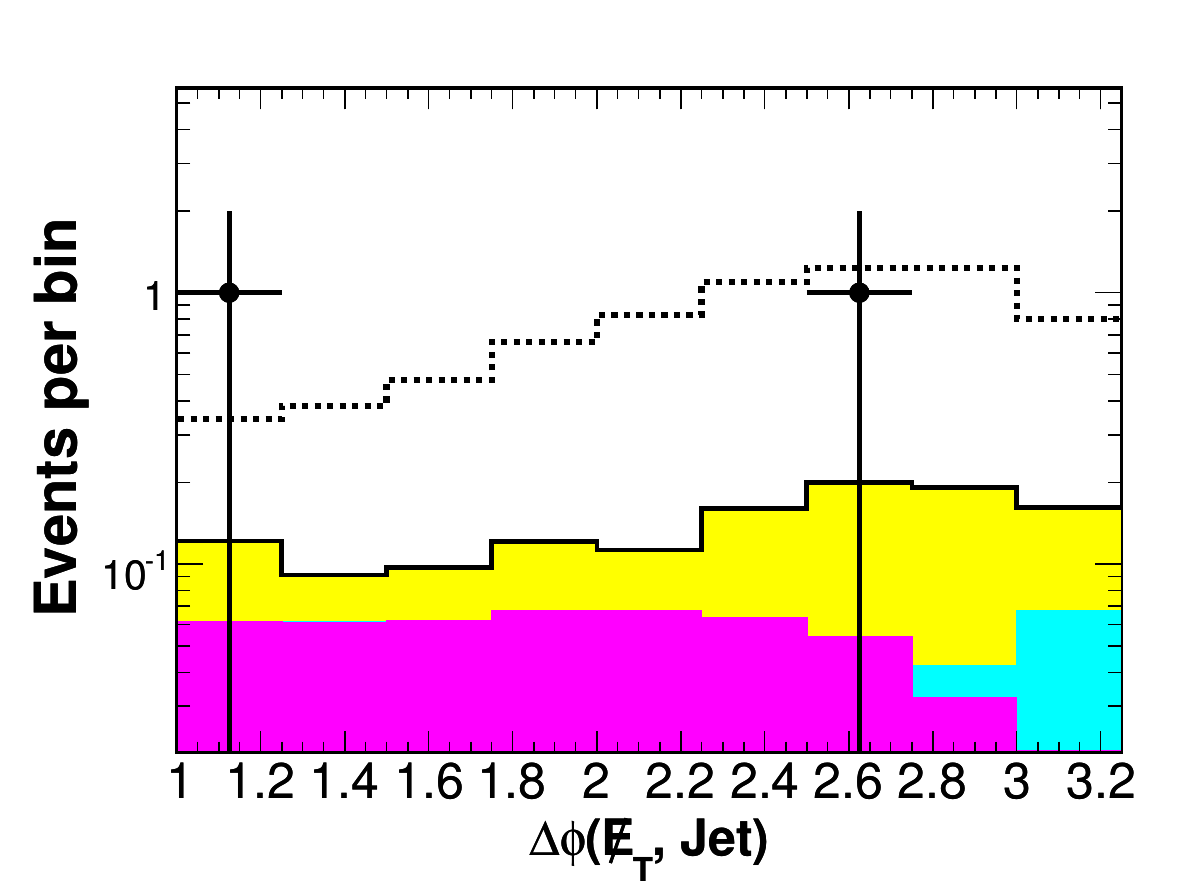}} \newline  \\
    \vspace{-14pt}
  \caption[The same as Fig.~\ref{fig:sig_bkg_kine}, but including the
  data in the signal \newline region.]{\small The predicted and 
  observed  photon $\et$, jet $\et$, \met, and \dphi distributions
  for the signal region after the 
  final  event selection requirements. The GMSB distributions are for 
    \mN~=~100~\munit and \tauN~=~5~ns.  
    There is no evidence for new physics. }
  \label{fig:sig_bkg_kine_data}
\end{figure*}

There are 2 events  in the final signal region,
$2<\tcorr<10$~ns, consistent with the background expectation of 
1.3$\pm$0.7 events. 
Figure~\ref{fig:sig_time_window}(b) shows in detail the
time window immediately around the signal region.  The data is consistent with
 background expectations.  The two events have \tcorr of 2.2~ns and
2.6~ns respectively.  Figure~\ref{fig:sig_bkg_kine_data} shows the
distributions for the background and signal expectations along with the data as
functions of the photon $\et$, jet $\et$, \met, and \dphi requirements.
There is no distribution that hints at an excess. 

\begin{table*}[tbp]
  \begin{center}
    {\small \addtolength{\tabcolsep}{1em}
    \begin{tabular}{l|c|c|c|c|c|c}
      \hline \hline  Requirement & \multicolumn{4}{c|}{Expected Background} &  Expected & Data \\ 
      & SM & Beam Halo & Cosmics & Total & Signal & \\ \hline
      Photon, \met, jet pre- & & & & & & \\
      \ \ \ \ selection cuts and & & & & & & \\
      $\ \ \ \ 1.2\le\tcorr\le10$~ns & 490.74$\pm$295.40 & 0.27$\pm$0.12 & 1.30$\pm$0.49 & 492.3$\pm$295.4 & 11.7$\pm$1.4 & 398 \\
      $\met>40$~GeV & 162.96$\pm$76.19 & 0.24$\pm$0.12 & 1.17$\pm$0.46 & 164.4$\pm$76.2 & 10.2$\pm$1.2 & 99 \\
      Jet $\et>35$~GeV & 154.52$\pm$72.96 & 0.12$\pm$0.08 & 0.79$\pm$0.37 & 155.4$\pm$73.0 & 9.4$\pm$1.1 & 97 \\
      $\dphi>1.0$ & 13.07$\pm$11.57 & 0.10$\pm$0.07 & 0.52$\pm$0.30 & 13.7$\pm$11.6 & 8.5$\pm$1.0 & 8 \\
      $2\le\tcorr\le10$~ns & 0.71$\pm$0.60 & 0.07$\pm$0.05 & 0.46$\pm$0.26 & 1.3$\pm$0.7 & 5.7$\pm$0.7 & 2 \\
      \hline \hline
    \end{tabular}
    }
    \vspace{1em}
    \caption[Summary of the expected and observed number of events
    from the background estimate after the event preselection and each
    requirement from the optimization, separated for each background,
    and the expected number of signal
    events.]{\label{tab:expevts}\small Summary of the expected and
      observed number of events from the background estimate after the
       preselection sample requirements and each requirement from 
      the optimization,
      separated for each background, and the expected number of signal
      events. The expected signal numbers
      are for a GMSB example point at \mN~=~100~\munit and
      \tauN~=~5~ns. Note that the additional requirement
      $1.2<\tcorr<10$~ns is applied at the top line to allow the
      background estimation methods to use the prompt control region to make
      predictions at each stage. The preselection sample cuts are listed in
      Table~\ref{tab:idcuts}. The background predictions match well
      with the observed number of events for each requirement
      indicating the background estimation methods are reliable. There
      is no evidence of new physics.}
  \end{center}  
\end{table*}

A model-independent exclusion limit can be assigned based on this
non-observation.  The two observed events and the background and its
uncertainty give a 95\% C.L. upper limit ($N^{\rm obs}_{95}$) on the
number of events produced of $N^{\rm obs}_{95} = 5.2$ events.  
 Any model of new physics that predicts more than this number of 
 delayed \gmetjet events is excluded. 
 To make our results useful for future model builders to calculate cross section limits for other acceptance models, we calculate
a correction factor, $C_{\rm sys}$, that takes into account the systematic uncertainties on the acceptance, efficiency and 
the luminosity, which are also fairly model-independent. Using the relation
\begin{equation}
  \sigma^{\rm obs}_{95} = \frac{N_{95}^{\rm obs}\cdot C_{\rm sys}}{\mathcal{L}\cdot (A\cdot\epsilon)} 
\label{eq:n95}
\end{equation}
 and the methods to calculate $\sigma_{95}^{\rm obs}$, 
we find $N_{95}^{\rm obs}\cdot C_{\rm sys}$ = 5.5 events.

\subsection{Cross Section Limits and Exclusion Regions for GMSB Production}
\label{sec:limits}
\label{sec:parametrization}


To compare our results to GMSB models we calculate the 95\% C.L. upper limits and compare to GMSB production cross sections. 
To allow for a more detailed comparison to
production cross sections for any other model that predicts heavy, long-lived,
 neutral particles that produce the \gmetjet final state~\cite{addfeng} we parametrize the acceptance using variables that are largely independent of
the GMSB specific dynamics. 

There are several effects that
cause the acceptance to vary as a function of both the \none mass and
lifetime.  The dominant ones are the probability that (a) at least
one \none of the two decays in the detector volume 
to produce a photon that passes the kinematic sample selection criteria
($P_{\mathrm{vol}}$) and that (b) \tcorr is within the signal time
window ($P_{\mathrm{t}}$).  
We find these are roughly independent of each other,  
and define $A \cdot\epsilon=P_{\mathrm{vol}}\cdot P_{\mathrm{t}}\cdot P_{\mathrm{corr}}$, where $P_{\mathrm{corr}}$ is a minor 
correction described below. 
We find:
\begin{eqnarray}
P_{\mathrm{vol}}  =  ( -0.254 + 6.85\cdot10^{-3}\mN - 1.54\cdot10^{-5}\mN^2 ) \nonumber &\\ 
\cdot\ ( 1 - e^{-\frac{-0.625+0.0647\cdot\mN}{\tauN+0.842}}) \\ \label{eq:modind2}
P_{\mathrm{t}} = ( -0.0449 + 8.69\cdot10^{-3}\mN \nonumber - 3.49\cdot10^{-5}\mN^2 ) \\
\cdot\ ( 1 - ( 1 - e^{\frac{-4.78}{\tauN+1.21}})^2 )  \label{eq:modind1}
\end{eqnarray}
where each function consists of two multiplicative terms: a
mass-dependent term that determines the overall scale, and a lifetime
dependent term. Here \mN is in \munit and \tauN is in ns.  The small
mass dependency of the  
overall scale and of the exponential term in $P_{\mathrm{vol}}$ both
come from variations in the \none boost with its
mass in production~\cite{prospects}. A higher \none boost can cause
the \none to leave the detector with a higher probability given its
lifetime and cause the photon to be emitted at smaller angles relative
to the \none direction such that its arrival time becomes similar to a promptly
produced photon. A variation in the boost is caused by a change in the
shape of the $p_{T}$ distribution as a function of 
the \none mass. Another important, but non-dominant, factor is the
lifetime term 
in the denominator of both exponentials. This takes into account the
effect that both the acceptance and efficiency are not zero at low \none
lifetimes but have a finite contribution due to the resolution of the \tcorr
measurement. This causes prompt photons to fluctuate into the signal
time window.
An additional lifetime dependent correction term, $P_{\mathrm{corr}}$, is
 introduced to compensate for remaining small deviations in 
$A \cdot\epsilon$: 
\begin{equation}
  P_{\mathrm{corr}} = 1.04 - \frac{0.2}{55.0}\ \tauN - \frac{0.011}{0.06+(1-\tauN)^2}, 
\end{equation}
 where \tauN is in ns. This simple parameterization well characterizes the acceptance for any GMSB model to better than 4\% and gives us confidence that it can be of use to future model builders.

\begin{figure}[tp]
  \centering
  \vspace{-12pt}
  \subfigure[]{
    \includegraphics[scale=0.45]{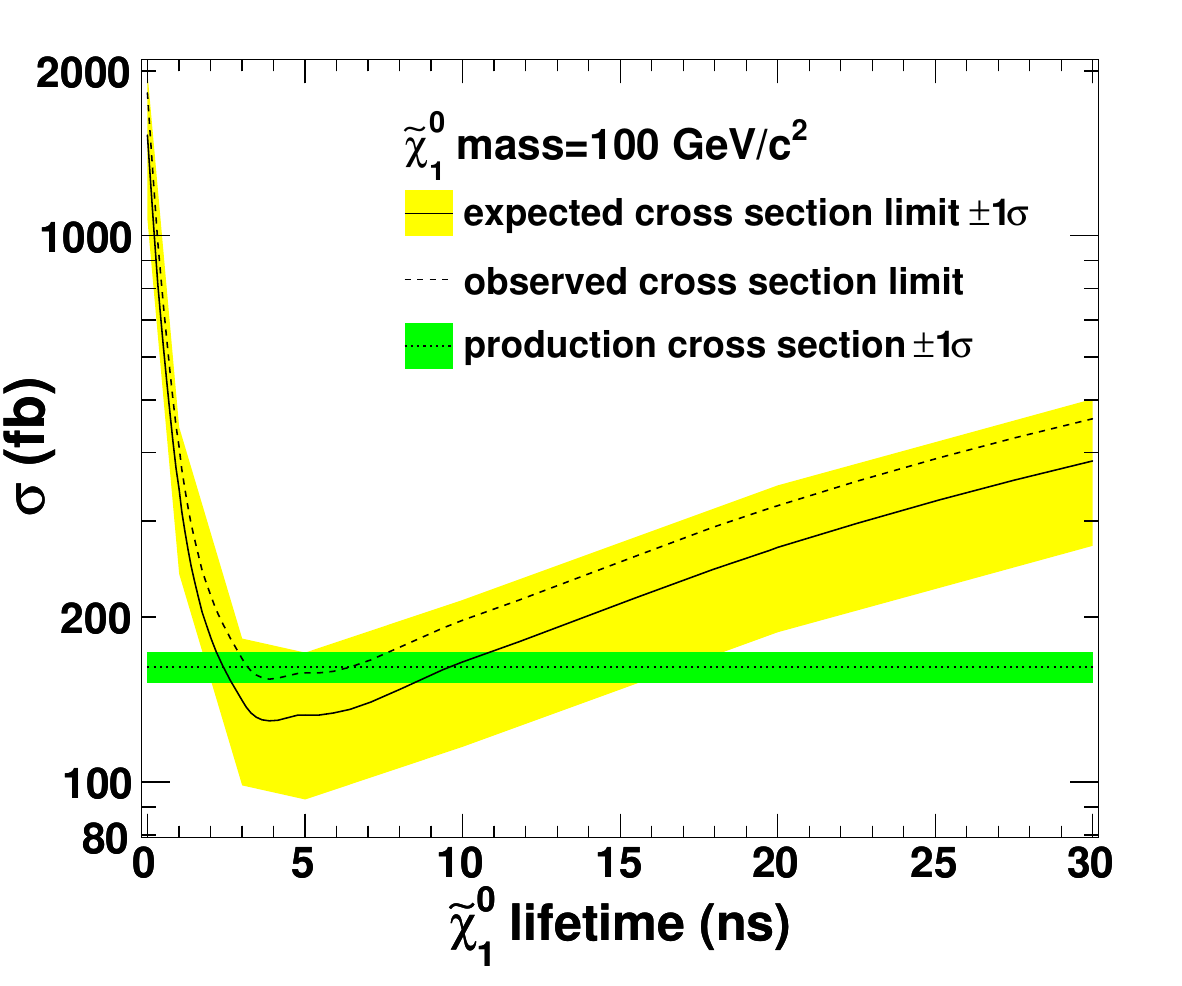}}\\
  \vspace{-19pt}
  \subfigure[]{
    \includegraphics[scale=0.45]{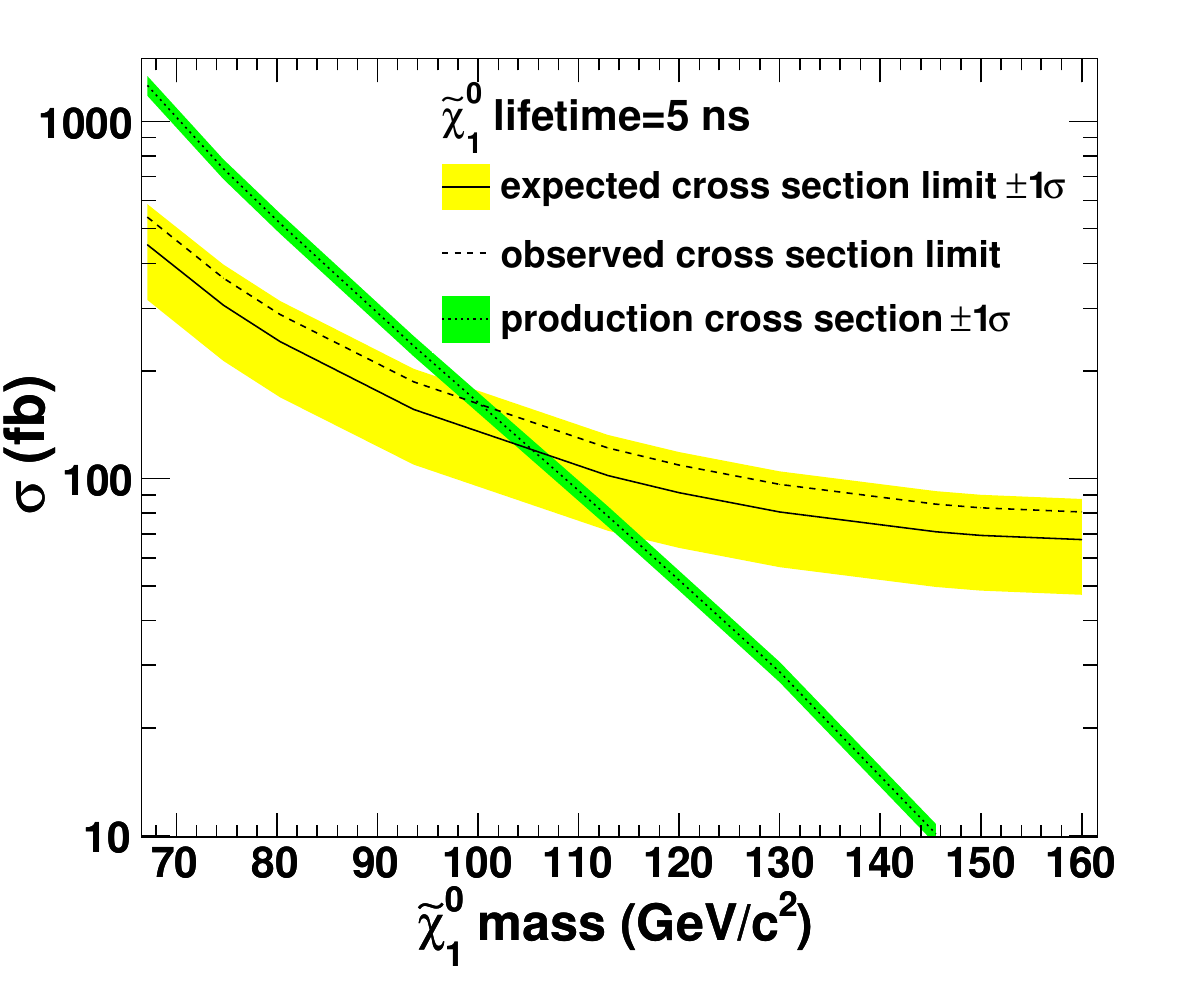}}
  \vspace{-14pt}
  \caption[The expected and observed cross section limits as a
  function of the \none lifetime at a mass of 100~\munit\ (a) and as a
  function of the \none mass at a lifetime of 5~ns
  (b).]{\label{fig:excl_1dim}\small The expected and observed cross
    section limits as a function of the \none lifetime at a mass of
    100~\munit\ (a) and as a function of the \none mass at a lifetime
    of 5~ns (b). Shaded green (darker shading) is the 6.5\%
    uncertainty-band for 
    the production cross section. 
    The yellow shaded region (lighter shading) is the variation in the
    expected limit due to 
    the statistical variation on the number of background events in the
    signal region ($\sim$30\%).}
\end{figure}

\begin{figure}[tb]
  \centering
  \vspace{-4pt}
  \hspace{20pt}
    \includegraphics 
[scale=.44]{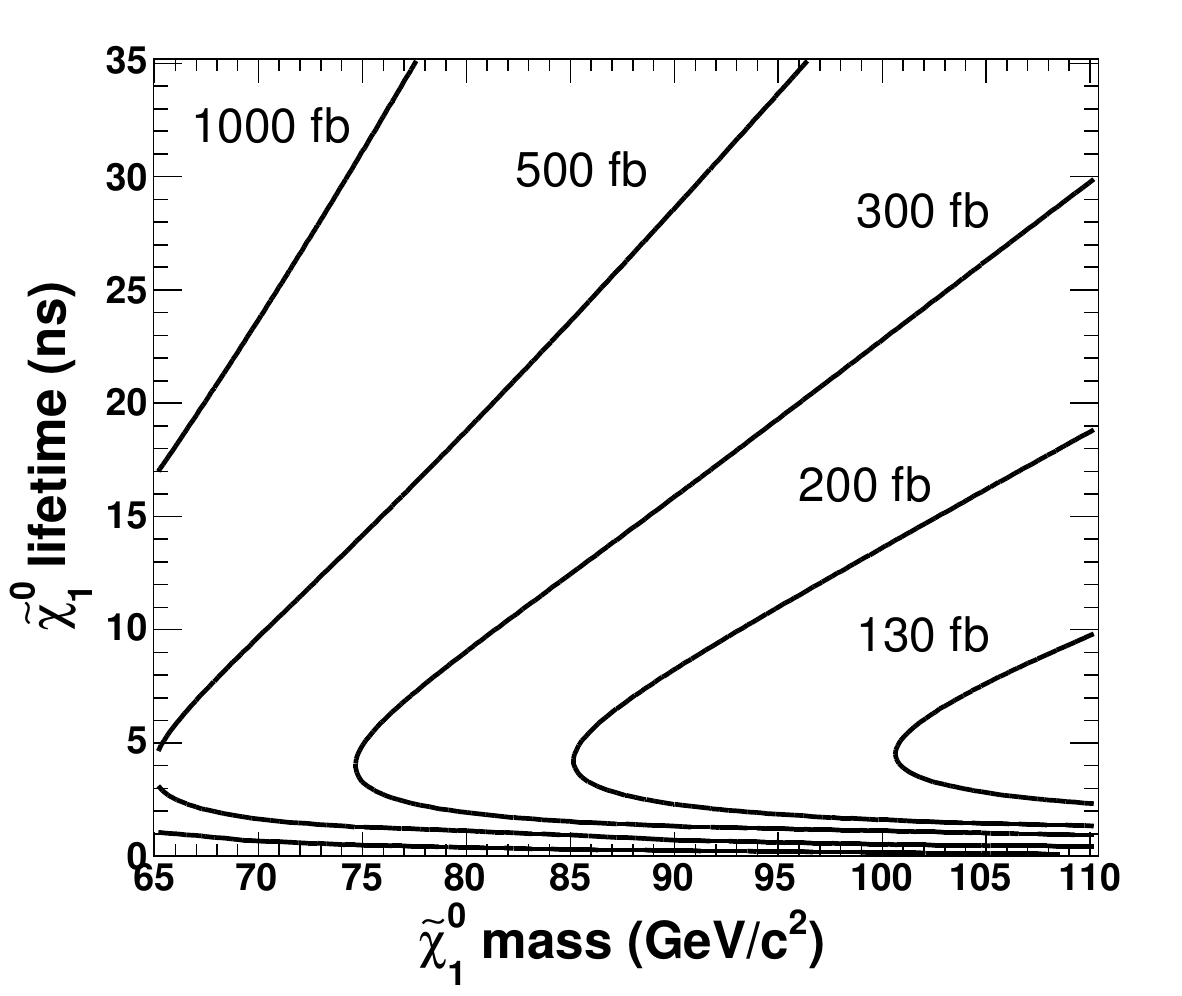}
    \caption[The contours of constant 95\% C.L. cross section upper limit 
      for the observed number of
      events.]{\label{fig:excl_cont}\small The contours of constant
      95\% C.L. cross section upper limit  
      for the observed number of
      events in the detector.}
\end{figure}

\begin{figure}[tb]
  \centering
  \hspace{12pt}
    \includegraphics
[scale=.43]
{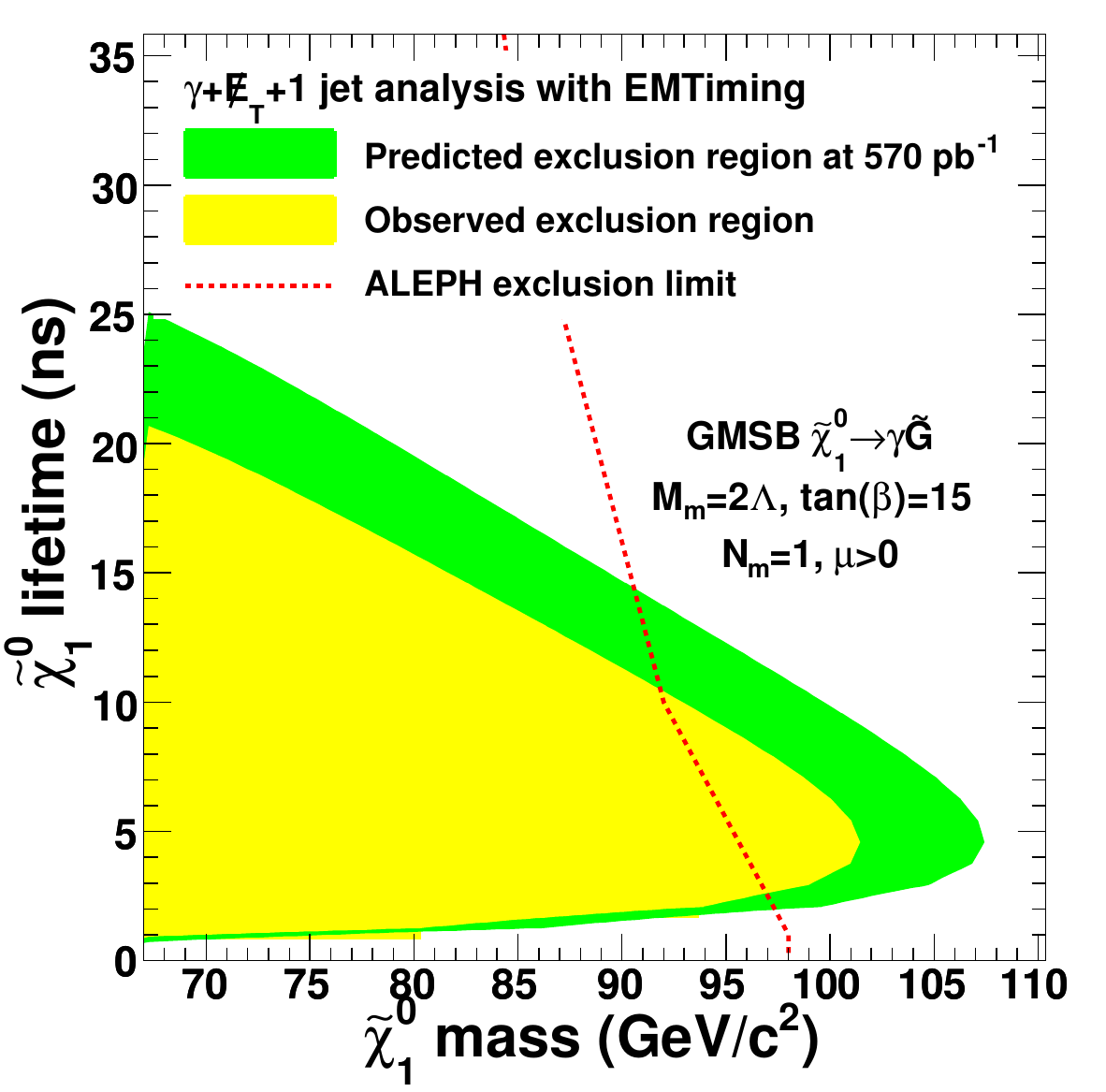}
    \caption[The expected and observed 95\% C.L. exclusion region using the
    interpolation function, along with the most stringent published
    LEP limits from ALEPH~\cite{lep}.]{\label{fig:excl_fit}\small The
      expected and observed 95\% C.L. exclusion region 
       along with the most stringent published LEP limits
      from ALEPH~\cite{lep}. The highest mass reach of 108~\munit
      (expected) and 101~\munit (observed) is achieved at a lifetime
      of 5~ns.}
\end{figure}

Figure~\ref{fig:excl_1dim} shows the expected and observed cross
section limits along with the NLO production cross section as a
function of \none lifetime at a mass of 100~\munit and as a function
of \none mass at a lifetime of 5~ns, close to the limit of the
expected sensitivity. Indicated  is the 6.5\% uncertainty-band
on the production cross section. The
band also shows the $\pm1\sigma$ statistical variations of the
expected cross section limit. 
Figure~\ref{fig:excl_cont} shows the contours of constant 95\% C.L.
cross section upper limit based on the two observed 
data events and  has its best sensitivity
 for lifetimes of $\sim$5~ns.
Figure~\ref{fig:excl_fit} shows the 95\% C.L.
exclusion region for \sigprod$>$~\sigexp and \sigprod$>\sigma^{\rm obs}_{95}$.
 Since the number of observed events is above expectations, 
 $\sigma^{\rm obs}_{95}$ is slightly larger than $\sigma^{\rm exp}_{95}$.
 The \none
mass reach, based on the expected (observed) number of events, is
108~\munit (101~\munit) at a lifetime of 5~ns. 
There is no exclusion of GMSB models with
\none lifetimes less than $\sim$1~ns as only few of the \none have a
long enough lifetime to produce delayed photons. However, most of the
parameter space there is already excluded by searches in
\ggmet~\cite{ggmet,lep}.  The large mass limits extend beyond
those of the LEP searches~\cite{lep} (using photon ``pointing'' methods)
and are currently the world's best.

\subsection{Future Prospects}

\begin{figure}[tb]
  \centering
  \includegraphics
[scale=0.43]
{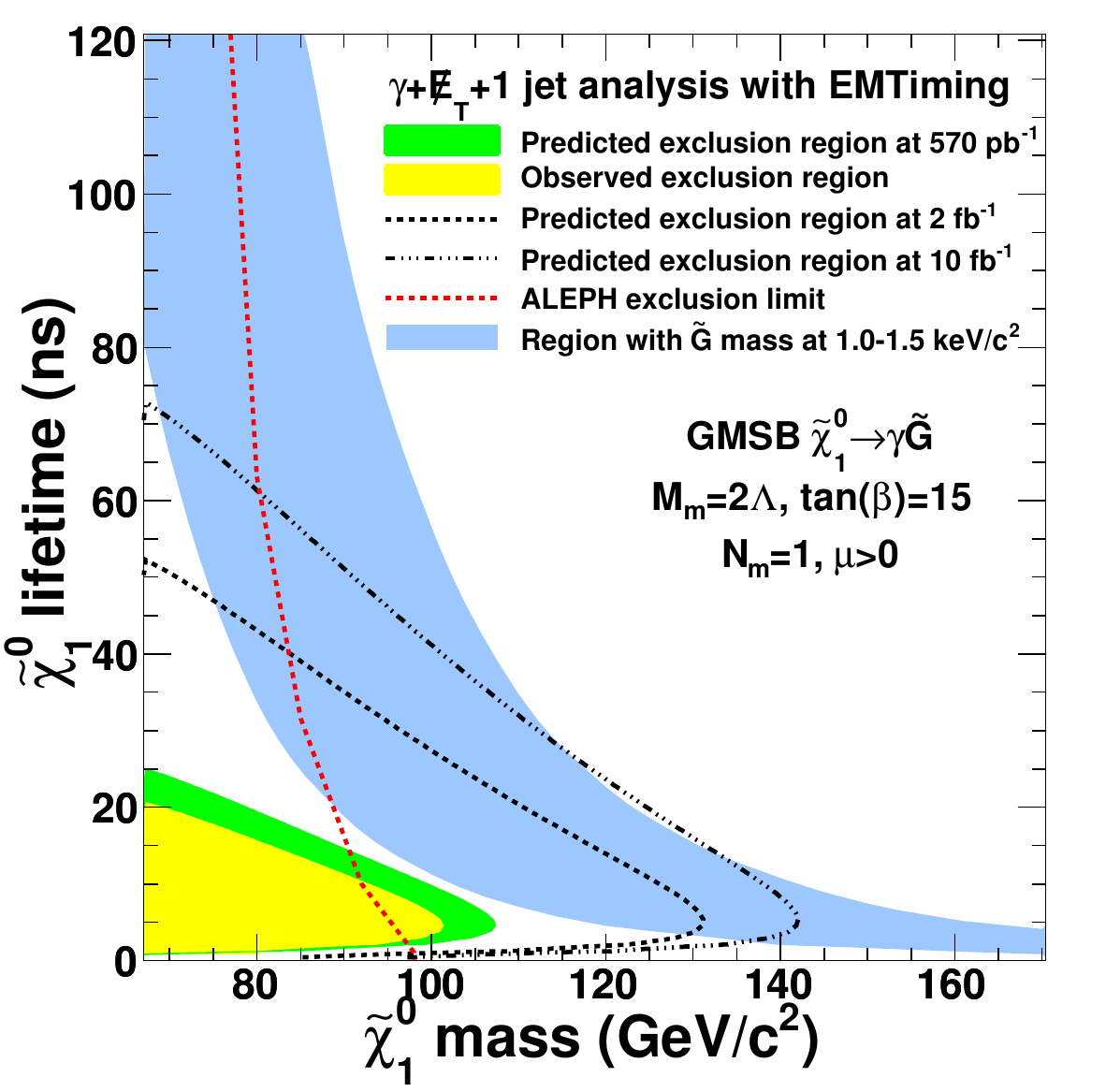}
    \caption[The dashed lines show the prediction of the exclusion
    limit after a scaling of the background prediction and the
    uncertainties for a luminosity of 2~\invfb and 10~\invfb
    respectively.]{\label{fig:excl_fit_lumi}\small 
      The expected 95\% C.L. exclusion region after a scaling of
      the background prediction and the uncertainties for a luminosity
      of 2~\invfb and 10~\invfb respectively. The shaded 
      band shows the
      parameter space where  
      $1~\leq m_{\grav}~\leq 1.5~\mkunit$. 
    }
\end{figure}

This search extends the exclusion region 
 close to the most important region 
of GMSB parameter space
where the \grav is predicted to be thermally produced in the early
universe with a mass of 1-1.5~\mkunit~\cite{baltz}. 
  With a higher luminosity this search
technique will be sensitive to this mass range. 
To investigate the prospects of such a search we calculate the
expected cross section limit assuming, for simplicity, 
that all backgrounds scale linearly 
with luminosity (the uncertainties remain a constant fraction of the 
 background).  While this assumption allows for a quick estimate, it
does not reflect the probable improvements in the background rejection
methods or the worsening effects due to the higher instantaneous
luminosity that could cause a higher fraction of background 
events with a wrong-vertex selection. As these effects would tend to
 balance each other, it can be considered to provide a reasonably 
balanced estimate. The resulting cross section limit
improvement, along with the expected 95\% C.L event limit, $N_{95}^{\rm
  exp}$, are shown in Table~\ref{tab:lumidetail} for 
our example point at \mN~=~100~\munit and \tauN~=~5~ns.
Figure~\ref{fig:excl_fit_lumi} shows the expected exclusion region for
a luminosity of 2 and 10~\invfb  
 along with the parameter space where $1\leq m_{\grav} \leq 1.5~\mkunit$. 
 The figure suggests that this search technique will be sensitive to
 all of this important parameter space at 10~\invfb luminosity for
 \none masses of less than $\sim$140~\munit and lifetimes of less than 30~ns.

\section{Conclusion}
\label{sec:conclusion}

We have presented a search for heavy, 
long-lived neutralinos that decay via $\gamma$\grav in a sample of \gmetjet
events from $p\bar{p}$ collisions at $\sqrt{s}=1.96$~TeV using the CDF~II 
detector. Candidate events were primarily selected based on the
delayed arrival time of the photon at the calorimeter as measured with
the newly installed EMTiming system. In 570~\invpb of data collected
during 2004-2005 at the Fermilab Tevatron, two events were observed,
consistent with the background estimate of 1.3$\pm$0.7 events. As
the search strategy does not rely on event properties specific to GMSB
models, any delayed \gmetjet signal (that passes our kinematic sample
cuts) is excluded at 95~\% C.L if it produces more than 5.5 events. This
result allows for setting both quasi model-independent cross section limits
and for an exclusion region of  GMSB 
models in the \none lifetime vs. mass plane, with a mass
reach of 101~\munit at \tauN~=~5~ns. These results extend the 
sensitivity to these models beyond those from LEP~II~\cite{lep} and are the world's best at masses $> 90$~\munit. 
By the end of Run~II, an integrated luminosity on the order of 10~\invfb\ 
might be collected, for which we estimate a mass reach of $\simeq
140$~\munit at a lifetime of 5~ns by scaling the expected number of background
events.

\begin{table}[tb]
  \begin{center}
    {\small \addtolength{\tabcolsep}{0.1em}
    \begin{tabular}{cccc}
      \hline \hline 
      Luminosity  & Expected  & Factor of & $N_{95}^{\rm exp}$ \\ (\invfb) &  Background & Improvement \\ & & on $\sigma_{\rm exp}$  \\\hline
      0.570 & \ \ \ \ \   1.3$\pm$0.7 (2) & 1 & \ \ \ \ \ \ \  4.6 (5.5)  \\
      2 & 4.3$\pm$2.3 & 0.46 & 7.4  \\
      10 & 21.9$\pm$11.6 & 0.0308 & 24.8  \\
 
     \hline \hline
    \end{tabular}
    }
    \vspace{1em}
    \caption[The expected sensitivity improvement for various
    luminosities for a 
    GMSB example point at \mN~=~100~\munit and \tauN~=~5~ns assuming
    all backgrounds and their uncertainty fractions scale linearly
    with luminosity.]{\label{tab:lumidetail}\small The expected
      search sensitivity 
      improvement for various luminosities for a GMSB example point at
      \mN~=~100~\munit and \tauN~=~5~ns assuming all backgrounds and
      their uncertainty fractions scale linearly with luminosity. 
      The numbers in parentheses reflect the observed values in this search. 
      The resulting exclusion region is shown in
      Fig.~\ref{fig:excl_fit_lumi}.}
  \end{center}  
\end{table}

\begin{acknowledgments}
We thank the Fermilab staff and the technical staffs of the
participating institutions for their vital contributions. This work
was supported by the U.S. Department of Energy and National Science
Foundation; the Italian Istituto Nazionale di Fisica Nucleare; the
Ministry of Education, Culture, Sports, Science and Technology of
Japan; the Natural Sciences and Engineering Research Council of
Canada; the National Science Council of the Republic of China; the
Swiss National Science Foundation; the A.P. Sloan Foundation; the
Bundesministerium f\"ur Bildung und Forschung, Germany; the Korean
Science and Engineering Foundation and the Korean Research Foundation;
the Science and Technology Facilities Council and the Royal Society,
UK; the Institut National de Physique Nucleaire et Physique des
Particules/CNRS; the Russian Foundation for Basic Research; the
Comisi\'on Interministerial de Ciencia y Tecnolog\'{\i}a, Spain; the
European Community's Human Potential Programme; the Slovak R\&D Agency;
and the Academy of Finland.  
\end{acknowledgments}


\begin{table*}[!t]
    {\small \addtolength{\tabcolsep}{0.2em}
    \begin{tabular}{ll}
      \hline \hline 
     
\multicolumn{2}{c}{Photon and Electron Identification Variables} \\ \hline

      Ficucial & $|X_{\mathrm{CES}}|<21$~cm and $9<|Z_{\mathrm{CES}}|<230$~cm for the calorimeter cluster centroid \\
      $E_{\mathrm{Had}}/E_{\mathrm{EM}}$ & The ratio of the energy deposited in the  hadronic calorimeter behind the cluster to the energy \\ & \ \ \ \ in the cluster as measured in the EM calorimeter  \\ 
      $E^{\mathrm{Iso}}$~(GeV) &    Energy in a cone of $\Delta$$R=\sqrt{\Delta\phi^{2}+\Delta\eta^{2}}=0.4$ around the   object, excluding the cluster energy \\
      Ntracks & Number of tracks pointing at photon cluster \\
      \sumpt (\punit) & Total $p_{T}$ of tracks in a cone of $\Delta$$R=0.4$ around  the cluster\\
      $E^{\mathrm{2^{nd} cluster}}$~(GeV) & Energy of a second EM cluster, if any, as identified in the shower-maximum detector  \\
      ${\rm A_{P}}$ & $\frac{|E_{\mathrm{PMT1}}-E_{\mathrm{PMT2}}|}{E_{\mathrm{PMT1}}+E_{\mathrm{PMT2}}}$ where $E_{\mathrm{PMT1}}$ and $E_{\mathrm{PMT2}}$ are the two PMT energies  \\
      $E/p$ & Ratio of the electron energy as measured in the  calorimeter to the momentum as  measured \\ & \ \ \ \  by the COT \\
      $\chi^{2}_{\mathrm{Strip}}$ & A $\chi^{2}$ comparison of the shower-maximum  profile to test beam data expectations \\ 
      $Lshr$ &  A comparison of the energy deposition of the electron, in adjacent towers, to expectations \\
      $\Delta x \cdot q$~(cm) & The comparison between the extrapolated   track position into the shower-maximum detector  \\ & \ \ \ \  and the measured cluster  centroid position, taking into account the track charge \\ 
       $z_{0}~(cm)$ & The measured $z$ position of the electron track along the beam line \\
      $\Delta z~(cm)$ & Difference between the measured electron track $z_{0}$ and the measured $z_{0}$ of the vertex \\
  
   \hline
       \hline
\multicolumn{2}{c}{Good Track Selection Variables} \\ \hline
      Slow Proton rejection & The ${dE}/{dx}$ for the track as it traverses the COT is not consistent with being from a \\ &  \ \ \ \ slow proton.  We require the timing variable to be less than 20~ns. For a description \\ & \ \ \ \   of the relationship between the timing measurement and ${dE}/{dx}$ see~\cite{slow_proton}. \\
      $z_{0}$~(cm) & The measured $z$ position of the track along the beam line \\
      $\mathrm{Err}(z_{0})$~(cm) & Uncertainty on the $z_{0}$ measurement \\ 
      $t_{0}$~(ns) &  The measured time of the track origin \\
      $\mathrm{Err}(t_{0})$~(ns) & Uncertainty on the $t_{0}$ measurement \\
      \hline \hline
    \end{tabular}
    }
    \vspace{1em}
    \caption[A list of all the ID variables used in this analysis.]
    {\label{tab:appendix}\small A description of the identification variables 
    used in this analysis for electrons, photons, and tracks.
}
\end{table*}

\appendix*

\section{ID Variables}
\label{app:appendix}

 In Table~\ref{tab:appendix} we provide a description of the
    identification variables  used in this analysis for electrons,
    photons, and tracks.

\vspace{15pt}


\centerline{REFERENCES} 

\vspace{-5pt}

\def\thebibliography#1{                                                               
\vskip .2truein                                                                       
                                                              
\bigskip                                                                              
\list    
  {[\arabic{enumi}]}{\settowidth\labelwidth{[#1]}\leftmargin\labelwidth               
    \advance\leftmargin\labelsep                                                      
    \usecounter{enumi}}                                                               
    \def\newblock{\hskip .11em plus .33em minus -.07em}                               
    \sloppy                                                                           
    \sfcode`\.=1000\relax}                                                            
                                                                                      
\let\endthebibliography=\endlist





\begin{thebibliography}{99}                                                     
\bibitem{gmsb2}   M.~Dine, and A.~Nelson, Phys.~Rev.~D
   \textbf{~48}, 1277 (1993);
  S.~Ambrosanio \etal, \Journal{\PRD}{54}{1996}{5395};
  C.~Chen and J.~Gunion, \Journal{\PRD}{58}{1998}{075005}. 
\bibitem{martin} S.~Martin, arXiv:hep-ph/9709356.
\bibitem{ref2} P.~Bode, J.~Ostriker and N.~Turok, Astrophys.~J.~556, 93 (2001).
\bibitem{eeggmet} F. Abe \etal\ (CDF Collaboration), 
  \Journal{\PRL}{81}{1998}{1791} and \Journal{\PRD}{59}{1999}{092002}. 
\bibitem{rapdef} In this paper we assume a cylindrical coordinate system that defines
  $z$ as the longitudinal axis along the proton beam, 
  $\theta$ as the polar angle, $\phi$ as the azimuthal angle relative to
  the horizontal plane and
  $\eta=-\ln \tan(\theta/2)$. 
  We take $E_T=E\sin\theta$ and $p_T=p \sin \theta$ where $E$ is the energy
  measured by the calorimeter and $p$ the momentum measured in the
  tracking system.  If no vertex is reconstructed we use $z_{\rm collision}=0$.
  We define $\vec{\met}=-\sum_i E_T^i \vec{n}_{i}$ where
  $\vec{n}_{i}$ is a unit vector in the transverse plane that points 
  from the interaction vertex
  to the $i^{\rm th}$ calorimeter tower.  $\met$ is
  the magnitude of $\vec{\met}$.                                              
\bibitem{feng2}   J.~Feng, S.~Su, and F.~Takayama, Phys.~Rev. D\textbf{~70}, 075019 (2004), and
  references therein.
\bibitem{pwt}  A.~Abdulencia \etal\ (CDF Collaboration), 
  \Journal{\PRL}{99}{2007}{121801}; 
  P.~Wagner, Ph.D. Thesis, Texas A\&M University,  
  FERMILAB-THESIS-2007-14.
\bibitem{CDFII} T.~Aaltonen \etal (CDF Collaboration), 
   arXiv:hep-ex/0708.3642v1, accepted for publication in \PRD.
\bibitem{ggmet} D.~Acosta \etal\ (CDF~Collaboration),
  \Journal{\PRD}{71}{2005}{031104}. V.~Abazov \etal (D0 Collaboration),
  \Journal{\PRL}{94}{2005}{041801}.

\bibitem{lep} A. Heister \etal\ (ALEPH Collaboration), Eur.~Phys.~J.
  C\textbf{~25}, 339 (2002); also see M.~Gataullin \etal, for the 
  L3~Collaboration, arXiv:hep-ex/0611010; G.~Abbiendi \etal\
  (OPAL~Collaboration), Proc.~Sci. HEP2005 346
  (2006); 
  J.~Abdallah \etal\ (DELPHI~Collaboration), Eur.~Phys.~J.
  C\textbf{~38} 395 (2005).



\bibitem{nlsp} J.~L.~Feng and T.~Moroi, 
  Phys.~Rev. D\textbf{~58}, 035001 (1998).
\bibitem{snowmass} B.~Allanach \emph{et al.}, Eur.~Phys.~J.
  C\textbf{~25}, 113 (2002).
\bibitem{baltz} E.~Baltz \etal, J. High Energy Phys.\textbf{~0305},
  067 (2003). 
\bibitem{prospects} P.~Wagner and D.~Toback, Phys.~Rev.~D\textbf{~70},
  114032 (2004).
\bibitem{jets} For a description of how clusters of energy, for example  
   $\tau^\pm\to\pi^\pm\pi^\mp\pi^\pm\nu_\tau$ decays, are identified as 
   jets, see 
   F.~Abe \etal (CDF Collaboration),
  \Journal{\PRL}{68}{1992}{1104}.  For a discussion of the jet energy
   measurements, see 
   T.~Affolder \etal (CDF Collaboration),
  \Journal{\PRD}{64}{2001}{032001}.   For a discussion of standard jet 
  correction systematics, see A.~Bhatti \etal, 
   Nucl.~Instrum.~Methods, A 566, 375 (2006).  We use jets with 
   cone size $\Delta$$R=0.7$.
\bibitem{nim} M.~Goncharov \emph{et al.}, Nucl.~Instrum.~Methods
  Phys.~Res., Sect. A\textbf{~565}, 543 (2006).
\bibitem{photoncuts} Y.~Liu, Ph.D. Thesis, Universit\'e de Gen\`eve,
  FERMILAB-THESIS-2004-37.
\bibitem{pythia} T.~Sj\"{o}strand \etal, \Journal{Comput. Phys.
    Commun.}{135}{2001}{238}. We use version 6.216.
\bibitem{alpha_explain} The incident
  angle \ag variation is due to the vertex position $z_0$ variation. 
  The calorimeter towers point to the
  center of the detector, so the angle extends out to
  $\sim$$18^{\circ}\approx \arctan(\frac{60\mathrm{\ cm}}{183\mathrm{\ cm}})$,
  where 60~cm is the maximal \zzero variation and 183~cm is 
  the radius from the beamline to the photon shower maximum in the detector.
\bibitem{ref9}  R.~Brun \etal, CERN-DD/EE/84-1 (1987). The EMTiming
  system is done separately as a parametrized simulation. 
\bibitem{slow_proton} D. Tonelli, Ph.D. Thesis, Scuola Normale Superiore, Pisa,
  FERMILAB-THESIS-2006-23.
\bibitem{vertalg} D.~Acosta \etal (CDF Collaboration), 
  \Journal{\PRD}{71}{2005}{052003}.
\bibitem{vertex} Note that throughout this section a
  ``vertex'' refers to the true underlying collision, whereas the term
  ``cluster'' refers to the clustered set of tracks that constitute the
  reconstructed vertex. 
\bibitem{emax} G.~McLachlan and D.~Peel, Wiley-Interscience,
  ISBN~0471006262. 
\bibitem{onyisi}  D.~Acosta \etal (CDF Collaboration), \Journal{\PRL}{89}{2002}{281801}. 
\bibitem{note}  We note that this occurs in 
  $\sim$14\% of the cases for this sample.  While this is an
  interesting number, the 
  fraction of events where the wrong vertex is picked is very sample
  dependent as it depends only on the probability that the
  highest-\sumpt vertex in the event is the same vertex that produces
  the photon. 
  For the luminosity for the data taking period there are, on average,
  between 0.4 and 4.4 min-bias collisions per event. As a comparison 
  SM $Z\gamma\rightarrow\nu\nu\gamma$ production produces   
  very few tracks in association with the photon so there is a 
  lower probability that the highest-\sumpt vertex produces the photon
  at high luminosity. As a second example, in
  $t\bar{t}\gamma\rightarrow\gamma+{\rm jets}$ events the photon is very
  likely to come from the highest-\sumpt vertex.
\bibitem{isajet} F.~Paige and S.~Protopopescu, BNL Report BNL38034,
  1986; F.~Paige, S.~Protopopescu, H.~Baer and X.~Tata,
  hep-ph/0001086. We used version 7.64 to generate the SUSY masses.
\bibitem{WZcross} A.~Abulencia \etal\ (CDF Collaboration), 
  J.~Phys.~G.\textbf{~34}, 2457 (2007); J.~Pumplin {\it et al.}, J. High Energy
  Phys.\textbf{~0207}, 012 (2002); J.~Huston {\it et al.}, J. High
  Energy Phys.\textbf{~0310}, 046 (2003). 
\bibitem{simeon} P.~Simeon and D.~Toback, J. of Undergrad. Research in
  Phys.\textbf{~20}, (2007).  
\bibitem{multiple_collisions} Multiple collisions in the event can produce extra vertices, one of
which can be picked incorrectly as the event vertex and cause \tcorr to be systematically mismeasured. To take this
effect into account in the acceptance calculation, an extra vertex is
simulated for each event with a $z_{0}$ and $t_{0}$ that are randomly
selected from Gaussians with $\sigma_{z}=30$~cm and
$\sigma_{t}=1.3$~ns.
To estimate the impact on the acceptance from a variation of the wrong-vertex fraction depending on the event requirements, the wrong-vertex
fraction is varied between 0\% and 10\%, obtaining an acceptance variation of
$<1.5\%$.

\bibitem{prospino} See  W.~Beenakker \etal, Nucl.~Phys.
  B\textbf{~492}, 51 (1997); We use \prospino.0.
\bibitem{limitcalc_cdf} J.~Conway, CERN Yellow Book
  Report No. CERN 2000-005 (2000), p.~247.
\bibitem{limitcalc} E.~Boos, A.~Vologdin, D.~Toback and J.~Gaspard,
  Phys.~Rev. D\textbf{~66}, 013011 (2002). 
\bibitem{kfactors} W.~Beenakker \etal, Phys.~Rev.~Lett.\textbf{~83},
  3780 (1999).
\bibitem{addfeng} J.~L.~Feng, A.~Rajaraman and F.~Takayama,
  \Journal{\PRD}{68}{2003}{063504};
  M.~J.~Strassler and K.~M.~Zurek, arXiv:hep-ph/0605193.
\end{thebibliography}
\end{document}